\documentclass{dune}
\pdfoutput=1

\input{common/preamble}

\begin{document}

\pagestyle{titlepage}


\pagestyle{titlepage}

\date{}

\title{
\rightline{\small DESY-22-112} \vspace{1cm}
\scshape\Large Dark Sector Studies with Neutrino Beams \\
\normalsize NF03 Contributed White Paper to Snowmass 2021
\vskip -10pt
}


\renewcommand\Authfont{\scshape\small}
\renewcommand\Affilfont{\itshape\footnotesize}

\author[1]{Brian~Batell$^{\thanks{Lead Editors}}$}
\author[2]{Joshua~Berger}
\author[3,4]{Vedran~Brdar}
\author[3]{Alan~D.~Bross}
\author[5]{Janet M. Conrad}
\author[6]{Patrick~deNiverville}
\author[7]{Valentina~De~Romeri}
\author[8]{Bhaskar~Dutta}
\author[9]{Saeid Foroughi-Abari}
\author[10,11,12]{Matheus~Hostert}
\author[3]{Joshua Isaacson}
\author[13]{Ahmed~Ismail}
\author[14]{Sudip~Jana}
\author[15]{Wooyoung~Jang}
\author[5]{Nicholas W. Kamp}
\author[16]{Kevin~J.~Kelly}
\author[8]{Doojin~Kim}
\author[17]{Felix~Kling}
\author[18]{Mathieu~Lamoureux}
\author[19]{David~McKeen}
\author[20]{Jong-Chul~Park}
\author[21]{Gianluca Petrillo}
\author[9]{Adam~Ritz}
\author[22]{Seodong~Shin}
\author[23]{Tyler~B.~Smith}
\author[24,25]{Sebastian~Trojanowski}
\author[23]{Yu-Dai~Tsai}
\author[21]{Yun-Tse~Tsai}
\author[6]{Richard~Van~de~Water}
\author[26]{Jason~Wyenberg}
\author[27]{Guang~Yang}
\author[* 15]{Jaehoon~Yu}

\vspace{-0.5cm}
\affil[1]{PITT PACC, Department of Physics and Astronomy, University of Pittsburgh, Pittsburgh, PA 15260, USA}
\affil[2]{Department of Physics, Colorado State University, Fort Collins, Colorado 80523, USA}
\affil[3]{Fermi National Accelerator Laboratory, Batavia, IL 60510, USA}
\affil[4]{Northwestern University, Department of Physics and Astronomy, Evanston, IL 60208, USA}
\affil[5]{Massachusetts Institute of Technology, Cambridge, MA, USA}
\affil[6]{Los Alamos National Laboratory, Los Alamos, NM 87545, USA}
\affil[7]{Instituto de F\'{i}sica Corpuscular, CSIC-Universitat de Val\`{e}ncia, 46980 Paterna, Spain}
\affil[8]{Mitchell Institute for Fundamental Physics and Astronomy, Department of Physics and Astronomy, Texas A\&M University, College Station, TX 77843, USA}
\affil[9]{Department of Physics and Astronomy, University of Victoria, Victoria, BC V8P 5C2, Canada}
\affil[10]{Perimeter Institute for Theoretical Physics, Waterloo, ON N2J 2W9, Canada}
\affil[11]{School of Physics and Astronomy, University of Minnesota, Minneapolis, MN 55455, USA}
\affil[12]{William I. Fine Theoretical Physics Institute, School of Physics and Astronomy, University of Minnesota, Minneapolis, MN 55455, USA}
\affil[13]{Department of Physics, Oklahoma State University, Stillwater, OK 74078, USA}
\affil[14]{Max-Planck-Institut f\"{u}r Kernphysik, Saupfercheckweg 1, 69117 Heidelberg, Germany}
\affil[15]{Department of Physics, University of Texas, Arlington, TX 76019, USA}
\affil[16]{CERN, Theoretical Physics Department, Geneva, Switzerland}
\affil[17]{Deutsches Elektronen-Synchrotron DESY, Notkestr. 85, 22607 Hamburg, Germany}
\affil[18]{Dipartimento di Fisica, INFN Sezione di Padova and Universitá di Padova, I-35131, Padova, Italy}
\affil[19]{TRIUMF, 4004 Wesbrook Mall, Vancouver, BC V6T 2A3, Canada}
\affil[20]{Department of Physics, Chungnam National University, Daejeon 34134, Republic of Korea}
\affil[21]{SLAC National Accelerator Laboratory, Menlo Park, CA, 94025, USA}
\affil[22]{Department of Physics, Jeonbuk National University, Jeonrabuk-do 54896, Republic of Korea}
\affil[23]{Department of Physics and Astronomy, University of California, Irvine, CA 92697-4575, USA}
\affil[24]{Astrocent, Nicolaus Copernicus Astronomical Center Polish Academy of Sciences, ul. Rektorska 4, 00-614, Warsaw, Poland}
\affil[25]{National Centre for Nuclear Research, ul. Pasteura 7, 02-093 Warsaw, Poland}
\affil[26]{Department of Physics and Astronomy, Dordt University, Sioux Center, IA 51250, USA}
\affil[27]{Department of Physics, University of California, Berkeley, California 94720, USA}

\vspace{2cm}

\maketitle

\vspace{2cm}

\begin{abstract}
An array of powerful neutrino-beam experiments will study the fundamental properties of neutrinos with unprecedented precision in the coming years. Along with their primary neutrino-physics motivations, there has been growing recognition that these experiments can carry out a rich program of searches for new, light, weakly-coupled particles that are part of a dark sector. In this white paper, we review the diverse theoretical motivations for dark sectors and the capabilities of neutrino beam experiments to probe a wide range of models and signatures. We also examine the potential obstacles that could limit these prospects and identify concrete steps needed to realize an impactful dark sector search program in this and coming decades.
\end{abstract}

\vspace{2cm}

\begin{center}\rule[-0.2in]{\hsize}{0.01in}\\\rule{\hsize}{0.01in}\\
\vskip 0.1in Submitted to the  Proceedings of the US Community Study\\ 
on the Future of Particle Physics (Snowmass 2021)\\ 
\rule{\hsize}{0.01in}\\\rule[+0.2in]{\hsize}{0.01in} \end{center}

\pagestyle{plain} 
\clearpage

\vspace{1cm}

\vspace{1cm}

\clearpage

\textsf{\tableofcontents}




\clearpage


\pagestyle{fancy}

\fancyhead{}
\fancyhead[RO]{\textsf{\footnotesize \thepage}}
\fancyhead[LO]{\textsf{\footnotesize \rightmark}}

\fancyfoot{}
\fancyfoot[RO]{\textsf{\footnotesize Snowmass 2021}}
\fancyfoot[LO]{\textsf{\footnotesize Dark Sector Studies with Neutrino Beams}}
\fancypagestyle{plain}{}

\renewcommand{\headrule}{\vspace{-4mm}\color[gray]{0.5}{\rule{\headwidth}{0.5pt}}}



\clearpage

\section*{Executive Summary}
\addcontentsline{toc}{section}{Executive Summary}
\label{sec:summary}


In this whitepaper, we survey the promising opportunities and remaining challenges for a rich program of dark sector searches at neutrino beam experiments in the next decade and beyond.  We review the diverse theoretical motivations for dark sectors, highlight the unique advantages and complementary role of neutrino beam facilities for dark sector searches, discuss the current and future experimental landscape, present the novel signatures and future experimental prospects for a variety of dark sector models, and consider the simulation and analysis tools needed to realize a robust search program. 

The idea of a dark sector (or hidden sector, secluded sector, etc.) comprising new states that are not charged under the known strong, weak, and electromagnetic forces, yet are weakly coupled to the Standard Model via a  portal interaction, is well motivated from a variety of perspectives. Dark sectors may provide novel answers to some of the big open questions in particles physics, including explaining 
dark matter and 
neutrino masses, among others. The portal concept, based on straightforward effective field theory reasoning, provides a systematic framework for the theoretical and experimental investigation of dark sectors. Furthermore, a number of creative dark sector models have been proposed to explain a variety of experimental anomalies.

An expansive experimental program is emerging to explore the dark sector, and neutrino beam experiments have a critical role to play in these investigations. Modern accelerator-based neutrino beam facilities feature enormous proton beam-target collision luminosities, which can supply copious secondary forward fluxes of dark sector particles. Contemporary neutrino detectors benefit from large active masses and volumes, excellent particle identification and reconstruction capabilities, and capacities for precision energy, spatial, and timing measurements, which can be leveraged to detect a variety of rare dark sector signals and distinguish them from beam-related and cosmic backgrounds. These experiments are particularly well suited to the study of hadrophilic and neutrinophilic dark sector interactions. The coming decade and beyond promises to be an exciting era for dark sector research, with a number of neutrino beam experiments already in operation and several ambitious planned projects on the horizon. 

The past decade has witnessed intense theoretical exploration of dark sector models and their phenomenology, including the novel signatures and promising search prospects at neutrino beam experiments. We review the status of a broad range of dark sector models, including scenarios featuring the vector portal, Higgs portal, neutrino portal, axion-like-particle (ALP) portal, dark neutrinos, and neutrinophilic interactions.  Collectively, these theoretical scenarios motivate a broad suite of searches, including  long-lived particle decays to a variety of visible final states, elastic and/or inelastic scattering with detector electrons or nuclei, neutrino up-scattering to dark neutrinos followed by visible decays, and modifications to neutrino scattering processes. As we highlight in specific case studies, current and future neutrino experiments will be able to probe large regions of uncharted parameter space and complement the energy frontier experiments. 

An effective and robust dark sector search program necessitates accurate simulation tools for both the myriad dark sector particle production channels and the rich array of detectable signatures. Several challenges must be met in the development of these tools, including modeling the complex target geometry and/or focusing horn, an accounting of nuclear physics effects in production and detection, the capability for fast detector simulation, and the identification/reconstruction of unique signal topologies, to name a few. 
Thus far, phenomenological studies have utilized a combination of publicly available event generators and home-grown codes to simulate signals and backgrounds, design mock analyses, and derive sensitivity estimates. 
However, in most cases this approach is not suitable for experimental analyses, and the development of packages that can be readily integrated into the existing simulation frameworks used by the collaborations is one key direction that calls for immediate effort. Furthermore, the development of novel reconstruction and analyses methods, perhaps including the use of modern machine learning methods, is another important arena where improvement can be anticipated. An action plan is presented containing recommendations for overcoming the numerous challenges in developing the simulation frameworks required to realize an impactful program of dark sector searches.

The search for dark sector particles at neutrino beam experiments
requires an accurate and precise understanding of the backgrounds induced from neutrino-nucleus interactions. It is therefore imperative to make improvements in
neutrino-nucleus interaction models using experimental measurements as input and reflect these refinements in the simulation tools. 
To accomplish this, an effective way of collaborating with the nuclear physics community must be sought and implemented in a timely fashion. 
Finally, in consideration of future neutrino beam experiments, it is evident that a thriving and powerful search program relies on a well-equipped near detector complex to leverage the full beam power and maximize the experimental sensitivity to feeble dark sector  interactions. 

Neutrino beam experiments represent an important front in the quest to explore the dark sector. Experiments currently in operation and those coming online over the next decade hold the promise to significantly advance these studies, yet there is still much important work to be done to realize their full physics potential, perhaps most notably in the development of robust and versatile simulation tools and improved neutrino-nucleus interaction modeling. Given the exciting array of opportunities outlined in this whitepaper, dark sector particle searches will form a vital part of the broader physics program at existing and future neutrino beam facilities and complement the experiments at the energy frontier.


\cleardoublepage

\section{Introduction
}
\label{sec:introduction}


Over the past decade or more, the study of dark sectors -- theories containing new elementary particles that do not experience the familiar electromagnetic, weak, and strong interactions -- has become a main line of physics beyond the Standard Model (BSM) research. There are a number of motivations underlying this development. First of all, a dark sector may provide a path towards cracking some of the outstanding puzzles in particle physics and cosmology. For instance, dark matter (DM) may reside in a dark sector, or the origin of neutrino mass may be linked to new neutral fermions within a dark sector. 
Secondly, dark sector models predict a rich variety of novel experimental and observational phenomena, which has inspired the development of innovative search strategies at existing or planned experiments as well entirely new experiments. 
In a related direction, numerous dark sector models have been developed as possible explanations for experimental anomalies. These include the muon anomalous magnetic moment, the MiniBooNE low energy excess, various excesses in astrophysical observations, among numerous others. Last but not least, another impetus comes from general effective field theory logic that motivates a systematic exploration of dark sector interactions through the so-called renormalizable and higher-dimensional portals, which connect gauge invariant operators of the Standard Model (SM) and the dark sector. 

It is clear that the dark sector paradigm is extremely rich both from a theoretical point of view and a phenomenological one. On the one hand, the dark sector could be very minimal, with only a few new particles, and many studies in the literature operate under this assumption. However, it is also quite plausible that the dark sector is as complex as the Standard Model, with a zoo of matter particles, Abelian and/or non-Abelian dark forces, confinement, spontaneous symmetry breaking, and so on. It is therefore evident that a wide array of experimental approaches must be pursued in order to probe the full range of phenomena possible within this framework. These include searches using accelerator- and reactor-based neutrino beam experiments, high-luminosity medium-energy $e^+e^-$ colliders, electron and muon beam fixed-target missing energy/momentum experiments, new low mass dark matter direct detection experiments, rare particle decay experiments, the experimenta at high-energy colliders such as the LHC, astrophysical observatories, among numerous others. Collectively, these experiments will provide powerful sensitivity to a broad range of dark sector scenarios in the MeV-GeV mass range~\cite{Alexander:2016aln,Battaglieri:2017aum,Beacham:2019nyx,BRNreport}.

Our goal in this white paper is to advance the special role played by accelerator-based neutrino beam experiments in dark sector studies. These experiments generically entail high-intensity proton beam collisions with a fixed target/beam dump coupled to a sensitive, large mass near detector positioned downstream of the collisions. The merits afforded by these experiments to neutrino studies naturally extend to the search for dark sector particles.  
These include the substantial collision luminosities and forward lab-frame kinematics intrinsic to proton fixed target experiments, creating of a ``beam'' of dark particles pointed towards the detector, in an analogous fashion to the neutrino beam. Being weakly coupled to ordinary matter, these dark sector particles can, like neutrinos, penetrate any intermediate shielding or dirt, enter the detector, and leave a distinctive signature through their decays or scattering. Alternatively, neutrinos may directly experience interactions with dark sector particles, which can leave their footprint through novel secondary interactions of the neutrino beam in the detector.  
Modern neutrino detectors enjoy excellent particle identification and reconstruction capabilities which can be exploited to characterize and discriminate dark sector signals from beam-related neutrino backgrounds, as well as those of cosmic origin. 

A number of analyses from past neutrino beam experiments have been reinterpreted to derive strong limits on various dark sector models. Furthermore, several modern neutrino beam experiments have recently produced world-leading experimental limits from dedicated searches for a variety of well-motivated dark sector particles, including light DM~\cite{Aguilar-Arevalo:2017mqx,MiniBooNEDM:2018cxm}, heavy neutral leptons~\cite{MicroBooNE:2019izn,ArgoNeuT:2021clc}, and Higgs portal scalars~\cite{MicroBooNE:2019izn,ArgoNeuT:2021clc,MicroBooNE:2021usw}. These studies demonstrate the efficacy of the basic experimental approach and motivate a comprehensive program of dark sector particle searches at neutrino beam experiments in the kinematic regime complimentary to the energy frontier experiments.
With several existing experiments already taking data, along with a number of near-term and future experiments on the horizon, the stage is set for a broad experimental exploration of dark sectors at neutrino beam experiments in the coming decade and beyond.

This whitepaper is structured as follows. We begin in Section~\ref{sec:theory}
 with a discussion of the broad theoretical motivations underlying the dark sector paradigm. Section~\ref{sec:advantages}
 highlights the advantages of neutrino beam experiments in the search for dark sector particles, while Section~\ref{sec:landscape}
 provides an overview of the exciting experimental landscape, both in terms of the current and planned near-term/future neutrino beam experiments. Section~\ref{sec:models}
 provides an in-depth examination of specific portals and dark sector models, including the Higgs portal, vector portal, neutrino portal, dark neutrinos/dipole portal, and $\nu$-philic interactions. 
Section~\ref{sec:tools} provides an overview of the simulation tools that are currently available, as well as the developments needed to maximize the physics output of the experiments. Finally, our conclusions and outlook are presented in Section~\ref{sec:outlook}.

\section{Theory Overview and Motivation 
} 
\label{sec:theory}


\subsection{Effective Field Theories and Portals 
} \label{subsec:Theory:Portals}
Nearly a decade on from the discovery of the Higgs boson at the LHC, arguably the dominant empirical motivations for new physics beyond the Standard Model (SM) concern neutrino mass and dark matter. These motivations have a common feature that their phenomenology points primarily to weak coupling, but not to a specific mass scale for new physics. This allows for a vast model-building parameter space, but systematic progress can be made on assuming that the explanation lies in a dark (or hidden) sector weakly coupled to the Standard Model. Effective field theory arguments can then be applied to systematize the exploration of the couplings between the SM and a potentially highly-complex dark sector. Fixed target high-luminosity experiments producing neutrino beams are well-suited to this task, and efforts have been renewed in this direction over the past decade. 

The dark sector framework can be characterized via the following Lagrangian,
\begin{equation}
 {\cal L} = {\cal L}_{\rm SM} + {\cal L}_{\rm DS} + \sum_{d=i+j} \frac{1}{\Lambda^{d-4}} {\cal O}_i^{\rm SM} {\cal O}_j^{\rm DS}, \label{L}
\end{equation}
where ${\cal L}_{\rm SM}$ and ${\cal L}_{\rm DS}$ refer to the Standard Model and dark sector respectively. By construction, interaction of dark sector degrees of freedom with the SM is only via SM-neutral operators ${\cal O}^{\rm DS}_j$ of dimension $j$ built from dark sector fields. $\Lambda$ characterizes the lowest mass scale at which degrees of freedom charged under both sectors can generate these interactions. The framework implicitly allows this scale to be very high, in which case attention is focused on the privileged interactions that are relevant or marginal, known as `portals', for which the couplings are not power-suppressed by the large mass scale,
\begin{equation}
\sum_{d=i+j} \frac{1}{\Lambda^{d-4}} {\cal O}_i^{\rm SM} {\cal O}_j^{\rm DS} = {\cal L}_{\rm portals} + {\cal O}(1/\Lambda) .
\end{equation}
Under the assumption that the dark sector operators are truly SM-neutral, as is now well known, there are only three UV-complete interactions of this type that are consistent with the SM electroweak symmetry-breaking structure. All require the presence of new light degrees of freedom. These renormalizable portals for the SM are characterized as follows \cite{Holdom:1985ag,Patt:2006fw,Minkowski:1977sc,Okun:1982xi}:
\begin{align}
 {\cal L}_{\rm neutrino\, portal}^{d=4} &= -\sum y_\nu^{\alpha I} (\bar{L}_\alpha H)N_I \longrightarrow -\frac{1}{\sqrt{2}} \sum vy_\nu^{\alpha I}  \bar{\nu}_\alpha N_I + \cdots \\
  {\cal L}_{\rm Higgs\, portal}^{d=3,4} &= - (\mu S + \lambda S^2)H^\dagger H \longrightarrow - \frac{\mu v}{m_h^2-m_S^2} S J_h + \cdots \\
 {\cal L}_{\rm vector\, portal}^{d=4} &= - \frac{\epsilon}{2\cos\theta_W} B_{\mu\nu} F'_{\mu\nu} \longrightarrow  \epsilon e A'_\mu J_{\rm EM}^\mu +\cdots \,.
\end{align}
These portals introduce new degrees of freedom in the form of dark neutral leptons (or right-handed neutrinos) $N_I$, dark singlet scalars $S$, and a dark photon $A'_\mu$ that can gain a mass via a hidden sector Higgs or Stueckelberg mechanism. The first two portals involve the SM Higgs doublet in a nontrivial manner, and below the scale of electroweak symmetry breaking where the Higgs field acquires a vev $v=246$~GeV, the arrows indicate the low energy phenomenology respectively in the form of neutrino mixing (neutrino portal) and  Higgs-scalar mixing (Higgs portal) so that $S$ couples to the scalar operator $J_h$ sourced by the physical Higgs field. The low energy manifestation of the vector portal is the $\epsilon$-suppressed coupling of the dark photon to the SM electromagnetic current $J_{\rm EM}^\mu$. These portals provide the simplest UV-complete approach to model building for dark sectors. Indeed, these dark sector degrees of freedom alone can explain Dirac and Majorana neutrino mass, a variety of models for low mass non-thermal dark matter, and the minimal mechanism of leptogenesis. Other couplings to the dark sector require operators that are higher-dimensional when expressed in terms of the full SM chiral structure. On general effective field theory grounds, we would thus expect the portals to play the leading role, if these vector, scalar or dark fermion degrees of freedom are present in nature.

On allowing for higher dimensional operators in (\ref{L}), a plethora of new interactions opens up. However, one case at dimension five has a special status, due to its connection with spontaneous symmetry breaking. In particular, pseudoscalar (pseudo)-Goldstone fields of axion-like particles (ALPs) are naturally light \cite{Peccei:1977hh,Weinberg:1977ma,Wilczek:1977pj}, and can in general interact via a series of dimension five operators. If we restrict attention to flavour diagonal operators, the interactions take the form,
\begin{equation}
 {\cal L}_{\rm axion\, portal} = \frac{a}{4 f_G} {\rm Tr}\, G^{\mu\nu} \widetilde{G}_{\mu\nu} + \frac{a}{4f_\gamma} F^{\mu\nu} \widetilde{F}_{\mu\nu} + \frac{1}{f_q} \partial_\mu a \sum_q \bar{q} \gamma^\mu \gamma^5 q + \frac{1}{f_l} \partial_\mu a \sum_l \bar{l} \gamma^\mu \gamma^5 l\,,
\end{equation}
where the decays constants $f_{G,\gamma,q,l}$ set the scale $\Lambda$ in this case. The non-renormalizability of these operators implies that UV-sensitivity will generically enter into the associated physical quantities.

Applications of this theoretical framework to various dark sector phenomenologies are discussed in later sections. The specific opportunities available with neutrino beam experiments will be discussed in Sections~\ref{sec:advantages} and \ref{sec:landscape}. The role of portals as new force mediation channels for dark matter models which go beyond the WIMP paradigm are described in Sec.\ref{subsec:Theory:DM} and explored in more detail in Sec.\ref{sec:models}. Neutrino mass is discussed in Sec.~\ref{subsec:Theory:NeutrinoMass}, and potential explanations of experimental anomalies are considered in Sec.\ref{subsec:Theory:Anomalies}.

\subsection{Dark Matter
}\label{subsec:Theory:DM}

Theories of dark sectors provide natural frameworks for light dark matter (DM). Perhaps most notably, it is well-known that standard thermal freezeout of DM annihilating to the SM through weak-scale interactions leads to overclosure of the universe for DM masses larger than a few GeV~\cite{Lee:1977ua}. However, in the presence of new light particles that can enhance DM annihilation, sub-GeV thermal DM can become viable. Each of the portals in the previous subsection provides a mediator that can serve in this role, coupling a DM particle $\chi$ living in the dark sector Lagrangian $\mathcal{L}_\mathrm{DS}$ to the SM through the portal operator. For instance, in the vector portal model, DM annihilation can proceed via the dark photon. Alternatively, mechanisms other than thermal freezeout can also set the relic density of light DM, including freeze-in, asymmetric DM, and strongly interacting DM. We focus here on the thermal scenario, describing the theoretical considerations that are most relevant for light DM studies at neutrino experiments; for a comprehensive review of DM in light dark sectors, see Ref.~\cite{Battaglieri:2017aum}.

With the inclusion of DM in the dark sector Lagrangian, the parameters of interest for DM physics are: $m_\chi$, the DM mass; $m_M$, the mediator mass; $g_\chi$, the coupling between the DM and the mediator; and $g_\mathrm{SM}$, the coupling of the mediator to the SM. As the mediator is typically light, the coupling $g_\mathrm{SM}$ must be small, while $g_\chi$ is relatively unconstrained. In describing DM annihilation through dark sector portals, then, there are two broad possibilities: \emph{direct} annihilation of DM through the mediator to SM particles, and \emph{secluded} annihilation of DM to a pair of mediators, with subsequent decays of the mediators to SM particles. In the secluded annihilation case, which requires $m_\chi > m_M$, the annihilation cross section scales as $\langle \sigma v \rangle \sim g_\chi^4 / m_\chi^2$, and is purely independent of $g_\mathrm{SM}$. For direct annihilation, by contrast, annihilation goes as $\langle \sigma v \rangle \sim g_\chi^2 g_\mathrm{SM}^2 m_\chi^2 / m_M^4$. While in principle both types of DM annihilation can occur simultaneously, because $g_\chi$ can be $\mathcal{O}(1)$ it is reasonable to expect that secluded annihilation dominates where it is kinematically possible. On the other hand, if $m_M > m_\chi$, direct annihilation is solely responsible for setting the thermal relic abundance.

In many cases where direct annihilation governs thermal freezeout, there is a direct connection between the relic density and potential DM signals at neutrino experiments. The proton collisions that eventually lead to neutrino beams can also produce dark sector mediators; the specific production mechanisms are model-dependent, and will be explored further in subsequent sections. If $m_M > 2 m_\chi$, so that the mediator can decay to DM, there is an effective beam of DM produced simultaneously with the neutrinos. Like the neutrinos, the DM can scatter inside downstream detectors. Furthermore, given a particular portal and DM Lagrangian, the combination of couplings and masses that result in the observed relic density can often be translated into an expected number of scattering events at a neutrino experiment. In the minimal scenario of light DM coupled through the vector portal, for example, thermal freezeout has been probed by MiniBooNE~\cite{MiniBooNEDM:2018cxm}, and examined phenomenologically for a variety of current and future neutrino facilities~\cite{deNiverville:2011it, deNiverville:2016rqh, Buonocore:2019esg, Berlin:2020uwy, Batell:2021aja}.

It should be noted that the relation between the DM thermal relic abundance and neutrino detector signals is model-dependent. Non-minimal examples include theories where the DM annihilation would have been enhanced in the early universe relative to today, e.g.~through $p$-wave annihilation or because of more complicated dark sector content as in theories of inelastic DM. In addition, dark sectors with mediators that have different SM coupling structures than the minimal portals, such as new $U(1)$ gauge bosons, can yield different constraints from neutrino experiments relative to other accelerator-based tests such as those based on electron beams~\cite{Batell:2014yra, Coloma:2015pih}. This model dependence underscores the importance of thoroughly exploring the landscape of dark sector models which include DM at neutrino experiments.

\subsection{Neutrino Mass
}\label{subsec:Theory:NeutrinoMass}
The discovery that neutrinos oscillate, and therefore have distinct mass and flavor eigenstates, has proven to be one of the most definitive pieces of evidence for physics beyond the Standard Model in the last two decades. Given that the Standard Model does not predict neutrino masses, and that the observed masses are at most $10^{-6}$ times that of the lightest charged fermion, the electron, it is reasonable to catalogue the types of new physics that can give rise to this phenomenon. This new physics is typically organized in terms of the new particles and/or interactions that are necessary to explain small, nonzero neutrino masses. Included in this categorization are the famous ``seesaw mechanisms,'' which invoke new physics at high scales (${\sim}$TeV -- Planck scale) in different ways -- see, e.g., Refs.~\cite{Minkowski:1977sc,GellMann:1980vs,Yanagida:1979as,Glashow:1979nm,Mohapatra:1979ia,Schechter:1980gr,Foot:1988aq}.

One feature present in many theories that explain neutrino masses is the addition of one or more SM-gauge-singlet fermions $N_i$ that interact with the SM fields via a Yukawa interaction,
\begin{equation}\label{eq:NeutrinoMassLagrangian}
    \mathcal{L} \supset -y_\nu^{\alpha i} \left(L_\alpha H\right) N_i,
\end{equation}
where $y_\nu$ is a dimensionless matrix, $\alpha = e,\, \mu,\, \tau$, and $i$ ranges between $1$ and the number of new fermions added to the theory. An additional mass term of $N_i$ can also be incorporated, and, depending on whether the accidental $U(1)$ global symmetry of Lepton number is preserved or broken, it may be of the Dirac or Majorana type. After electroweak symmetry breaking and the Higgs field $H$ acquires its vacuum expectation value $v$, various mass terms (some linking the SM neutrinos inside $L_\alpha$ and the new fields $N_i$, and some amongst the $N_i$) are generated. Once the post-EWSB mass terms are diagonalized, the masses of the SM-like and new-physics fermions may be determined. Depending on the $N_i$ mass terms and the values in the matrix $y_\nu^{\alpha i}$, these mostly-SM neutrinos can have the sub-eV masses observed in experiments today. 

The form of Eq.~\eqref{eq:NeutrinoMassLagrangian} is one of the renormalizable portals outlined in Section~\ref{subsec:Theory:Portals}. This gives rise to new phenomenology that will be described throughout the remainder of this white paper, specifically in Sections~\ref{sec:advantages}, \ref{sec:landscape}, and~\ref{subsec:Benchmark:NeutrinoPortal}. Not only does Eq.~\eqref{eq:NeutrinoMassLagrangian} give rise to light neutrino masses, it also leads to mixing between the light, mostly-SM neutrinos and the new heavy states $N_i$. The new phenomena that will occur include production of $N_i$ in neutrino-beam environments as well as scattering processes where an incoming neutrino scatters off some target to produce heavier $N_i$. Section~\ref{subsec:Benchmark:NeutrinoPortal} will explore how current and upcoming experiments are well-poised to explore these various phenomena.

\subsection{Experimental Anomalies
}\label{subsec:Theory:Anomalies}

The  explanation of the MiniBooNE excess of electron-like events at $4.8\sigma$~\cite{MiniBooNE:2008yuf,MiniBooNE:2018esg,MiniBooNE:2020pnu} requires new physics since the nuclear and hadronic physics uncertainties can not solely account for such an observation \cite{Brdar:2021ysi}. Further, one of the recent MicroBooNE results~\cite{MicroBooNE:2021zai} constrains the $\Delta \to N\gamma$ background more stringently, disfavoring the possibility of its $\sim\!3$ times more enhanced branching ratio (BR) that would explain MiniBooNE anomaly~\cite{MiniBooNE:2020pnu}. This observation advocates the need for new physics to explain such an excess.
Various   new physics scenarios discussed to date include: $(i)$ neutrino-based solutions with light sterile neutrinos~\cite{Karagiorgi:2009nb,Collin:2016aqd,Giunti:2011gz,Giunti:2011cp,Gariazzo:2017fdh,Boser:2019rta,Kopp:2011qd,Kopp:2013vaa,Dentler:2018sju,Abazajian:2012ys,Conrad:2012qt,Diaz:2019fwt,Asaadi:2017bhx,Karagiorgi:2012kw,Pas:2005rb,Doring:2018cob,Kostelecky:2003cr,Katori:2006mz,Diaz:2010ft,Diaz:2011ia,Gninenko:2009ks,Gninenko:2009yf,Bai:2015ztj,Liao:2016reh,Carena:2017qhd,Moss:2017pur,deGouvea:2019qre,Moulai:2019gpi,Dentler:2019dhz}; $(ii)$ with heavy sterile neutrinos and new gauge and Higgs sectors~\cite{Bertuzzo:2018itn,Ballett:2018ynz,Fischer:2019fbw,Datta:2020auq,Dutta:2020scq,Abdallah:2020biq,Abdullahi:2020nyr,Brdar:2020tle,Abdallah:2020vgg,Kamp:2022bpt} and $(iii)$ dark sector based solutions ~\cite{Dutta:2021cip}.

The ongoing beam based CE$\nu$NS experiments, COHERENT, CCM,  JSNS$^2$ etc. provide interesting platforms to investigate the existing anomalies. However, the neutrino based solutions requiring heavier sterile neutrinos  are mostly outside the realm of investigation of the CE$\nu$NS experiments using 800 MeV to 3 GeV proton beams. For testing such solutions, experiments like CHARM-II, MINERvA and T2K \cite{Arguelles:2018mtc,Brdar:2020tle} are more powerful. On the other hand, the oscillation solutions of the MiniBooNE anomaly with a sterile neutino mass $\sim1$ eV can be probed at the CE$\nu$NS experiments since $L/E$ for these experiments are similar to the MiniBooNE experiment. Further, since both muon and electron neutrinos emerge from the stopped pion and muon decays, it is possible to probe $\sin^22\theta_{\mu e}$ vs $\Delta m_{41}^2$ parameter space at CE$\nu$NS experiments.

The dark sector solutions involving scalar/pseudoscalar (axion like particles) with mass $\sim$ few MeV and dark matter with mass 10 MeV provide appealing solutions to the MiniBooNE excess \cite{Dutta:2021cip}. The scalar/pseudo scalar undergoes inverse Primakoff scattering at the detctor producing a photon that appears as an electron-like event in the detector. The dark matter solution, on the other hand, involves a vector mediator emerging from the three body decay of charged pion which then upscatters in the detector; the produced state then decays  into a pair of collimated $e^+e^-$ that can also yield a single-shower signal at MiniBooNE.  Both angular and energy spectra can be fit using these solutions.  The scalar, pseudoscalar and vector mediators emerge from three body decays of charged pions and kaons. Two important points regarding this production are: $(i)$ the three body decay modes of the charged pions/kaon are not helicity suppressed like in the case of the two body processes; $(ii)$ the magnetic horn at the Booster beam facility focuses and thereby enhances the flux of the charged pions/kaons compared to the neutral ones.  Both these realizations allow the dark sector interpretation to be allowed even after considering the constraints from the MiniBooNE off-target mode \cite{MiniBooNE:2017nqe} (see also recent proposal for a new off-target experiment \cite{Bhattarai:2022mue}). The dark sector based solutions are in agreement with the MicroBooNE observation; taking 320 excess events below 300 MeV visible energy \cite{MiniBooNE:2020pnu},  the dark sector models predict that MicroBooNE  should expect $\sim\!18$ event excess in the $1\gamma0p$ analysis at low energy.  It is interesting that the MicroBooNE reports a 18 event (2.7$\sigma$) excess for the $1\gamma0p$ sample in the 200-250 MeV visible energy bin \cite{MicroBooNE:2021zai}, demonstrating consistency with the predictions and the possibility for checking this solution with future data. 

There exists also the LSND anomaly: a 3.8 {$\sigma$} excess of electron antineutrino interactions over standard backgrounds, observed by the LSND Collaboration in a beam dump experiment with 800 MeV protons~\cite{LSND:2001aii}. Above discussed dark sector (pseudo)scalar based realizations \cite{Dutta:2021cip}, as well as neutrino-based solutions with both eV-sacle \cite{Dentler:2018sju} and somewhat heavier sterile neutrinos (see e.g. \cite{Abdallah:2020vgg}) can  explain the LSND excess. However, in the case of light sterile neutrino, there is a tension  with global neutrino data for explaining both MiniBooNE and LSND, stemming chiefly from MINOS \cite{MINOS:2017cae} and IceCube muon neutrino disappearance searches \cite{IceCube:2020tka}.

For completeness, let us also point out a mild excess in the first round of COHERENT data release. For the CsI detector, standard neutrino backgrounds can be reduced by invoking appropriate timing and energy cuts in order to remove prompt neutrino events from pion decays and delayed events from muon decays. After using this search strategy of imposing a combination of energy and timing cuts to the existing CsI data of the COHERENT, a mild excess beyond known backgrounds was reported~\cite{Dutta:2019nbn, Dutta:2020vop}. The new data release from the COHERENT experiment needs to be investigated in order to understand this excess.

\subsection{Other Theoretical Motivations
}

Along with the obvious rationale of explaining dark matter and/or neutrino masses, the dark sector/portal paradigm is motivated by other unsolved problems in particle physics and cosmology. 
 
A prime example is the Higgs naturalness problem. The strong limits from the LHC on traditional solutions to this puzzle, including supersymmetry and compositeness, have motivated new radical theoretical approaches, many of which employ a hidden/dark sector and a portal coupling. For instance, the Twin Higgs model posits an entire copy of the SM, related to our visible sector through a $Z_2$ exchange symmetry, along with a Higgs portal coupling~\cite{Chacko:2005pe}. The Twin Higgs construction leads to mirror top and gauge partners, which are neutral under the SM gauge interactions and thus unconstrained by direct searches at the LHC. In this way, the Twin Higgs provides a solution to the little hierarchy problem, along with a motivated example of a rich dark sector, with potentially far-reaching implications for dark matter, neutrino masses, astrophysics and cosmology~\cite{Batell:2022pzc}. Another novel approach to the Higgs naturalness problem is the cosmological relaxation of the electroweak scale~\cite{Graham:2015cka}. The basic idea is that a light ``relaxion'' field with a small Higgs portal coupling scans the Higgs mass in the early universe. Once electroweak symmetry breaking is triggered, feedback from the relaxion potential stops its evolution, fixing the effective Higgs mass to a value that is parametrically small compared to the UV cutoff. A variety of novel probes of this scenario exist, including searches for sub-GeV relaxions at high-intensity neutrino experiments~\cite{Flacke:2016szy}.

Inflation provides another interesting example. While it is usually assumed that the inflationary sector is decoupled from the SM at low energies (below the inflationary Hubble scale), this does not necessarily need to be the case. An interesting counterexample is the case of a light inflaton coupled via the Higgs portal~\cite{Bezrukov:2009yw}, which leads to a rich phenomenology at intensity frontier experiments, including neutrio beam experiments. 

There are also a number of interesting proposals for the generation of the matter-antimatter asymmetry at temperatures at or below the weak scale, which feature new light neutral particles with portal interactions. A prominent example is the ARS mechansim for leptogenesis~\cite{Akhmedov:1998qx} as exhibited in the $\nu$MSM~\cite{Asaka:2005pn}, which relies on right-handed neutrinos and the neutrino portal. Other interesting examples include baryogenesis through heavy meson oscillations~\cite{Elor:2018twp}. 



\FloatBarrier
\section{Advantages of Neutrino Beam Experiments
}
\label{sec:advantages}

Precision measurements for neutrino oscillation parameters at the next generation neutrino experiments require high-power proton beams and 
sensitive near detectors, as well as large mass far detectors, to minimize both the statistical and systematic uncertainties.  The last few decades have seen a dramatic  
increase in neutrino beam intensity. Typical proton intensities of $\sim 10^{21-22}$ POT per year have been achieved by BNB, NuMI, SNS, and Lujan. 
Coupled with state of the art neutrino detectors and  significant advances in dark sector theory/phenomenology, 
this has opened up new searches for an array of BSM phenomena, including new states produced in the high intensity proton beam-target interactions  
and exotic secondary neutrino interactions in the near detector. The opportunities for BSM studies are wide-ranging and include searches for
sterile neutrinos, HNLs, dark matter, axion-like particles, and dark neutrinos; see Section~\ref{sec:models} for further details.  
Future neutrino beam-lines such as LBNF will continue this trend.
It is worth noting that the light dark sector particles (masses below $\mathcal O({\rm GeV}))$ expected in various BSM theories 
generally have feeble interactions with SM particles and in many cases fall outside the experimental sensitivities of 
conventional high energy colliders such as the LHC. 
Intensity frontier experiments, including neutrino beam experiments, thus have a critical role to play in searches for new light states. 
In this section, we briefly summarize the advantages of neutrino beam experiments in studying dark sector particles in terms of beam line capabilities, detector capabilities, and facility for further background mitigation.

 \subsection {Beam Line Capabilities}

Low energy proton beams below 10 GeV have certain advantages.  If the physics processes under consideration are accessible at such low energies, 
then the best strategy is to 
employ the beam power to increase the proton intensity, thereby enhancing the collision luminosity and the production rate of BSM particles. 
Furthermore, as beam energy increases, additional SM particle production channels open up, 
potentially giving rise to 
new backgrounds to a rare event search. 
For instance, at the 8 GeV BNB, charm production is small and does not pose a significant background.  At a 800 MeV stopped pion source, Kaon production is not a concern.  
However, the clear advantage of higher energy beams is access to higher mass particle production, extending the kinematic reach 
in BSM particle mass, though typically at the expense of reach in 
the small BSM particle coupling.

Pulse-type beams in low energy neutrino experiments searching for CE$\nu$NS and sterile neutrinos, such as COHERENT, CCM, and JSNS$^2$, have the capability to utilize
the beam arrival timing information in detectors.
Along with kinematic measurements (e.g., recoil energy information), the timing information provides an additional important handle to discriminate BSM signatures from neutrino backgrounds, as demonstrated in the case of 
light dark matter searches~\cite{Dutta:2019nbn,Dutta:2020vop}.
More details regarding this unique advantage are explained in Sec. 5.6.

The precision measurements of neutrino oscillation parameters require a high statistics sample of neutrinos and highly capable detectors with as low of energy threshold as possible.  High intensity protons beams incident to a well designed target and focusing horn assembly enables access to the neutrino energy range that fits the baseline of the experiment with a high neutrino flux.
The unprecedented high intensity protons at the beam facility for future neutrino experiments, e.g. the 1.2MW up-gradable to 2.4MW beam power at the the Long Baseline Neutrino Facility (LBNF)~\cite{DUNE:2016hlj,DUNECollaboration2015}
presents opportunities to probe physics beyond the Standard Model, as will be explored in great depth in Section~\ref{sec:models}.

\subsection {Detector Capabilities}

Technologies for neutrino detectors are also well suited 
for BSM particle as they often share similar signatures to neutrino interactions, e.g., scattering or production of energetic visible SM particles in the final state. 
Large volumes are necessary for rare event searches, bigger is typically better though smaller detectors can sometimes have better capabilities such as speed, resolution, etc.   Choice of type of detector volume is important to enhance certain physics process, e.g. argon for coherent scattering, oil/water for Cerenkov light production,  etc.  High resolution tracking capabilities (e.g., as enabled by TPCs) allow for  
event type and topology identification, which can be different for neutrinos and BSM particle interactions (e.g. lack of final state hadronic particles from axion-like particle scattering).
Instrumentation is crucial in determining detector speed, energy resolution, etc.

Neutrino experiments with large volume far detectors, such as DUNE and HK, can take prominent roles in searching for signatures of BSM particles of cosmogentic origin
with relatively low fluxes.
An example of detector capabilities of large volume LArTPC far detectors of DUNE in identifying multi-particle tracks signatures coming from energetic light dark matter is analyzed in Ref.~\cite{DeRoeck:2020ntj}.
It is proven that the technology adopted in the detectors is particularly suited for searching for complicated signatures, e.g., inelastic boosted dark matter~\cite{Kim:2016zjx}, due to good resolution, particle identification, and dE/dx measurements to recognize merged tracks.

\begin{figure}
    \centering
    \includegraphics[width=0.49\textwidth]{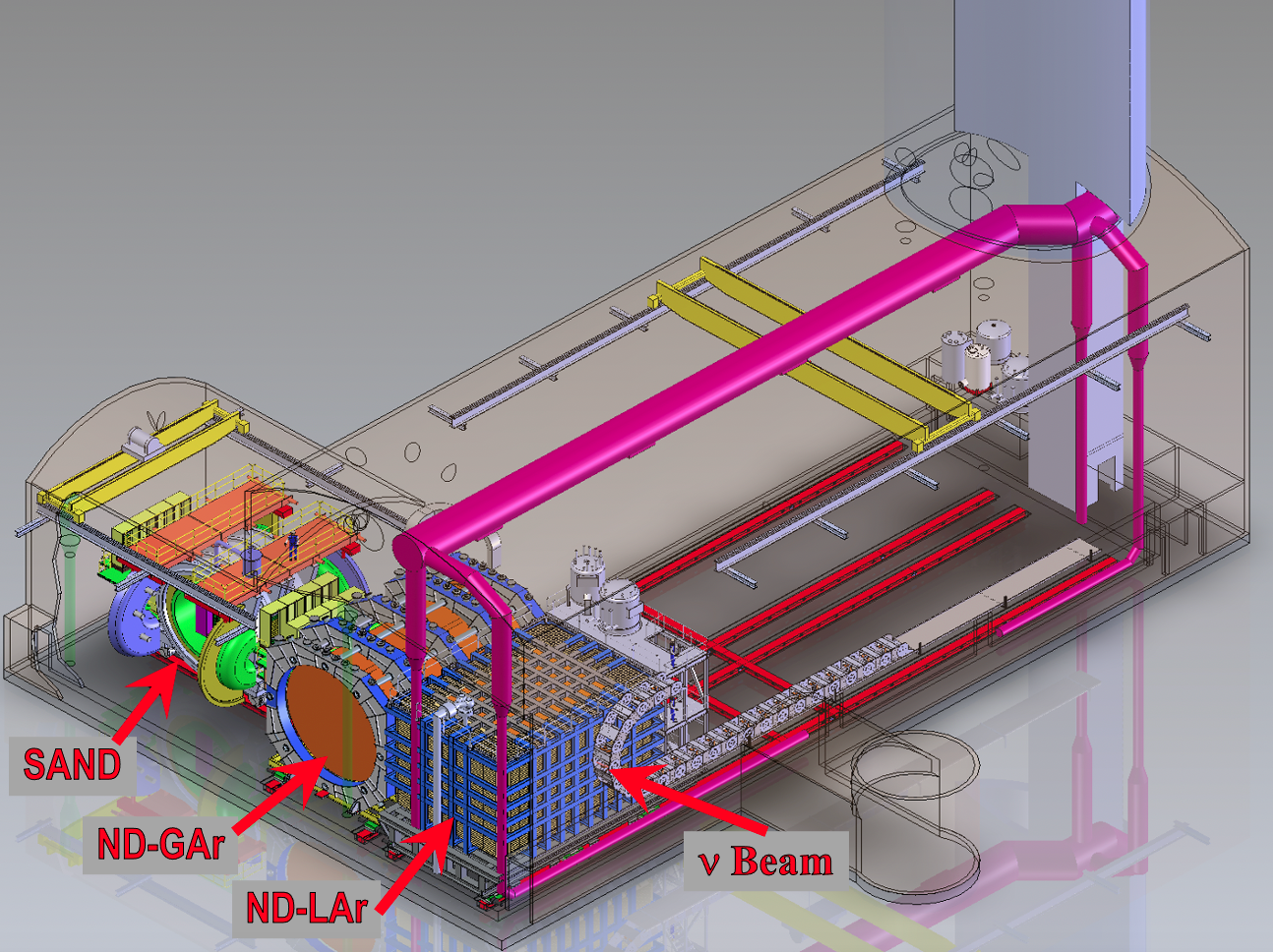}
    \includegraphics[width=0.49\textwidth]{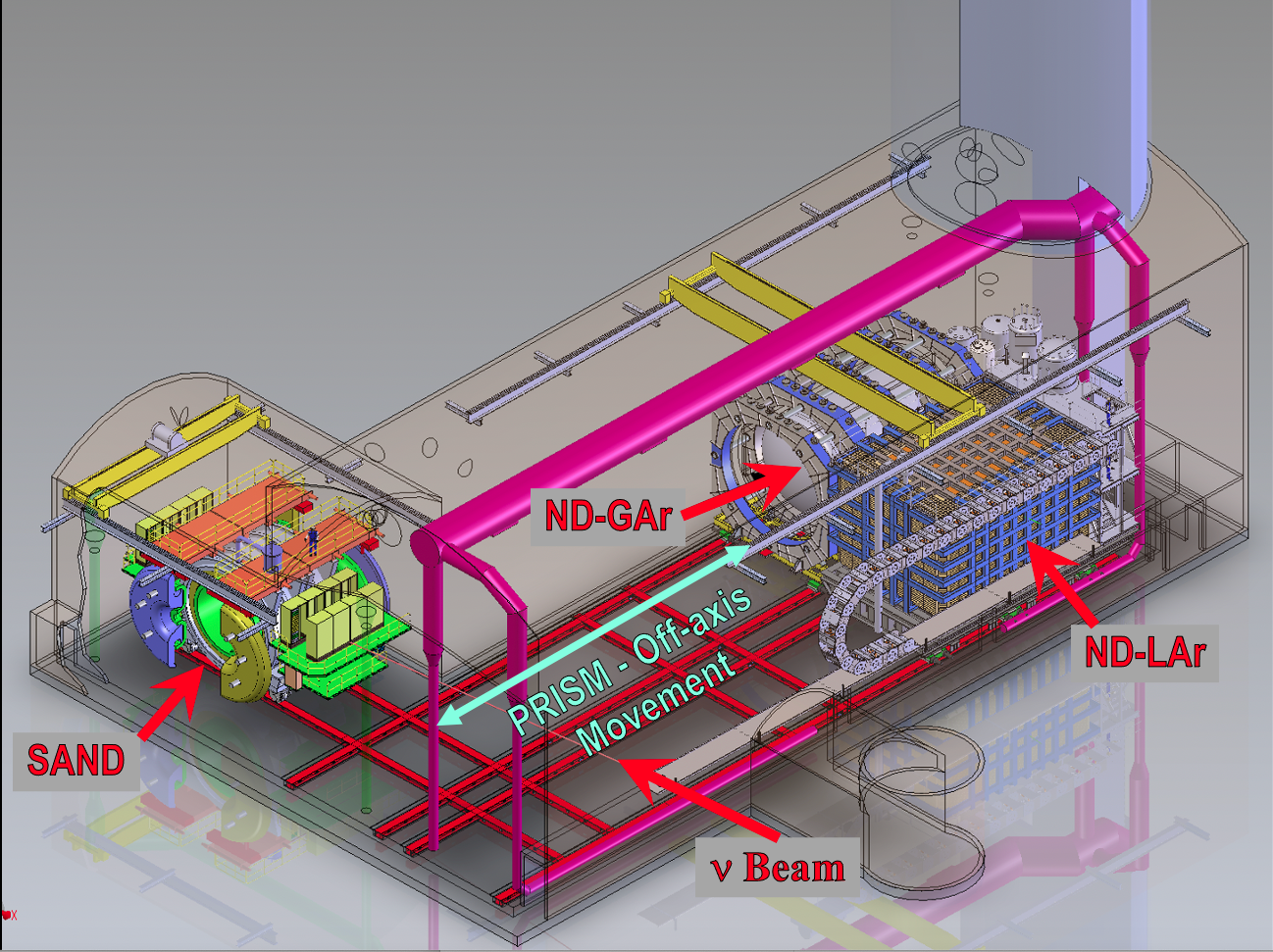}
    \caption{Left: Current conceptual design of the DUNE near detector complex which consists of a LArTPC, followed by a high pressure GArTPC and the straw tube detector based on axis beam monitor called SAND (System for on-Axis Neutrino Detection).  Right: The DUNE near detector complex with the LArTPC and GArTPC placed in its farthest 30.5m off-axis position or 53 mrad off-axis.}
    \label{fig:dune-nd}
\end{figure}

The high proton intensities and powerful near detector components at future neutrino experiments enables direct searches for dark sector particles that can be produced through a variety of portal couplings. 
A rich variety of dark sector particle signatures at neutrino near detectors can be envisioned, including 
scattering of the dark sector particles with the detector target nucleus or their decays to visible SM final states.
Precision imaging detectors such as the DUNE LArTPC near detector complex as shown in Fig.~\ref{fig:dune-nd} are well suited for dark sector particle scattering signatures whereas the high pressure GArTPC provides a large detection volume for dark sector particle decays,  with lower backgrounds from $\nu$-N interactions than in LArTPC near detector.

\begin{figure}
    \centering
    \includegraphics[width=0.9\textwidth]{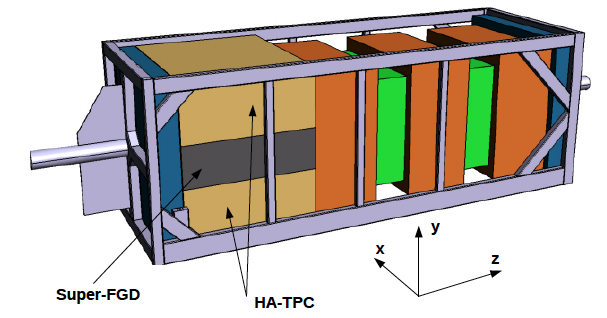}
    \caption{CAD 3D Model of the ND280 upgrade detector. In the upstream part (on the left in the drawing) two High-Angle TPCs (brown) with the scintillator detector Super-FGD (gray) in the middle will be installed. In the downstream part, the tracker system composed by three TPCs (orange) and the two FGDs (green) will remain unchanged. The TOF detectors are not shown in this plot. The detector is mechanically mounted on the basket, a steel beam structure (light gray), supported at both ends. The beam is approximately parallel to the z axis, the magnetic field is parallel to the x axis.}
    \label{fig:nd280}
\end{figure}

In addition, the PRISM concept employed by future neutrino experiments can significantly reduce neutrino flux uncertainties and help mitigate neutrino backgrounds to BSM signatures.
Since the flux of neutrinos in the beam decreases faster as a function of the off-axis angle than the dark sector particle flux in many models, taking data at different off-axis angles helps in extending the sensitivity reach in various dark sector particle searches.
Hyper-K utilizes the existing ND280~\cite{T2K:2019bbb,2022} which is also equipped with the PRISM capability in vertical direction, as illustrated in Fig~\ref{fig:nd280}. 

\subsection {Facility for further background mitigation}

Dark sector particle searches can typically occur parasitically to the main neutrino physics program by taking advantage of the characteristic production mechanisms, final state topologies, kinematics, and timing properties of the signal.  
However, in certain cases these methods may become limited by the copious neutrino backgrounds and have reduced sensitivity. 
For certain searches, e.g.,  vector portal dark matter scattering, it is advantageous to turn this neutrino background off by impacting the beam on a dump to suppress charged pion decay. 
The MiniBooNE beam off target run~\cite{MiniBooNEDM:2018cxm} is an example of re-tooling a neutrino beam line for a beam dump run.  However, compromises are made.  If a neutrino beamline can be designed up front with a dedicated parallel beam dump, it significantly expands BSM searches with minimal cost. 

One possible idea for enabling a neutrino and dark sector beam line to coexist is to utilize a neutral rich horn focusing system~\cite{yu-nrfh}.
The idea is to exploit the fact that neutrinos result from pion decay in flight. A three-dimensional dipole is placed immediately downstream of the horn string to take the cylindrical focused secondary charged particles and bend them with minimal loss toward the direction of the neutrino experiment facility consisting of the decay pipe, dump area, and finally detector complex(es).
The remaining high-energy beam protons that have not interacted and the charge neutral particles, traveling in the same direction as the incident primary protons, proceed toward a beam dump.
A high-precision detector to enable detection of these new particles can then be placed closely behind the dump 
to reduce the loss of the flux.
Given that these particles have very small interaction rates 
with SM particles, one can envision an array of multiple experiments to be 
placed in a dark matter experiment facility (DMEF).
Since most of the charged particles are bent away before they decay to neutrinos, the beams in the DMEF would be enriched with these neutral particles, with a greatly reduced background neutrino component, except for low-energy neutrinos resulting from pion decays upstream of the 3D dipole.
A back-of-the-envelop calculation suggests that a 5 orders of magnitude reduction of the neutrino flux directed towards the DMEF is possible, under the assumption of 100\% bending efficiency.
In addition, since the dump and the detector(s) in the DMEF could be placed immediate downstream of the dipole while the neutrino facility would require a long decay region followed by a dump and additional range out system, an additional gain of the neutral particle flux by two orders of magnitude is possible.  
These results suggest a signal to background ratio enhancement of four to five orders of magnitude, a substantial and worthwhile gain.

Furthermore, the use of absolute beam timing relative to the detector response can significantly reduce backgrounds. Typically fast $\sim ns$ timing can reduce cosmogenic and other random external backgrounds~\cite{MiniBooNE:2020pnu}, as well as sub-luminal backgrounds such as neutrons generated at the target~\cite{CCM:2021leg}.  This is achieved with fast detectors as well as efforts to reduce beam spill timing of accelerators.  This is best done at the design phase in both bases as reverse engineering can be expensive, though not impossible.  

 
\section{The Experimental Landscape
}
\label{sec:landscape}


\subsection{Current and Near-Term Experiments}
There are a number of short baseline experiments that are operating
or soon to be online that are addressing the short baseline 
neutrino oscillation anomalies. 
These experiments are also ideal tools to perform searches for other dark sector particles such as sub-GeV dark matter, ALPs, heavy neutral leptons, etc.  Furthermore, the short baseline anomalies themselves 
could also be a manifestation of a dark sector coupled through a portal.  Short baseline experiments with low energy beams ($\lesssim {\cal O}(10\,\rm GeV)$) provide powerful probes of light dark sector states (masses below $ {\cal O}(1\,\rm GeV)$) due to their intense proton beams, short baselines ($\sim 100$s meters), and large sensitive detectors. 
For heavier dark sector states (masses near or above $ {\cal O}(1\,\rm GeV)$), it is necessary to utilize near detectors with higher energy beams, such as the 
the LBNF 120\,GeV proton beam. 
Finally, stopped pion sources at J-PARC, LANSCE, SNS, and future PIP-II have high power coupled with extremely short baselines ($\lesssim 20$ meters) which enable very sensitive searches at masses below a few hundred MeV.  
A variety of dark sector models
can be tested at these current and near-term neutrino beam experiments, making the prospects for future discoveries with a reasonable timeline and cost an exciting possibility.

\subsubsection{Long Baseline Neutrino Experiments}

Long-baseline neutrino experiments typically comprise both far detectors and near detectors, among which the near detectors have either the same technology or the same target element as the far detectors.

{\bf NO$\nu$A}~\cite{NOvA:2007rmc}: 
The currently operating NOvA (NuMI Off-axis electron-neutrino Appearance) experiment uses the 120 GeV NuMI (Neutrinos from the Main Injector) proton beam at Fermilab, which collides with a carbon target with an intensity of up to 700 kW, and can deliver up $6 \times 10^{20}$ POT per year. The NOvA Near Detector (ND) is a  4m x 4m x 15m, 300 t active mass, a fine-grained liquid scintillator detector, located along the Fermilab NuMI beam line 900~m downstream of the target. The ND is stationed 11 meters to the left of the beam center (14 mrad off-axis), and as a result is exposed to a narrow band neutrino beam peaked at 2 GeV. The distinct energy and angular profile of the neutrino beam at the near detector location can help to discriminate dark sector signatures in some cases.

{\bf DUNE}~\cite{tdr-vol-1,DUNE:2021tad}: The Fermilab LBNF beam, with its 1.2~MW, will deliver 1.1$\times 10^{21}$~POT per year and neutrinos with wideband energy and a peak roughly at 2.5~GeV.
This will be the most intense neutrino beam in the world and is upgradable to multi-watt power. 
Consisting of a modular LArTPC detector (ND-LAr), a high pressure gaseous argon TPC (ND-GAr), and a beam monitor (System for on-Axis Neutrino Detection, SAND), the near detector (ND) complex is to be located about 574~m downstream of the LBNF neutrino producton target.

The ND-LAr is located most upstream in the ND complex and utilizes the same technology as the FD to reduce systematics from different nuclear effects than FD.  
It measures the $\nu$-Ar interactions in the total active mass of 150t.
ND-GAr immediately follows the ND-LAr. It is a magnetized (0.5~T) large volume high pressure (10atm) gas TPC with an electromagnetic calorimeter surrounding the TPC. It is capable of low threshold $\nu$-Ar event tracking and with precision $p_\mu$ measurements in its 1~t fiducial mass.
ND-LAr and ND-GAr are on a transport system that can move them off-axis, forming the DUNE-PRISM.  This system takes the two detectors up to 30.5~m off-axis ($\sim53$mrad or $3^\circ$) to control $\nu$ flux systematic uncertainties.
Finally, SAND is located at the most downstream in the complex and provides a constant monitoring of the on-axis $\nu$ beam flux to the FD. It consists of a straw tube tracker and an electrimagnetic calorimeter and is equipped with a 0.6T magnet.

ND-LAr has dimensions of 5~m$\times$7~m$\times$3~m, corresponding to 67~tonne fiducial mass.
Its modular design and the pixelated charge readout system allow, respectively, isolated photon detection and full 3D reconstruction of tracks, enhancing the capabilities of coping with a high multiplicity environment from the intense neutrino flux and the high event rate.
The charge signal will be read out by 4~mm pixels and 2.5~$\mu$s time-binning, while the scintillation light signal will be isolated in each module with a volume of 1~m$\times$1~m$\times$3~m, and be detected by the light detector with $\sim$10~ns time resolution.

{\bf T2K, T2K-II, T2HK}~\cite{T2K:2011qtm,Hyper-Kamiokande:2018ofw,T2K:2019bbb}: 
The Tokai-to-Kamioka (T2K) experiment is a long-baseline neutrino experiment in Japan that has been in operation since 2010. The T2K neutrino beam is produced by directing the 30 GeV proton beam at JPARC on a graphite target. The T2K-II phase includes an upgrade of the beam, increasing its power from ~500 kW to 1.3 MW over the next several years. A near detector complex is located 280 m from the target, which houses several near detectors, including ND280. ND280 comprises several subdetectors (central tracker with three TPCs, scintillator based fine-grained detector (FGDs), and $\pi^0$ detector (P0D)) along with a 0.2 T magnet, and is surrounded by an electromagnetic calorimeter and a Side Muon Range Detector. ND280 has excellent track and momentum reconstruction and particle identification capabilities. An upgrade of ND280 allows for better detection efficiency over the full solid angle and improved reconstruction capabilities. 

\subsubsection{Short/Very Short Baseline Neutrino Experiments}

Currently, many short/very short baseline neutrino experiments are in operation or under construction.
Such experiments are typically equipped with a low-energy but high-intensity proton beam of $\lesssim \mathcal{O}$(10 GeV) and $\sim10^{21}-10^{24}$ POT per year, hence producing low-mass dark-sector particles copiously. 
Moreover, most of their on-going/projected detectors are featured by a multi-ton scale in mass and located $\mathcal{O}(10-100\, {\rm m})$ from the target, and thus a large amount of exposure to the dark-sector particle flux is expected even with a small amount of duty time.
In this section, we provide a brief review on these experiments, summarizing their key specifications

{\bf MicroBooNE, ICARUS, SBND}~\cite{MicroBooNE:2015bmn}: Individual experiments in the Short-Baseline Neutrino (SBN) program of three LAr-TPC detectors are located along the Booster Neutrino Beam (BNB) at Fermilab.
\begin{itemize}
\item
MicroBooNE: MicroBooNE operated from 2015 to 2020, collecting events produced by BNB (on-axis) and NuMI ($7^{\circ}$ off-axis).
Composed of single TPC with 2.5 m drift and 32 PMTs (8 inch) in acrylic support coated with TPB,
the 170~t-LAr (active: 87~t) detector is located 470~m from the BNB target.
Cosmic-ray tagger (CRT) covers top and side of the detector.
\item
ICARUS: The ICARUS detector was activated in August 2020 and its commissioning is expected to be completed by 2022. 
The detector is located 600 m from the BNB target and has been also exposed to neutrino beams coming $\sim 6^{\circ}$ off-axis from the NuMI target, and has collected a large number of neutrino events.
The detector consists of 760 t of LAr (active: 476 t), 4 TPCs with 1.5 m drift, and 360 PMTs (8 inch) coated with TPB, which has almost full CRT coverage.
\item
SBND: The SBND detector is under construction and will be fully commissioned by early 2023.
The detector is a 260 t  LAr (active: 112 t) detector with 120 PMTs (8 inch, 96 coated with TPB), 192 X-ARAPUCA modules, and TPB coated reflector foils on the cathode, which is located 110 m from the target and has $4\pi$ CRT coverage.

\item The BNB has the capability to run in off-target mode concurrent with neutrino mode - kicker magnet to select protons on- or off-target on a per pulse basis.  This would allow all the above LAr experiments to perform sub-GeV dark matter searches similar to that done by MiniBooNE~\cite{MiniBooNEDM:2018cxm}, but with higher sensitivity.

\end{itemize}

{\bf JSNS$^2$}~\cite{Ajimura:2017fld, Ajimura:2020qni}: The J-PARC Sterile Neutrino Search at the J-PARC Spallation Neutron Source (JSNS$^2$) experiment utilizes an 1~MW beam of 3~GeV protons incident on a mercury target.
\begin{itemize}
\item
First detector: The first detector of JSNS$^2$ has started data taking in June 2020.
The detector located 24~m from the target has 17 t fiducial mass of a gadolinium(Gd)-loaded liquid-scintillator (LS) and an acrlyic vessel with 120 PMts (10 inch), which is surrounded by $\sim30$ t of unloaded LS for the gamma catcher and outside background veto.
\item
Second detector: A new detector is planned to be installed 48~m from the target and will start data taking in 2023.
The second detector will be composed of 35 t fiducial mass of Gd-LS and an acrlyic vessel with 240 PMTs, which will also have the gamma catcher and veto regions filled with $\sim130$ t of unloaded LS.
\end{itemize}

{\bf Coherent Captain Mills (CCM)}~\cite{CCM:2021leg, Aguilar-Arevalo:2021sbh}: 
CCM runs at Los Alamos National Laboratory, located 23\,m from the Lujan center target, where 800\,MeV protons at 100\,kW impact on a tungsten target with a 275\,nsec at 20\,Hz beam profile.
\begin{itemize}
\item 
The detector is a 5 t fiducial-volume LAr detector with high PMT coverage, and no TPC.
Event detection relies on observation of the 128 nm scintillation light, which is shifted to the visible via 1,1,4,4-Tetraphenyl-1,3-butadiene.
The detector has fast $\sim$1\,nsec time response with energy reconstruction from 10\,keV all the way up to 100\,MeV.
\item 
CCM120 with 120 PMTs ran in 2019.
CCM200 with 200 PMTs is running at present, and is planned to continue for a total of three years to collect $2.25 \times 10^{22}$\,POT, with its search capabilities for dark matter and ALP's detailed in \cite{Aguilar-Arevalo:2021sbh, CCM:2021leg, CCM:2021lhc}
\end{itemize}

{\bf IsoDAR@Yemilab}~\cite{Alonso:2021kyu,Alonso:2022mup}: The Isotope Decay At Rest (IsoDAR) experiment pairs a high-power, accelerator driven $\bar \nu_e$ source with the LSC detector at Yemilab.
\begin{itemize}
\item
A 60 MeV compact cyclotron extracts 10 mA of protons that are targets on beryllium.
Ejected neutrons are captured in an isotopically pure $^7$Li sleeve.
Neutron capture produces $^8$Li that $\beta-decays$, resulting in 1.15$\times 10^{23}$ $\bar \nu_e$/year with a well understood energy, peaking at 6 MeV.
In the process, excited nuclei are produced that decay mono-energetically to photons.
Dark-sector particles can be also be produced in these decays.  
\item 
The LSC detector consists of 3 layers of a target, buffer and veto, with target consisting of 2.3 kt of scintillator.
Path-lengths of particles from the source through the LSC extend 9.5 m to 26.5 m from the IsoDAR target.   
The complex is excavated and instrumentation of prototyping is underway.
Running is expected to begin in 2027.
IsoDAR new physics capability is quoted for 5 (4) years of calendar- (live-)time.
\end{itemize}

\subsubsection{Reactor-Based Neutrino Experiments}
An increasing number of reactor-based neutrino experiments are ongoing or will be operational in the near future. 
Various nuclear reactions allow for production of not only neutrinos but photons. As their power spans $\sim MW$ (research-purpose reactors) to $\sim GW$ (commercial reactors), any dark-sector particles interacting with neutrinos and/or photons can be copiously produced. 
Many of the reactor-based experiments are equipped with low-threshold detectors that are placed in the vicinity of the reactor core. We summarize key specifications of existing and future reactor neutrino experiments in Table~\ref{tab:listreactor}.

\begin{table}[t]
    \resizebox{\columnwidth}{!}{%
    \begin{tabular}{c|c c c c}
    \hline \hline
    Experiment~ & Thermal power [GW] & ~Detector~ & ~Mass~ & ~Distance [m] \\
    \hline
    CONNIE~\cite{FernandezMoroni:2014qlq,CONNIE:2016ggr}  & 3.95 & Skipper CCD & 52~g & 30 \\
    CONUS~\cite{Hakenmuller:2019ecb}  & 3.9 & Ge & 3.76~kg & 17.1 \\
    MINER~\cite{Agnolet:2016zir,Dent:2019ueq} & 0.001 & Ge$\,+\,$Si & 4~kg & 1 -- 2.25 \\
    NEON~\cite{neon} & 2.82 & NaI[Tl] & $\sim10/50/100$~kg (Ph1/2/3) & 24 \\
    $\nu$-cleus~\cite{Strauss:2017cuu}  & 4 & CaWO$_4\,+\,$Al$_2$O$_3$ & ~6.84~g$\,+\,$4.41~g~ & ~15/40/100 (N/M/F) \\
    $\nu$GeN~\cite{Belov:2015ufh}  & $\sim 1$ & Ge & 1.6 -- 10~kg & 10 -- 12.5\\
    RED-100~\cite{Akimov:2017xpp,Akimov:2017hee} & $\sim1$ & DP-Xe & $\sim100$~kg & 19 \\
    Ricochet~\cite{Billard:2016giu}  & 8.54 & Ge$\,+\,$Zn & 10~kg & 355/469  \\
    SBC-CE$\nu$NS~\cite{sbc,AristizabalSierra:2020rom}& 0.68 & LAr[Xe] & 10~kg &  3/10 \\
    SoLid~\cite{SoLid:2020cen}  & 40 -- 100 & PVT$\,+\,^6$LiF:ZnS(Ag)  & 1.6~t &  5.5 -- 12\\
    TEXONO~\cite{Wong:2015kgl}  & 2.9 & Ge & 1.06~kg & 28\\
    vIOLETA~\cite{Fernandez-Moroni:2020yyl} & 2 & Skipper CCD & 1 -- 10~kg & 8 -- 12 \\
    NCC-1701 at Dresden-II~\cite{Colaresi:2022obx} & 2.96 & Ultra low-noise Ge & 3~kg & 8\\
    \hline \hline
    \end{tabular}
    }
    \caption{Key specifications of existing and future reactor (neutrino) experiments. Table contents are imported from \cite{Fortin:2021cog}. [CCD: charge couple device, DP-Xe: dual-phase xenon, Ph1/2/3: phase1/phase2/phase3, N/M/F: near/medium/far]\label{tab:listreactor}}
\end{table}

\subsubsection{Neutrino and new physics searches in the forward region of the LHC}

\begin{figure}[th]
    \centering
    \includegraphics[width=0.90\textwidth]{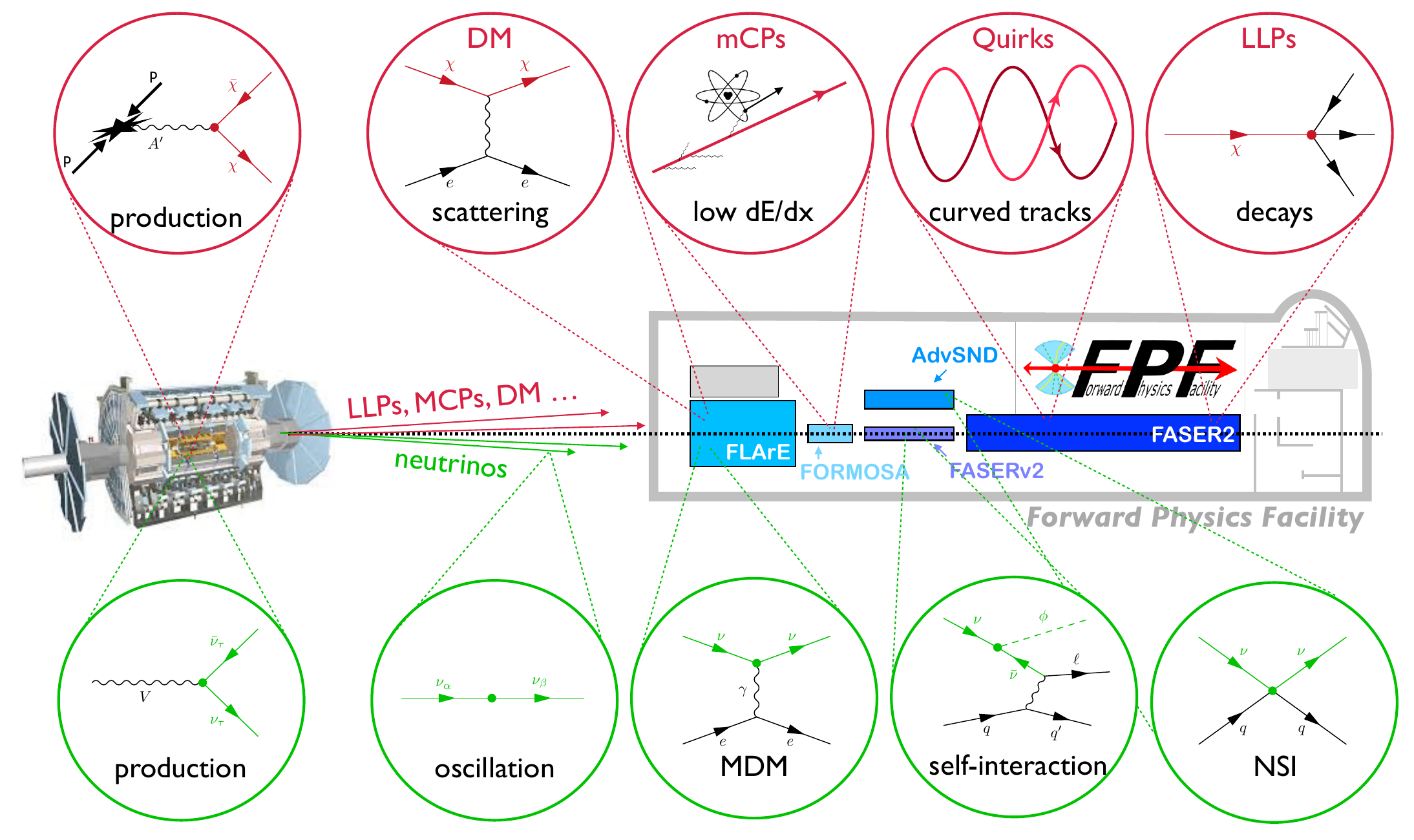}
    \caption{Schematic illustration of BSM signatures to be searched for at forward LHC experiments.  using the beam of dark sector particles (top) and neutrinos (bottom). \textbf{Top}: Signatures from i) scattering of DM in the neutrino detectors; ii) non-standard energy deposition of mCPs; iii) helical tracks in quirk models; and iv) decays of long-lived particles.  \textbf{Bottom:} BSM neutrino physics probes via i) additional BSM neutrino production mechanisms; ii) sterile neutrino oscillations; iii) neutrino magnetic moments; iv) neutrino NSI; and v) neutrinophilic particles produced in neutrino interactions. }
    \label{fig:fpf}
\end{figure}

A new neutrino and BSM physics program will soon be initiated during Run 3 of the LHC in the far-forward region of the collider. It has also been proposed to significantly extend this physics program toward the future high-luminosity (HL-LHC) era by building the dedicated Forward Physics Facility (FPF)~\cite{MammenAbraham:2020hex, Anchordoqui:2021ghd, Feng:2022inv}. We summarize them briefly below and in Table~\ref{tab:farforwardLHC}, in which we assume $14~\textrm{TeV}$ $pp$ collision energy and $150~\textrm{fb}^{-1}$ ($3~\textrm{ab}^{-1}$) of integrated luminosity for Run 3 (HL-LHC) experiments. The main detection channels of neutrinos and new physics particles are schematically illustrated in Fig.~\ref{fig:fpf}.

{\bf FASER}~\cite{FASER:2018ceo,FASER:2018bac}: The FASER detector has been proposed to search for highly-displaced decays of forward-going LLPs produced at the ATLAS IP during LHC Run 3~\cite{Feng:2017uoz}. The detector is placed at a distance of $480~\textrm{m}$ at the beam collision axis in the side tunnel of the LHC. It consists of a cylindrical decay volume with $1.5~\textrm{m}$ length and $R=10~\textrm{cm}$ radius, which is followed by the tracking stations. A magnetic field of 0.55~T is used to bend the tracks of charged products of LLP decays and the end calorimeter allows for measuring electromagnetic energy. Both FASER and its proposed successor FASER2~\cite{Feng:2022inv} to operate during the HL-LHC era have a great potential for finding long-lived beyond the Standard Model particles~\cite{FASER:2018eoc}, while high-energy CC neutrino interactions can also be detected based on their interactions in front of the spectrometer~\cite{Arakawa:2022rmp}.

{\bf FASER$\nu$}~\cite{FASER:2019dxq}: FASER$\nu$ is the pioneering experiment for detecting collider neutrinos, specifically neutrinos produced at the LHC. It is located in front of the FASER decay volume and is a $1.2$ tonne detector that contains tungsten plates and emulsion films. In the summer of 2021, the FASER collaboration reported the detection of the first neutrino events in FASER$\nu$~\cite{FASER:2021mtu}. During LHC Run 3 many more events are expected. What can we learn from that? One of the primary missions of this experiment is the measurement of neutrino cross sections at $\mathcal{O}(100-1000)$ GeV energies \cite{FASER:2019dxq,Brdar:2021hpy}. Further, FASER$\nu$ will detect $\mathcal{O}(10)$ tau neutrino events \cite{FASER:2020gpr}. The successor of FASER$\nu$ during the HL-LHC phase will be FASER$\nu$2 detector~\cite{SnowmassFASERnu2}.

\begin{table}
\begin{center}
\resizebox{\columnwidth}{!}{%
\begin{tabular}{ c||c|c|c|c||c|c|c } 
\hline
\hline
Experiment & \multicolumn{4}{c||}{Detector} & \multicolumn{3}{c}{Neutrino CC interactions}\\
\hline
 Name  & Mass / Size & Coverage & Technology & Time &  $\nu_e+\bar{\nu}_e$ & $\nu_\mu+\bar{\nu}_\mu$ &  $\nu_\tau+\bar{\nu}_\tau$\\
 \hline
 \hline
 FASER      & 0.05~m$^3$& $\eta\gtrsim 9$               & Tracking     & 2022-2025 & 
  \multicolumn{2}{c|}{See~\cite{Arakawa:2022rmp}} & \\
 FASER$\nu$ & 1 ton     & $\eta\gtrsim 8.5$             & Emulsion     & 2022-2025 & 
  901/3.4k & 4.7k/7.1k & 15/97\\
 SND@LHC    & 800 kg    & $7\lesssim\eta\lesssim 8.5$   & Emulsion     & 2022-2025 & 
  137/395 & 790/1k & 7.6/18.6\\
\hline\hline
 FASER2     & 31~m$^3$  & $\eta\gtrsim 7.1$             & Tracking     & 2030-2042 & 
 \multicolumn{2}{c|}{See~\cite{Arakawa:2022rmp}} & \\
 FASER$\nu$2& 20 ton    & $\eta\gtrsim 8.5$             & Emulsion     & 2030-2042 & 
  178k/668k & 943k/1.4M & 2.3k/20k\\
 AdvSND     & 2 ton     & $7.2\lesssim\eta\lesssim 9.2$ & Electronic   & 2030-2042 & 
  6.5k/20k & 41k/53k & 190/574\\
 FLArE      & 10 ton    & $\eta\gtrsim 7.5$             & LArTPC       & 2030-2042 & 
  36k/113k & 203k/268k & 1.5k/4k\\
 FORMOSA    & 5~m$^3$   & $\eta\gtrsim 7.4$             & Scintillator & 2030-2042 & 
 (19k/59k) & (106k/140k) & (781/2.1k) \\
 \hline
 \hline
 \end{tabular}}
\end{center}
\caption{Neutrino detection strategies in the far-forward experiments at the LHC. In the first five columns we give: the name of the experiment, its mass or size, pseudorapidity coverage, employed detection technique (see main text for more details), and the envisioned time of its operation. The FASER, FASER$\nu$, and SND@LHC experiments are currently running and will operate until the end of LHC Run 3 in 2025. The other listed detectors will run in the FPF at the HL-LHC phase from 2031 to 2042. The last three columns contain the expected number of the electron, muon, and tau neutrino-induced events, respectively. In each of these columns, the lower and upper estimate is provided to indicate the (current) expected uncertainty, as discussed in Ref.~\cite{Kling:2021gos}.\label{tab:farforwardLHC}}
\end{table}

{\bf SND@LHC}~\cite{SHiP:2020sos}: A similar search strategy employing nuclear emulsion technology will also be used by SND@LHC during LHC Run 3. The detector will be placed on the opposite side of the ATLAS IP with respect to FASER$\nu$. It will be placed off-axis and neutrinos interacting inside its fiducial volume would be coming from charm meson decays more preferentially than for on-axis detectors. This detector also has the potential to search for feebly interacting particles in yet unexplored parameter space \cite{SHiP:2020sos}. The proposed AdvSND detector to take data during the HL-LHC era might use electronic tracker layers instead of nuclear emulsion~\cite{Feng:2022inv}.

{\bf FLArE}~\cite{Batell:2021aja,Feng:2022inv}: FLArE is a proposed liquid argon detector with $\mathcal{O}(10~\textrm{ton})$ fiducial mass that is expected to be placed in the FPF and operate during the HL-LHC era. On top of studying interactions of high-energy neutrinos, this detector has a great potential to detect hidden sector particles; let us for instance stress searches for light dark matter \cite{Batell:2021aja,Batell:2021blf} and millicharged particles~\cite{Kling:2022ykt}.

{\bf FORMOSA}~\cite{Foroughi-Abari:2020qar,Feng:2022inv}: FORMOSA is a milliQan-type detector proposed to search for millicharged particles in the FPF during HL-LHC. It is a scintillator-based detector with the size of $1~\textrm{m}\times 1~\textrm{m}\times 5~\textrm{m}$. It will be sensitive to low ionization energy depositions expected from postulated BSM particles with small $\mathcal{O}(10^{-3})$ electric charges which can be abundantly produced in the far-forward region of the LHC. High-energy depositions from neutrino-induced interactions will also be detected in FORMOSA, while it remains to be studied how such events will be reconstructed. We provide the expected number of events for different neutrino flavors in Table~\ref{tab:farforwardLHC}.

\subsection{Future Experiments}

{\bf PIP2-BD: GeV Proton Beam Dump at Fermilab’s PIP-II LINAC}
\label{PIP2-BD}
The completion of the PIP-II superconducting LINAC at Fermilab as a proton driver for DUNE/LBNF in the late 2020s creates an attractive opportunity to build a dedicated beam dump facility at Fermilab (see Ref.~\cite{Toups:2022yxs}).  A unique feature of this Fermilab beam dump facility is that it can be optimized from the ground up for HEP. Thus, relative to spallation neutron facilities dedicated to neutron physics and optimized for neutron production operating at a similar proton beam power, a HEP-dedicated beam dump facility would allow for improved sensitivity to accelerator-produced dark matter. The Fermilab facility could be designed to suppress rather than maximize neutron production and implement a beam dump made from a lighter target such as carbon, which can have a pion-to-proton production ratio up to $\sim$2 times larger than the heavier Hg or W targets used at spallation neutron sources. The facility could also accommodate multiple, 100-ton-scale high energy physics experiments located at different distances from the beam dump and at different angles with respect to the incoming proton beam. This flexibility—which can constrain uncertainties in expected signal and background rates by making relative measurements at different distances and angles—would allow for sensitive dark sector and sterile neutrino searches.

The continuous wave capable PIP-II LINAC at Fermilab can simultaneously provide sufficient protons to drive megawatt-class $\mathcal{O}$(GeV) proton beams as well as the multi-megawatt LBNF/DUNE beamline. By coupling the PIP-II LINAC to a new Booster-sized, permanent magnet or DC-powered accumulator ring, the protons can be compressed into pulses suitable for a proton beam dump facility with a rich physics program. The accumulator ring could be located in a new or existing beam enclosure and be designed to operate at 800~MeV but with an upgrade path allowing for future operation in the GeV range. The accumulator ring would initially provide 100~kW of beam power, limited by stripping foil heating, and have a $\mathcal{O}(10^{-4})$ duty factor.  One variant of this accumulator ring would be a $\sim$100 m circumference ring operating at 1.2~GeV with a pulse width of 20~ns and a duty factor of $\mathcal{O}(10^{-6})$, which would greatly reduce steady-state backgrounds.  Another is an accumulator ring coupled to a new rapid cycling synchrotron replacing the Fermilab Booster with an increased proton energy of 2~GeV and an increased beam power of 1.3 MW~\cite{Ainsworth:2021ahm}.

PIP2-BD is a cylindrical, 100-ton, scintillation-only LAr detector placed on-axis, 18 m downstream from a carbon proton beam dump.  PIP2-BD's sensitivity to various dark sector models has been studied for the accumulator ring scenarios listed above, assuming a 5-year run and a 75\% uptime~\cite{Toups:2022yxs}.

{\bf T2HKK}~\cite{Hyper-Kamiokande:2016srs, Abe:2018uyc}: 
The strategy of the Hyper-Kamiokande (HK) experiment is to build two identical water-Cherenkov detectors of total 520 kt.
The first one of 260kt total mass is under construction in Japan at 295~km from the J-PARC neutrino beam with $2.5^\circ$ Off-Axis Angles (OAAs).
Building the second one in Korea is being promoted since the J-PARC neutrino beam passes through Korea with an $1-3^\circ$ OAA. 
Placing the second one in Korea rather than in Japan will enhance almost all physics capabilities, especially neutrino oscillation physics, due to the longer baseline ($\sim$1,100 km) and the larger overburdens of the Korean candidate sites ($\sim$1,000 m).

{\bf Nuclear Physics Facilities} 
One of the most advantageous features at the future neutrino experiments, that enables exploration of BSM physics is the necessary high intensity proton beams to produce large number of neutrinos. 
Many nuclear physics facilities also possess such features, in particular the rare nuclear isotope accelerator facilities, such as ISOLDE at CERN~\cite{Catherall:2013poa}, FRIB at the Michigan State University in the U.S.~\cite{frib} and the RAON~\cite{Jeon:2014oja}, the accelerator facility under construction in Korea.
While the energies of the beams in these facilities are necessarily low due to the just-rightness nature of producing a high flux rare nuclear isotopes, the high intensity of the beams, including protons and nuclei, typically in the range of a few 100s of MeV per nucleon, provides an opportunity to explore beam originating dark sector particles. A case study on an experiment called DAMSA (Dump-produced Aboriginal Matter Searches at Accelerators) was preformed recently for the RAON facility. 

The high intensity nature of the RAON beams, just like ISOLDE and FRIB, enables investigating dark sector particles, including axion-like particles and dark photons. The DAMSA at RAON places a detector system consisting of a 10 m long vacuum chamber followed by a fine grain, high speed, high resolution electromagnetic calorimeter (ECAL) immediate downstream of the tungsten beam dump at RAON. Thanks to the just-right nature of the RAON beams with 600~MeV protons, backgrounds from neutral particle are expected to be small.  For these reasons, DAMSA is capable of probing the region of ALP parameter space below the so-called “cosmological triangle'', benefiting from a high-intensity photon flux and maximizing the on-axis angular coverage. It is further found that the close proximity of the detector to the ALP production dump makes it possible to probe a high-mass region of ALP parameter space which has never been explored by the existing experiments.

\renewcommand{\tabcolsep}{1pt}
\begin{table}\footnotesize\centerline{
\begin{tabular}{|ccc|c|c|c|c|c|c|c|}
\hline
\multicolumn{3}{|c|}{Experiment} &
  $\mu$~BooNE &
  \begin{tabular}[c]{@{}c@{}}SBN \\ (ICARUS/SBND)\end{tabular} &
  NOVA &
  DUNE &
  Hyper-K &
  JSNS2 &
  CCM \\ \hline
\multicolumn{3}{|c|}{\begin{tabular}[c]{@{}c@{}}$L_\mathrm{base}$ (km)/\\ $L_\mathrm{T2ND}$ (km)\end{tabular}} &
  0.47 &
{\begin{tabular}[c]{@{}c@{}} 0.6/\\0.11 \end{tabular}}&
  810/0.9 &
{\begin{tabular}[c]{@{}c@{}} 1300/\\0.574 \end{tabular}}&
{\begin{tabular}[c]{@{}c@{}} 295/\\0.28/$\sim$1 \end{tabular}}&
{\begin{tabular}[c]{@{}c@{}} 0.024/\\0.048 \end{tabular}}& 
{\begin{tabular}[c]{@{}c@{}} 0.023/\\NA \end{tabular}}\\ \hline
\multicolumn{1}{|c|}{\multirow{5}{*}{\bf $\nu$ Beam}} &
  \multicolumn{2}{c|}{Ep (GeV)} &
  8 &
  8 &
  120 &
  80 -- 120 &
  30 &
  3 &
  0.8 \\ \cline{2-10} 
\multicolumn{1}{|c|}{} &
  \multicolumn{2}{c|}{\begin{tabular}[c]{@{}c@{}}Intensity \\ (MW)\end{tabular}} &
  0.03 &
  0.03 &
  0.75 &
  1.2 - 2.4 &
  1.3 &
  1 &
  0.1\\ \cline{2-10} 
\multicolumn{1}{|c|}{} &
  \multicolumn{2}{c|}{\begin{tabular}[c]{@{}c@{}}\textless{}$E_{\nu}$\textgreater \\ (GeV)\end{tabular}} &
  0.6 &
  0.6 &
  2 &
  3 &
  0.6 &
  0.04 &
  0.03 \\ \hline
\multicolumn{1}{|c|}{\multirow{5}{*}{\begin{tabular}[c]{@{}c@{}} {\bf Detector} \\ {\bf Parameters}\end{tabular}}} &
  \multicolumn{1}{c|}{\multirow{2}{*}{ND}} &
  Tech &
  \begin{tabular}[c]{@{}c@{}}LArTPC \end{tabular} &
  \begin{tabular}[c]{@{}c@{}}LArTPC \end{tabular} &
  \begin{tabular}[c]{@{}c@{}}Liquid \\ Scint.\end{tabular} &
  \begin{tabular}[c]{@{}c@{}}LArTPC \end{tabular} &
  \begin{tabular}[c]{@{}c@{}}Scint./H2O \\ Cerenkov\end{tabular} &
  \begin{tabular}[c]{@{}c@{}}Gd-Liquid \\ Scint.\end{tabular} &
  \begin{tabular}[c]{@{}c@{}}LArScint \end{tabular}\\ \cline{3-10} 
\multicolumn{1}{|c|}{} &
  \multicolumn{1}{c|}{} &
  $V_A$(t) &
  96 &
  112 &
  300 &
  147 &
  4/100 &
  17 &
  5  \\ \cline{2-10} 
\multicolumn{1}{|c|}{} &
  \multicolumn{1}{c|}{\multirow{2}{*}{FD}} &
  Tech &
  NA &
  \begin{tabular}[c]{@{}c@{}}LArTPC \end{tabular} &
  \begin{tabular}[c]{@{}c@{}}Liquid \\ Scint.\end{tabular} &
  \begin{tabular}[c]{@{}c@{}}LArTPC \end{tabular} &
  \begin{tabular}[c]{@{}c@{}}H2O \\Cerenkov \end{tabular} &
  \begin{tabular}[c]{@{}c@{}}Gd-Liquid \\ Scint. \\Cerenkov \end{tabular} &
  \begin{tabular}[c]{@{}c@{}} NA \end{tabular} \\ \cline{3-10} 
\multicolumn{1}{|c|}{} &
  \multicolumn{1}{c|}{} &
  \multicolumn{1}{c|}{$V_A$(t)} &
  NA &
  470 &
  14k &
  40k &
  188k &
  35 &
  NA \\ \hline
\multicolumn{1}{|c|}{\multirow{12}{*}{\begin{tabular}[c]{@{}c@{}}{\bf Detector} \\ {\bf Performance}\end{tabular}}} &
  \multicolumn{2}{c|}{$E_\mathrm{Th}$ (MeV)} &
  1 &
  0.5 &
  8 &
   {\begin{tabular}[c]{@{}c@{}} \textless{}5MeV for p;\\  \textless{}3MeV for $\pi^{\pm}$; \\  \textless{} 1MeV for e \end{tabular}}&
  $\sim$4 &
  $\sim$5 &
  $\sim$0.01\\ \cline{2-10} 
\multicolumn{1}{|c|}{} &
  \multicolumn{2}{c|}{\begin{tabular}[c]{@{}c@{}}$\sigma_E$/ \\ $\sigma_{p}$ \end{tabular}} &
  \begin{tabular}[c]{@{}c@{}}3\% $p_{\mu}$;\\  \textless{}2\% (9\%) $E_{p}$ \\ w/ KE \textgreater{}100MeV\\  (\textgreater{}40MeV); \\ =\textless{}12\% $E_{e}$\end{tabular} &
  3\% &
  \begin{tabular}[c]{@{}c@{}}$\sim$2\% @ 1 \\ GeV (range)\end{tabular} &
  4\% $E_\mathrm{reco}$ &
  \begin{tabular}[c]{@{}c@{}}\cal{O}(3\%) \\ @ 1\,GeV\end{tabular} &
  $\sim$2\% &
  $\sim$10\% \\ \cline{2-10} 
\multicolumn{1}{|c|}{} &
  \multicolumn{2}{c|}{\begin{tabular}[c]{@{}c@{}}Spatial \\ Resolution\end{tabular}} &
  \cal{O}(1\,mm) &
  \cal{O}(1\,mm) &
  1\,cm &
     {\begin{tabular}[c]{@{}c@{}} LAr: \cal{O}(1\,mm); \\  GAr: \cal{O}(250\,$\mu$m) lat. \\\cal{O}(1\,mm) drift\end{tabular}}&
  \cal{O}(20\,cm) &
  \cal{O}(10\,cm) & 
   \cal{O}(10\,cm) \\ \cline{2-10} 
\multicolumn{1}{|c|}{} &
  \multicolumn{2}{c|}{\begin{tabular}[c]{@{}c@{}}Pointing \\ Resolution\end{tabular}} &
  \cal{O}(1\,mR) &
  \cal{O}(1\,mR) &
  5\,mR &
  \cal{O}(1\,mR) &
  \cal{O}(20\,mR) &
  NA &
  NA \\ \cline{2-10} 
\multicolumn{1}{|c|}{} &
  \multicolumn{2}{c|}{\begin{tabular}[c]{@{}c@{}}Time \\ Resolution\end{tabular}} &
  \cal{O} (1\,ns) &
  \cal{O} (1\,ns) &
  \cal{O} (1\,ns) &
  \cal{O} (1\,ns) &
  \cal{O} (1\,ns) &
  \cal{O} (1\,ns) &
  \cal{O} (1\,ns)  \\ \hline
\end{tabular}}
\label{tab:det_properties}
\caption{Summary of existing and upcoming accelerator based $\nu$ experiments. Note that $L_\mathrm{T2ND}$ is the distance between the target and the near detector of the listed experiment, $V_A$ is the active volume of the detector measured in tons, $E_\mathrm{Th}$ is the energy threshold of the detector.}
\end{table}
\renewcommand{\tabcolsep}{6pt}


\section{Benchmark Models, Signatures, and Experimental Prospects
}
\label{sec:models}


Current and future neutrino beam experiments can search for a wide range of BSM signatures and probe a variety of dark sector particles, as summarized in~Table~\ref{tab:models}. In this section, we survey these exciting prospects in several classes of dark sector models, including the Higgs portal, vector portal, neutrino portal, ALP portal, dark neutrinos and dipole portal, and neutrino-philic mediators. 

\begin{table}[hbt]
\centering
\begin{tabular}{c c c}
\hline
Model & Production & Detection \\
\hline\hline
Higgs Portal & $K$, $B$ decay & Decay 
($\ell^+\ell^-$)
\\ \hline
\multirow{3}{*}{Vector Portal} & $\pi^{\pm,0},\ \eta.\ K^\pm$ Decay & Scattering ($\chi e^-$, $\chi X$, Dark Tridents) \\
 & Proton Bremmstrahlung & Decay 
 ($\ell^+\ell^-$, $\pi^+\pi^-$)
 \\ 
 & Drell-Yan & Inelastic Decay 
 ($\chi \to \chi^\prime \ell^+ \ell^-$) 
 \\ \hline
 Neutrino Portal & $\pi$, $K$, $D_{(s)}$, $B$ decay & Decay (many final states) 
 \\ \hline
ALP Portal & Meson Decay & Decay 
 ($\gamma \gamma$, $e^+e^-$, etc) \\ ($\gamma$-coupling dominant)
 & Photon Fusion & Inverse Primakoff process
 \\ 
 & Primakoff Process &  Conversion ($a\to \gamma$)
 \\ \hline
  Dark Neutrinos & SM Neutrino & Upscattering + Decay 
 ($\nu\to \nu_D$, $\nu_D \to \nu \ell^+ \ell^-$)
 \\ 
Dipole Portal & Dalitz Decay & Decay ($\nu_D \to \nu \gamma$)
  \\\hline
 $\nu$philic Mediators & SM Neutrino & Scattering (Missing $\slashed{p_T}$, SM Tridents) \\ \hline
\hline
\end{tabular}
\caption{A selection of models that can be probed by neutrino beam experiments.}\label{tab:models}
\end{table}

\subsection{Higgs Portal
}\label{subsec:Benchmark:HiggsPortal}

\subsubsection{Model}

The Higgs portal is a minimal renormalizable extension of the standard model Higgs sector. In the minimal model, the new scalar couples to the Higgs via 
\begin{equation}
    \mathcal{L} \supset -(AS+\lambda S^2)H^\dagger H ,
\end{equation}
where $S$ is a scalar singlet, $H$ is the SM Higgs doublet, and $A$ and $\lambda$ are dimensional and dimensionless portal couplings, respectively. Such scalars may be connected to solutions to big questions in particle physics and cosmology and could also naturally be relatively light, within the kinematic reach of neutrino beam experiments.
For instance, light scalars are thermal dark matter candidates~\cite{Silveira:1985rk,McDonald:1993ex,Burgess:2000yq}, may serve as a mediator to a dark sector\cite{Pospelov:2007mp,Krnjaic:2015mbs},
 could provide a solution to the electroweak hierarchy problem through cosmological relaxation~\cite{Graham:2015cka}, or facilitate baryogenesis in the early universe~\cite{Profumo:2007wc,Croon:2019ugf}. After the SM Higgs doublet acquires a vacuum expectation value, the $S$ and $H$ fields mix. Through this mixing, the $S$-like mass eigenstate inherits the interactions of the Higgs; the mixing angle is constrained to be quite small experimentally. Denoting this angle by $\theta$, the full set of interactions between $S$ and the SM fermions and gauge bosons is
\begin{equation}
    \mathcal{L} \supset \sin \theta \, S \left( \frac{2 m_W^2}{v} W_{\mu}^+ W^{\mu-} + \frac{m_Z^2}{v} Z_\mu Z^\mu - \sum_f \frac{m_f}{v} \bar{f} f \right),
\end{equation}
where $f$ runs over the SM fermions and $v$ is the usual Higgs vev. 

In addition, the $S$ could have interactions with a hidden sector comprising other SM singlet particles, such as a separate DM candidate. In the case where the scalar couples to a DM particle $\chi$, there are two broad regimes in which the relic abundance could be obtained through thermal freeze-out, as discussed more generally previously. The secluded annihilation case, where freeze-out proceeds through $\chi \chi \to S S$, is possible but not provide a sharp predictive target on the scalar-Higgs mixing angle, as the annihilation cross section depends only on the $S$ coupling to DM. 
Thermal freeze-out is compatible with a wide range of Higgs-mediator mixing angles in this scenario. 
By contrast, the case where freeze-out occurs through annihilation of DM to SM fermions through the scalar provides a sharp prediction for the mixing angle. In this regime, however, thermal dark matter is conservatively ruled
out by a combination of rare meson decays ($B$ and $K$ meson decays), direct detection,
and off-shell Higgs width measurements \cite{Krnjaic:2015mbs}.

Given the strong constraints on scalar mediators which decay to thermal DM, we will focus on the production and decay of minimal scalar mediators which have \emph{only} SM decay channels. With this assumption, the phenomenology of such scalars at neutrino experiments generally consists of long-lived particle searches for a light, very weakly interacting $S$ decaying to leptons or 
hadrons within the detector. In the following subsections, we describe the production mechanisms and experimental signatures of the scalar decay that are relevant for neutrino facilities.

\subsubsection{Production and decay}

Proton fixed-target collisions at neutrino experiments can produce scalar mediators through their mixing with the Higgs in three main ways: meson decay, bremsstrahlung, and the Drell-Yan process. The relative importance of these modes is solely a function of the scalar mass, as all production mechanisms involve one insertion of the mixing angle. We review these mechanisms here.

\paragraph{Meson decay}
The main source of scalars through meson decay is through the decay $K \to \pi S$. The dominant contributions to the decay rate in these flavor-changing processes come from penguin diagrams where a scalar is emitted from a top quark. For charged kaons, the decay width is approximately~\cite{Willey:1982mc,Leutwyler:1989xj,Gunion:1989we,Batell:2019nwo}   

\begin{equation}
    \Gamma(K^\pm \to \pi^\pm S) \approx \frac{\sin^2 \theta}{16 \pi m_K} \left| \frac{3 V_{td}^* V_{ts} m_t^2 m_K^2}{32 \pi^2 v^3} \right|^2 \lambda^{1/2} \left( 1, \frac{m_S^2}{m_K^2}, \frac{m_\pi^2}{m_K^2} \right)
\end{equation}
where the kinematic function is $\lambda(a, b, c) = a^2 + b^2 + c^2 - 2 a b - 2 a c - 2 b c$. Neutral $K_L$ decays can produce scalars in an analogous fashion. By contrast, $K_S$ decays are correspondingly less important because the significant SM $2 \pi$ decay width leads to a much smaller relative branching fraction to scalars than for other kaons.

For heavier scalars and with higher energy beams 
the decays of other mesons can also provide an important production mechanism. Notably, $B \to K S$ would proceed through similar penguin diagrams as the kaon decays above. However, the production of $B$ mesons is quite suppressed relative to kaons at the energies that are typical of fixed target neutrino experiments, due to the heavier bottom quark mass.
On the other hand, for TeV-scale proton collisions, such as at SHiP~\cite{Alekhin:2015byh}, or at the LHC, $B$ decays can be the dominant scalar production mode. Above the scalar mass threshold $m_K - m_\pi$ where scalars can no longer be produced through $K$ decays, other modes such as bremsstrahlung can be important.
The production of scalars through direct bremsstrahlung off quark lines in decays of other mesons, such as $\eta$ and $\eta'$ mesons, is also relatively small compared to the other $S$ production mechanisms discussed here. 

\paragraph{Bremsstrahlung}
Among other channels, forward bremsstrahlung of dark scalars is particularly important in the mass range of around a GeV, due to the possibility of enhancement via mixing with hadronic resonances with the same quantum numbers. However, computing this production rate in the forward region involves nonperturbative (diffractive scattering) QCD processes associated with the forward $pp$ cross section, and thus requires various approximations~\cite{Boiarska:2019jym,Foroughi-Abari:2021zbm}.

For sub-GeV mass dark sector states and low energy reactions the dark sector state couples with the proton coherently. Following directly from the portal interaction, one considers the induced coupling to protons,
\begin{equation}
 {\cal L}_{\rm eff} \supset - g_{SNN} \, \sin\theta \, S \bar{p} p 
\end{equation}
where 
the coupling to nucleons $g_{SNN} = 1.2 \times 10^{-3}$ can be obtained from QCD low energy theorems~\cite{Boiarska:2019jym}. For heavier states and at higher energies, one needs to extend this coupling to include a timelike scalar nucleon form factor, incorporating mixing with isoscalar $J^{PC}= 0^{++}$ scalar resonances through a sum of Breit-Wigner components~\cite{Batell:2020vqn,Foroughi-Abari:2021zbm}.

\begin{figure}
    \centering
    \includegraphics[width=0.49\textwidth]{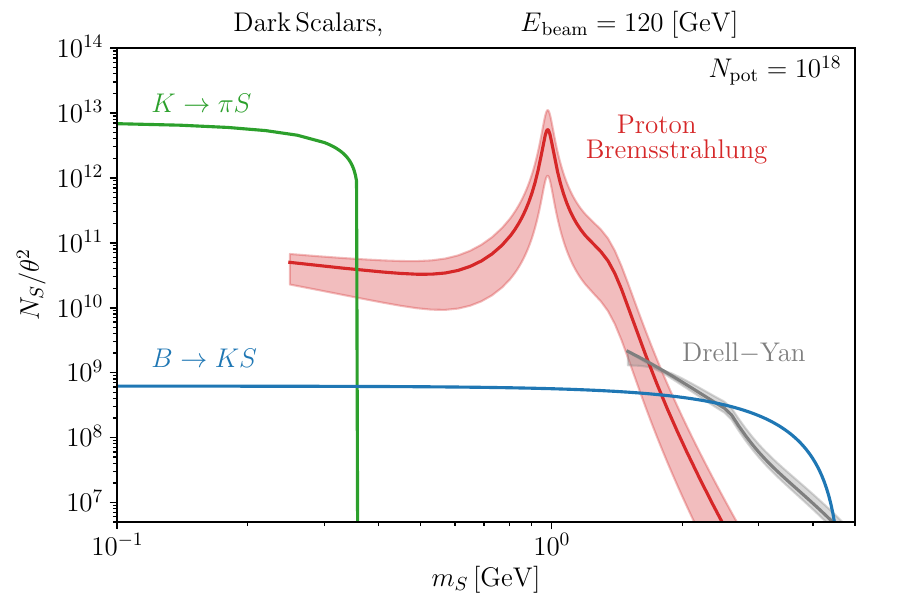}
    \includegraphics[width=0.49\textwidth]{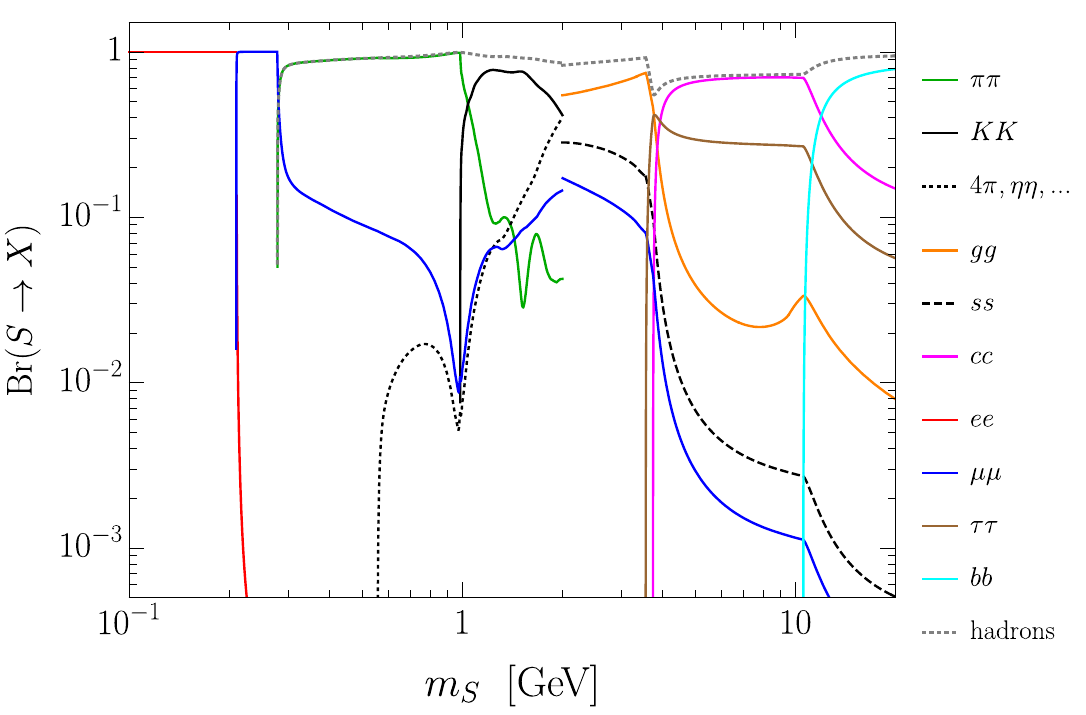}
    \caption{Left: Scalar production rates in 120 GeV proton fixed target collisions through kaon decay, $B$-meson decay, proton bremsstrahlung, and Drell-Yan; taken from Refs.~\cite{Foroughi-Abari:2021zbm}.
    Right: Branching fractions of a Higgs portal scalar as a function of scalar mass following Ref.~\cite{Winkler:2018qyg}.}
    \label{fig:prod_decay}
\end{figure}

In Fig. (\ref{fig:prod_decay}), the red curve denotes the proton bremsstrahlung rate using the quasi-real approximation in non-single diffractive scattering, and the uncertainty band results from varying a cut-off scale $\Lambda_p \in [1,2]$ GeV corresponding to the intermediate off-shell proton form-factor (where the central value is 1.5 GeV). 

\paragraph{Drell-Yan} 
Finally, for scalar masses at or above the QCD scale, a partonic description of the proton collisions is appropriate, and the $S$ can be produced through gluon fusion and Higgs-$S$ mixing. The cross section for direct production is~\cite{deNiverville:2012ij}
\begin{equation}
    \sigma(p p \to S) = \frac{\alpha_s^2 N^2 \sin^2 \theta}{576 \pi v^2} \sum_q \int_\tau^1 \frac{dx}{x} \tau f_g(x) f_g(\tau / x)
\end{equation}
where $N$ is the number of quarks heavier than $0.2 m_S$. While direct $S$ production is the only available mechanism for $S$ above the GeV scale, its cross section tends to be low compared to meson production or bremsstrahlung in the mass regions where each of those modes dominates.

The left panel of Fig.~\ref{fig:prod_decay} shows the total scalar production arising from meson decay, bremsstrahlung and Drell-Yan production as a function of the scalar mass for a 120 GeV proton beam, e.g.,~NuMI. We see that for light scalar masses, kaon decay is the main source of $S$ production. Beyond the mass threshold for kaon decay to produce scalars, bremsstrahlung takes over, followed by direct production at the GeV scale. Note that this figure shows the total production without regard to the angle at which the scalars are emitted relative to the initial proton beam. While the majority of the scalars are well-collimated, the evaluation of a signal at a neutrino detector involves simulating the full phase space distribution of the produced scalars, followed by their decays. We now turn to the latter question.

Assuming only SM decays, the scalar decays exactly as a light Higgs boson would, with the caveat that its decay widths are all suppressed by the square of  mixing angle. For light scalars with phenomenologically viable mixing angles, this typically results in long $S$ lifetimes, such that a scalar produced at the proton target can live well beyond its arrival at a downstream neutrino detector placed hundreds of meters away. When $m_S < 2 m_\pi$, the scalar can only decay to electrons or muons, with decay width
\begin{equation}
    \Gamma(S \to \ell^+ \ell^-) = \sin^2 \theta \frac{m_\ell^2 m_S}{8 \pi v^2} \left( 1 - \frac{4 m_\ell^2}{m_S^2} \right)^{3/2}
\end{equation}

When the $S$ is heavy enough to decay hadronically, however, the calculation of its decay width becomes more complicated. The decay width of the scalar to two pions is~\cite{Voloshin:1985tc,Donoghue:1990xh}
\begin{equation}
    \Gamma(S \to \pi \pi) = \sin^2 \theta \frac{3 |G_\pi(m_S^2)|^2}{32 \pi v^2 m_S} \left(1 - \frac{4 m_\pi^2}{m_S^2} \right)^{1/2}\ .
\end{equation}
Here, $G_\pi$ is a form factor which is in general derived from meson scattering data, though it can be approximated by chiral perturbation theory just above threshold. For scalars between the two pion threshold and 1.0 GeV, estimates of the width can be obtained using chiral perturbation theory and dispersion methods (see e.g., Ref.~\cite{Winkler:2018qyg}), while for even heavier scalars one can employ a perturbative partonic description.  
The decay modes of the Higgs portal scalar are summarized in the right panel of Fig.~\ref{fig:prod_decay}. As the scalar mass increases, new decay modes that become kinematically accessible rapidly dominate over existing ones, a consequence of the Higgs-like couplings of the scalar. 

\subsubsection{Experimental limits and prospects}

\begin{figure}
    \centering
    \includegraphics[height=3.3in]{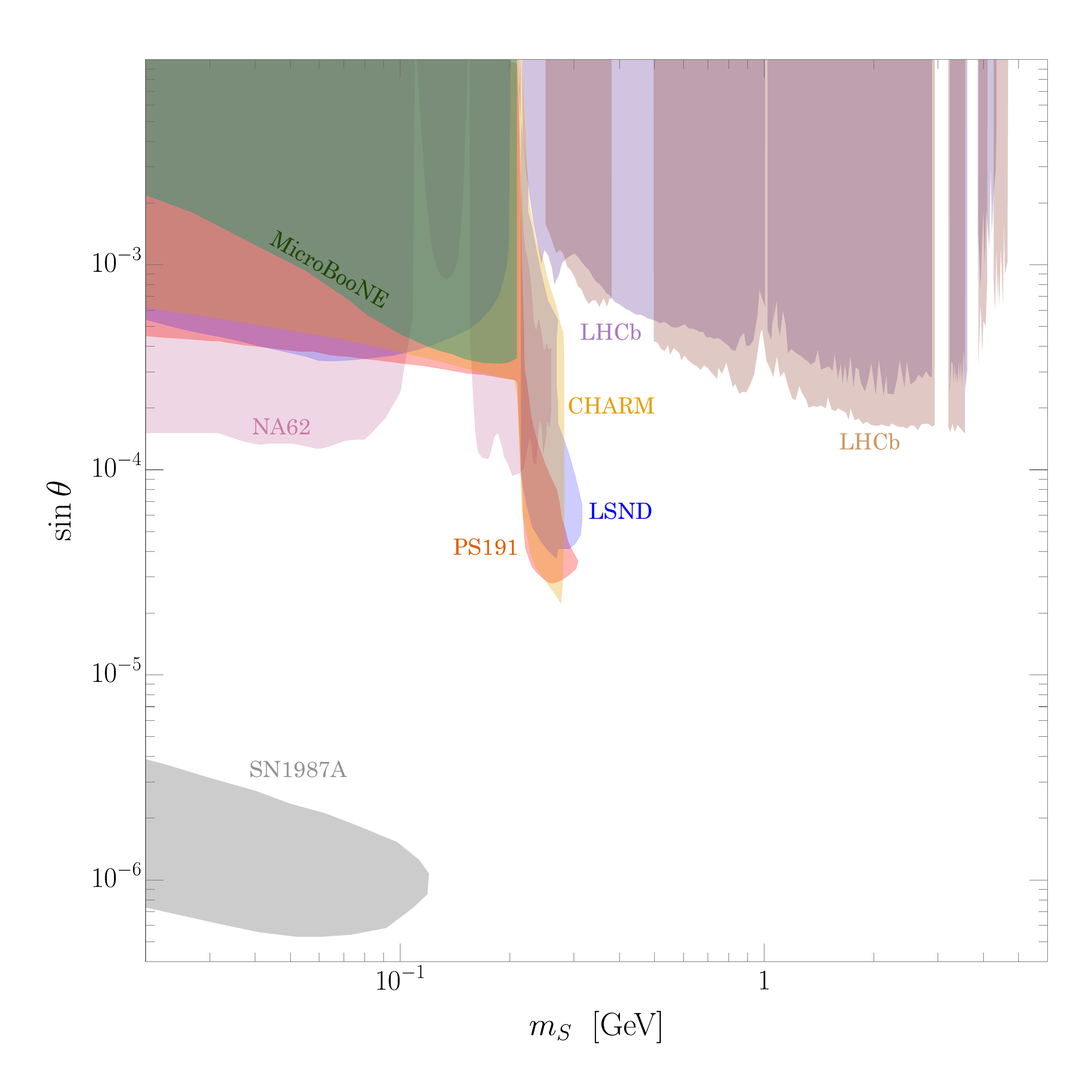}
        \includegraphics[height=3.3in]{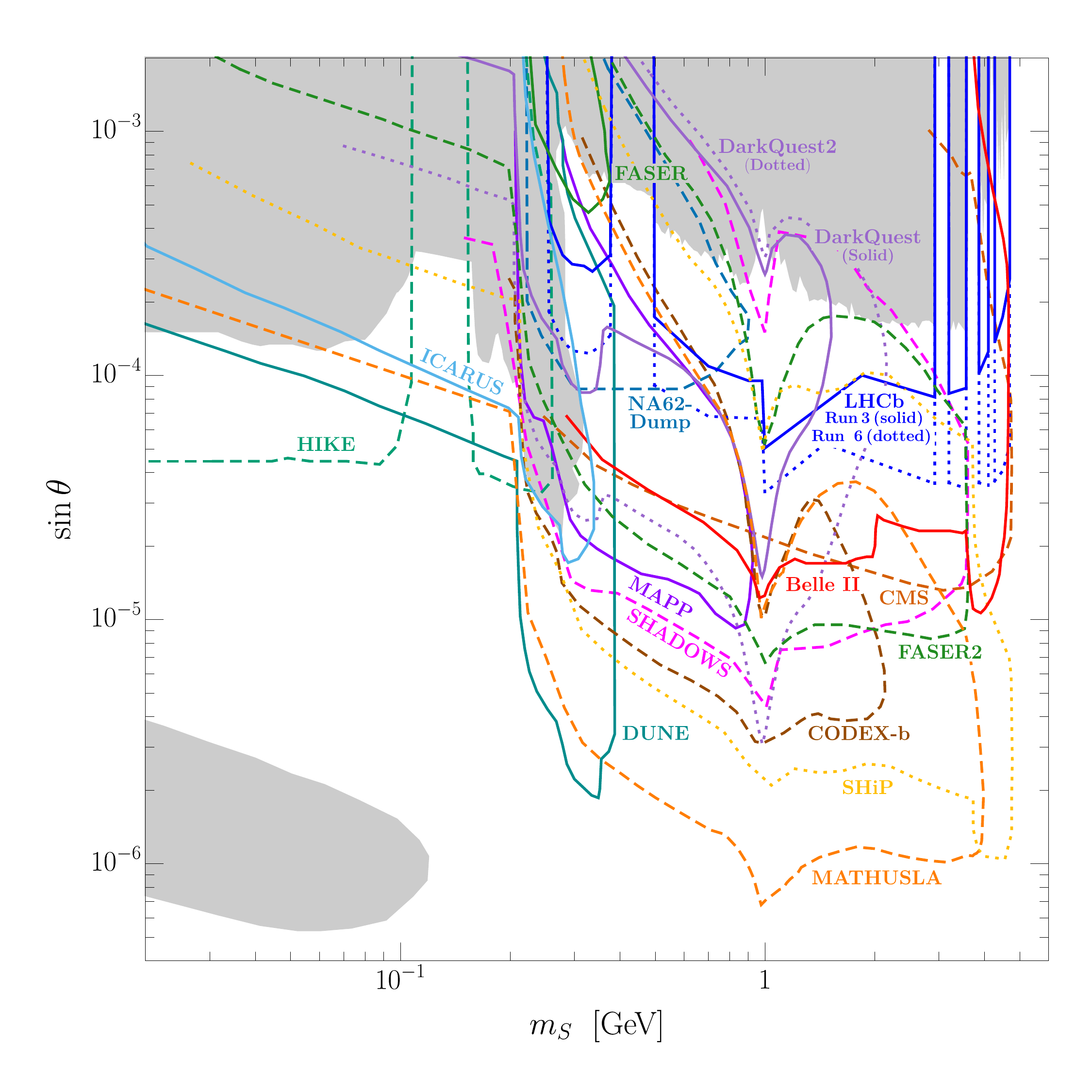}
\caption{Current limits (left panel) and projected reach (right panel) on a Higgs portal scalar. 
Existing bounds come from LHCb~\cite{Aaij:2015tna,Aaij:2016qsm}, LSND~\cite{LSND:1997vqj,LSND:2001aii,Foroughi-Abari:2020gju}, PS191~\cite{Bernardi:1985ny,Bernardi:1987ek,Gorbunov:2021ccu}, MicroBooNE~\cite{MicroBooNE:2021usw}, CHARM~\cite{Bergsma:1985qz,Winkler:2018qyg,Egana-Ugrinovic:2019wzj,Goudzovski:2022vbt}, NA62~\cite{NA62:2021zjw,NA62:2020xlg,NA62:2020pwi}, and SN1987A~\cite{Dev:2020eam}.
Also shown are projections from a number of existing and proposed future experiments, including 
ICARUS~\cite{Batell:2019nwo},
DUNE~\cite{Berryman:2019dme},
Belle II~\cite{Kachanovich:2020yhi,Filimonova:2019tuy}, 
LHCb~\cite{Craik:2022riw},
CMS~\cite{Evans:2020aqs},
DarkQuest~\cite{Batell:2020vqn,Berlin:2018pwi},
FASER and FASER2~\cite{Feng:2017vli},
MoEDAL-MAPP~\cite{Pinfold:2791293,MoEDAL},
CODEXb~\cite{Aielli:2022awh},
MATHUSLA~\cite{Curtin:2018mvb},
NA62-Dump~\cite{Beacham:2019nyx},
SHADOWS~\cite{SHADOWS}, 
HIKE~\cite{SHADOWS},
and SHiP~\cite{Alekhin:2015byh}.
}
    \label{fig:higgsportal_limits}
\end{figure}

Having discussed the production and decay of Higgs portal scalars, we now turn to the existing bounds and future sensitivity at neutrino beam experiments. A summary is presented in Figure~\ref{fig:higgsportal_limits}, which shows the constraints and projections in the scalar mass $(m_S)$ - mixing angle $(\sin\theta)$ plane. 

Some of the best existing constraints on dark scalars arise from past neutrino experiments. 
At LSND, a recent analysis~\cite{Foroughi-Abari:2020gju} incorporating the effects of proton bremsstrahlung found that LSND provides the leading current limits for some scalar masses, notably just above the dimuon threshold.
The MicroBoone collaboration has carried out a dedicated search~\cite{MicroBooNE:2021usw} for Higgs portal scalars produced by kaons decaying at rest, where the scalars decay in the detector to $e^+ e^-$ pairs, and the main region of new sensitivity is for scalar masses close to the pion mass, where rare kaon decay limits (see below) have a blind spot. Reinterpretations of analyses from past proton fixed target experiments such as CHARM~\cite{Bergsma:1985qz,Winkler:2018qyg,Egana-Ugrinovic:2019wzj,Goudzovski:2022vbt}, and PS191~\cite{Bernardi:1985ny,Bernardi:1987ek,Gorbunov:2021ccu}, MicroBooNE~\cite{MicroBooNE:2021usw}, also probe this parameter space. Other experimental bounds on GeV-mass scalars come from LHCb~\cite{Aaij:2015tna,Aaij:2016qsm}, PS191~\cite{Bernardi:1985ny,Bernardi:1987ek,Gorbunov:2021ccu}, CHARM~\cite{Bergsma:1985qz,Winkler:2018qyg,Egana-Ugrinovic:2019wzj,Goudzovski:2022vbt}, NA62~\cite{NA62:2021zjw,NA62:2020xlg,NA62:2020pwi},
Lastly, for smaller mixing angles than can be probed by terrestrial experiments, energy loss in SN 1987a provides a limit on Higgs portal scalars as well~\cite{Dev:2020eam}.

Looking ahead, the addition of SBND and ICARUS to MicroBooNE along the Booster beamline will eventually allow for further sensitivity to light scalars decaying to electrons, muons and pions. Notably, ICARUS is only a few degrees off-axis from the NuMI beamline, allowing for further searches for Higgs portal scalars. The potential of SBN detectors to search for such scalars has been considered phenomenologically~\cite{Batell:2019nwo}. In the future,  the DUNE near detector~\cite{Berryman:2019dme} will also be able to search for long-lived dark scalars and will have improved sensitivities by a factor of a few to ten over the SBN experiments.

At higher masses in the GeV range, proton fixed target experiments such as  DarkQuest at Fermilab will be able to powerfully probe Higgs portal scalars. During Run 3 of the LHC, FASER will also search for light scalars produced at high rapidity~\cite{Feng:2017vli,FASER:2019aik}. 
In the future, the DUNE near detector~\cite{Berryman:2019dme} will also be able to search for long-lived dark scalars and will have improved sensitivities by a factor of a few to ten over the SBN experiments. Other experiments which will extend the sensitivity to MeV-GeV-mass scalars include 
Belle II~\cite{Kachanovich:2020yhi,Filimonova:2019tuy}, 
LHCb~\cite{Craik:2022riw},
CMS~\cite{Evans:2020aqs},
FASER2~\cite{Feng:2017vli},
MoEDAL-MAPP~\cite{Pinfold:2791293,MoEDAL},
CODEXb~\cite{Aielli:2022awh},
MATHUSLA~\cite{Curtin:2018mvb},
NA62-Dump~\cite{Beacham:2019nyx},
SHADOWS~\cite{SHADOWS}, 
HIKE~\cite{SHADOWS},
and SHiP~\cite{Alekhin:2015byh}.

\subsection{Vector Portal
}\label{subsec:Benchmark:VectorPortal}
Numerous dark sector scenarios interact with the SM through the aforementioned the vector portal, often by introducing a new vector boson called a dark photon, $A^\prime$ or $V$. This section will provide a brief summary of the most popular vector portal models and how the new particles introduced by these models can be produced and detected at neutrino beam experiments. This section will focus on dark photons with couplings to protons and masses $m_{A^\prime} \in [2m_e, \mathcal{O}(1\,\mathrm{GeV})]$ as the range of greatest interest to neutrino beams, with some exceptions.

\subsubsection{Models}
\label{ssec:vector_portal_intro}
    
In the simplest variant, the \textbf{minimal dark photon}, the dark photon is the gauge boson of some novel $U(1)^\prime$ symmetry. SM particles are not charged under this new symmetry, but the dark photon couples to hypercharge when the $U(1)^\prime$ symmetry is broken:
\begin{equation}
    \mathcal{L}_{A^\prime} = -\frac{1}{4}A_{\mu\nu}^2 - \frac{\varepsilon}{2 \cos\theta_W}A^\prime_{\mu\nu}B^{\mu\nu} - \frac{1}{2} m_{A^\prime}A^\prime_\mu A^{\prime \nu},
\label{eq:Lagrangian_dark_photon}
\end{equation}
where $m_{A^\prime}$ is the dark photon mass, $\varepsilon$ is the kinetic mixing strength parameterising the strength of the coupling between the standard model and the dark sector, $A^\prime_{\mu\nu} \equiv \partial_\mu A^\prime_\nu - \partial_\nu A^\prime_\mu$ is the $U(1)^\prime$ field strength and $B^{\mu\nu}$ is the hypercharge field strength. 

The minimal dark photon predominantly decays into pairs of charged leptons or, at larger masses, into hadronic states, providing a clear experimental signature. The strongest limits on the model come from asymmetric electron-positron colliders, rare decay searches and both proton and electron beam dumps (see left panel of Fig. \ref{fig:dark_photon_limits}). Dark photons with very weak couplings can also be ruled out by supernova cooling \cite{Chang:2016ntp}. Dark photons that couple to muons alter the muon magnetic moment \cite{Pospelov:2008zw}, potentially resolving the existing 4.2$\sigma$ disagreement between theory and experiment \cite{Muong-2:2021ojo, Muong-2:2006rrc, Aoyama:2020ynm}. The dark photon explanation for the $(g-2)_\mu$ discrepancy has been excluded by existing searches for the minimal dark photon (see the left panel of Fig. \ref{fig:dark_photon_limits}), but could be explained by baryophobic and electrophobic variants such as models that gauge $L_\mu-L_\tau$ \cite{Baek:2001kca}.

A scalar or fermionic particle $\chi$ charged under the $U(1)^\prime$ symmetry may be added to the minimal dark photon model to create a thermal relic dark matter model,
\begin{align}
    \mathcal{L}_{\mathrm{Scalar}}& \supset |D_\mu \chi|^2 - m_\chi^2 |\chi|^2\\
    \mathcal{L}_{\mathrm{Fermion}}& \supset \chi(i \slashed{D}_\mu - m_\chi)\bar\chi,
\end{align}
where $m_\chi$ is the mass of the new dark matter candidate, $D_\mu \equiv (\partial_\mu - i g_D A^\prime_\mu)$ and $g_D$ is a dark coupling constant. $g_D$ is often rewritten as the dark coupling strength $\alpha_D \equiv \frac{g_D^2}{4\pi}$. By introducing a new annihilation channel into SM particles through the dark photon these scenarios are able to reproduce the observed dark matter relic density. If $2 m_\chi < m_{A^\prime}$, then the dominant decay mode of the dark photon will be to the invisible dark matter states. The primary model signatures are then either missing energy or momentum from dark photon production, the scattering of the dark matter off of electrons and nucleons in the detector. Limits on the minimal dark matter scenario are shown in Fig. \ref{fig:dark_matter_limits}.

The fermionic dark matter scenario requires a small mass splitting between dark matter states in order to evade constraints from the Cosmic Microwave Background. This splitting need not remain small, though. In the inelastic dark matter scenario, there are two fermionic dark sector particles $\chi_1$ and $\chi_2$, with $m_{\chi_2} > m_{\chi_1}$ and the fractional mass difference between the two states defined as $\Delta = \frac{m_{\chi_2} - m_{\chi_1}}{m_{\chi_1}}$. Depending on the parameters chosen, the $\chi_2$ can be sufficiently long-lived to reach a detector and be observed through the inelastic decay $\chi_2 \to \chi_1 \ell^+ \ell^-$, or through up(down) scattering.

Finally, the leptophobic dark photon $A_B$ gauges baryon number rather than $U(1)^\prime$, giving the interactions
\begin{equation}
    \mathcal{L}_B \supset - A^\mu_B (g_B J^B_\mu + g_\chi J^\chi_\mu + \varepsilon_B e J_\mu^\mathrm{EM}),
\end{equation}
where $J_\mu^B \equiv \frac{1}{3} \sum_i \bar q_i \gamma_\mu q_i$ is the SM baryonic current, $J_\mu^\chi$ is the dark current seen previously, $J_\mu^\mathrm{EM}$ is the SM electromagnetic current, and $\varepsilon_B \sim \frac{eg_B}{(4\pi)^2}$ is a small kinetic mixing generated at the loop level. The leptophobic scenario escapes many of the strong constraints on the previous scenarios that arise through lepton couplings to the dark photon, but has some model building complications. As the scenario is not anomaly free, the introduction of an "anomalon" to resolve the anomaly can create strong model dependent constraints \cite{Dror:2017ehi}. Other alternatives that are of interest to neutrino beam experiments include gauge bosons that couple to anomaly free combinations of baryon and lepton family numbers; see e.g., ~\cite{Heeck:2018nzc,Kling:2020iar,Batell:2021snh}.

\subsubsection{Production}
\label{ssec:vector_portal_production}
 In principle, any process in which photons participate at a neutrino facility can lead to the production of $A^{\prime}$ (or other vector bosons which may belong to dark sectors), since the dark photon mixes with the SM photon. The main production channels include:
\begin{description}
    \item[Meson Decays] An intense flux of mesons is produced at neutrino experiments where high-energy protons hit a target. As a consequence, low-mass dark photons may be produced from the decay of neutral mesons $\mathfrak{m} \to A^{\prime} \gamma$ (mainly $\mathfrak{m} = \pi$, $\eta$, $\eta^{\prime}$ and more rarely also $K$ and $D$) provided that they have sizeable coupling to quarks. (Off-shell decays $\mathfrak{m} \to \gamma A^{\prime *}$ can also be relevant, if the dark photon acts as a portal to a dark matter particle $\chi$ and $M_{\chi} > M_{A^\prime}/2$ or $M_{A^\prime} > m_\mathfrak{m} > 2M_{\chi}$).
    Three-body decays of charged mesons, $\pi^\pm/K^\pm \to \ell \nu_\ell A'$, are a good source of the dark matter flux. While their corresponding two-body decays are helicity-suppressed, the three-body decays can utilize the full phase space, hence the resulting branching fraction can be significantly enhanced. Moreover, in beam-focusing neutrino facilities, the associated dark matter flux can be effectively focused~\cite{Dutta:2021cip}. 

    \item[Bremsstrahlung] At fixed-target experiments, the incoming charged particle (a proton) can emit forward-peaked vector particles via the radiative process $p Z \to p Z A^{\prime}$, with resonant vector meson mixing. This production mechanism is relevant for dark photons with intermediate masses between hundreds of MeV and $1$ GeV \cite{Blumlein:2013cua,Gorbunov:2014wqa,deNiverville:2016rqh,Tsai:2019buq,Foroughi-Abari:2021zbm}.

    \item[Drell-Yan]  In fixed-target collisions, dark photons can also be produced via direct partonic production. At the lowest order in QCD, this includes the Drell-Yan process $q\overline{q} \to A^{\prime}$, in which a quark-antiquark pair annihilates into the dark photon~\cite{Batell:2009di}. This is the dominant production mechanism for heavier dark photons ($m_{A^{\prime}} \gsim 1$ GeV). 
    
\end{description}

\subsubsection{Signatures}
\label{ssec:vector_portal_signatures}

The presence of dark photons and other members of vector portal dark sectors leave 
various signatures, and this section will briefly summarize those relevant to neutrino beam experiments.
\begin{description}
    \item[Visible and Inelastic Decays:] If the dark photon is sufficiently long-lived, it may propagate to a decay volume or a neutrino detector before decaying. Such a dark photon can then decay dominantly to visible states such as lepton pairs ($A^\prime \to \ell^+ \ell^-)$ or hadrons if no other dark sector states are available. In these searches, a neutrino beam acts much like a beam dump experiment, producing weakly coupled particles that can travel freely through matter and then decay visibly in a decay volume. Inelastic decays, in which a dark sector particle decays into both invisible and visible particles provide an additional decay channel in inelastic dark matter model.
    
    \item[Elastic Scattering:] A dark photon that decays invisible dark matter states produces a dark matter beam alongside the neutrino beam. These dark matter states can then be detected in a similar manner to neutrinos, through elastic scattering with the nucleons and electrons of the neutrino detector. The greatest complication to these searches is that neutrinos are the primary background, and searches must either rely on raw statistics by observing excess events above the expected neutrino rate, or find some means of differentiating the dark matter and neutrino scattering signals. One possibility is to use timing, as the heavier dark matter particles will be created in-time with the neutrinos, but arrive at the detector out-of-time due to their slightly slower travel speed. 
    
    Dark matter, like neutrinos, can also scatter coherently off nuclei at lower energy experiments like COHERENT and CCM. Timing can once again serve as a tool, though the dark matter is expected to arrive before the majority of the neutrino signal; 
    the dark matter is produced nearly instantaneously through $\pi^0$ decay and the associated events typically fall in prompt timing bins, while neutrinos are produced through the decays of the longer-lived $\pi^+$ and $\mu^+$. 
    In such experiments, muon-induced neutrino events often populate in delayed timing bins within their timing resolution. By contrast, pion-induced neutrinos are single-valued in energy (i.e., $\sim30$~MeV), hence their typical deposit energy is smaller than that of dark matter.
    Therefore, a suitable combination of timing and energy cuts can reduce neutrino background events significantly, isolating dark matter events~\cite{Dutta:2019nbn,Dutta:2020vop}. 
    A similar strategy is applicable to JSNS$^2$ in which the electron scattering channel is available~\cite{Dutta:2019nbn,Dutta:2020vop}.
    
    \item[Inelastic Scattering and Dark Tridents:] Inelastic scattering provides more complicated signatures that may be easier to disentangle from existing backgrounds. For the kinetically mixed dark matter scenarios at lower energies this includes inelastic pion production $\chi N_i \to \chi N_j \pi^{0,\pm}$ (used in the MiniBooNE DM analysis \cite{MiniBooNEDM:2018cxm}) or dark tridents, where the $\chi$ emits a virtual $A^\prime$ that produces an electron-positron pair \cite{deGouvea:2018cfv}. DM induced deep inelastic scattering becomes possible at much higher energies, such as at far forward neutrino detectors at the LHC~\cite{Batell:2021blf,Batell:2021aja,Batell:2021snh}.
    
    Inelastic dark matter scenarios require inelastic up- or down-scattering by construction, though provided the outgoing $\chi$ state is sufficiently long-lived this could look a great deal like elastic scattering. A striking signature possible for some model parameter space is that of up-scattering from the $\chi_1$ to the $\chi_2$ state followed by the decay $\chi_2 \to \chi_1 \ell^+ \ell^-$ inside of a neutrino detector~\cite{Izaguirre:2014dua,Kim:2016zjx,Izaguirre:2017bqb,Berlin:2018pwi}. 
\end{description}

\subsubsection{Millicharge Searches}
\label{ssec:mCP_Search}

A variant on the kinetically mixed dark matter scenario is that of millicharged particles (mCP) \cite{Dobroliubov:1989mr}, where the dark matter candidate carries some fraction of an electron's charge $Q=\varepsilon e$. mCP can directly arise when the new particle has a hypercharge significantly smaller than that of an electron.
The dark photon in these scenarios is either massless or completely absent. The existing limits on mCP are shown in the right panel of 
Fig.~\ref{fig:dark_matter_limits}.
Almost all the past, existing, and future neutrino experiments have sensitivity to the mCP from electron scattering events. These experiments include LSND \cite{Magill:2018tbb}, ArgoNeuT \cite{Acciarri:2019jly}, BEBC~\cite{Marocco:2020dqu}, Super-K \cite{Plestid:2020kdm}, T2K/Hyper-Kamiokande \cite{Gorbunov:2021jog}, DUNE \cite{Magill:2018tbb,Harnik:2019zee}, and FLArE~\cite{Kling:2022ykt}. 
One can also add small dedicated detectors (similar to the milliQan detector \cite{Haas:2014dda}) to the neutrino facilities to improve the mCP sensitivity, which include FerMINI at Fermilab \cite{Kelly:2018brz}, SUBMET at J-PARC \cite{Choi:2020mbk}, and FORMOSA at LHC \cite{Foroughi-Abari:2020qar}.

\begin{figure}[htb!]
    \centering
    \includegraphics[width=0.45\textwidth]{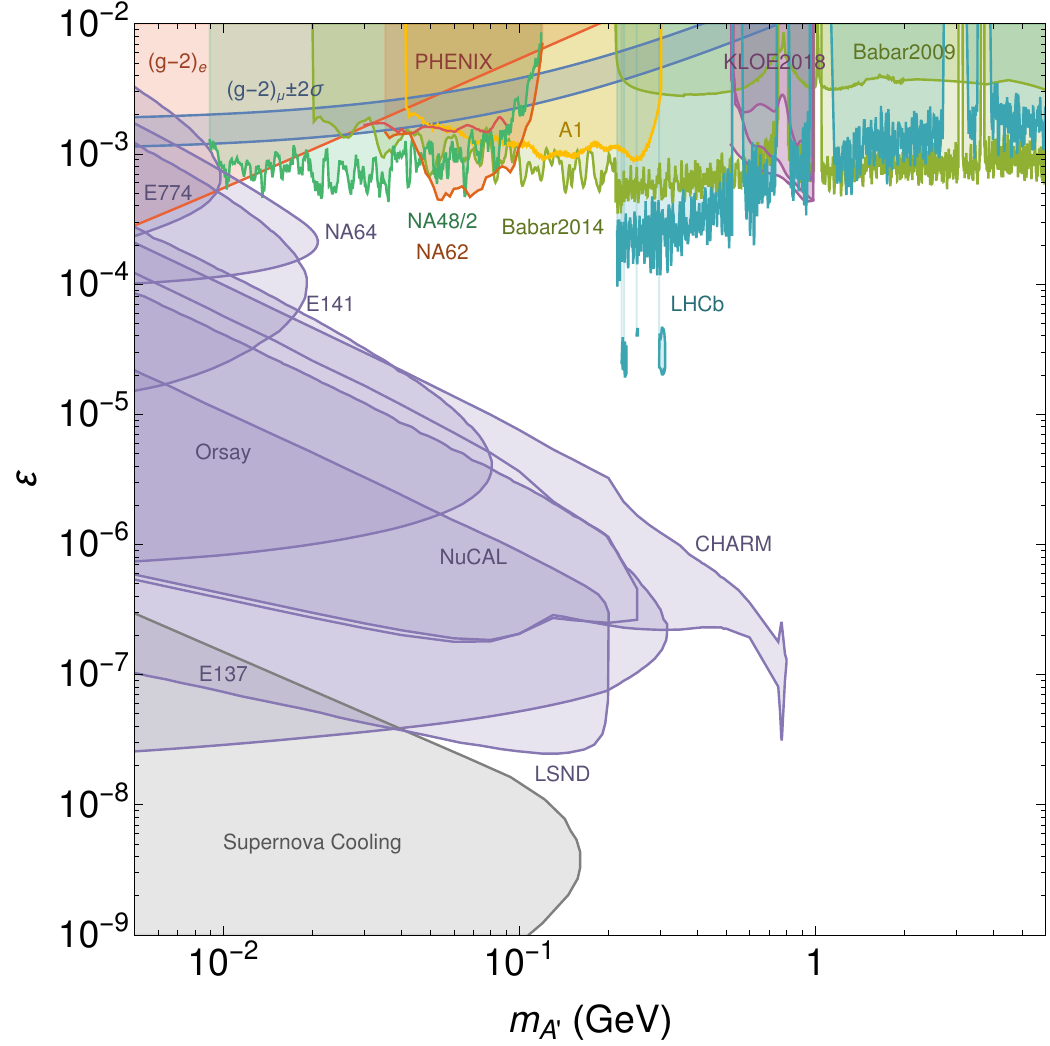} ~
    \centering
    \includegraphics[width=0.425\textwidth]{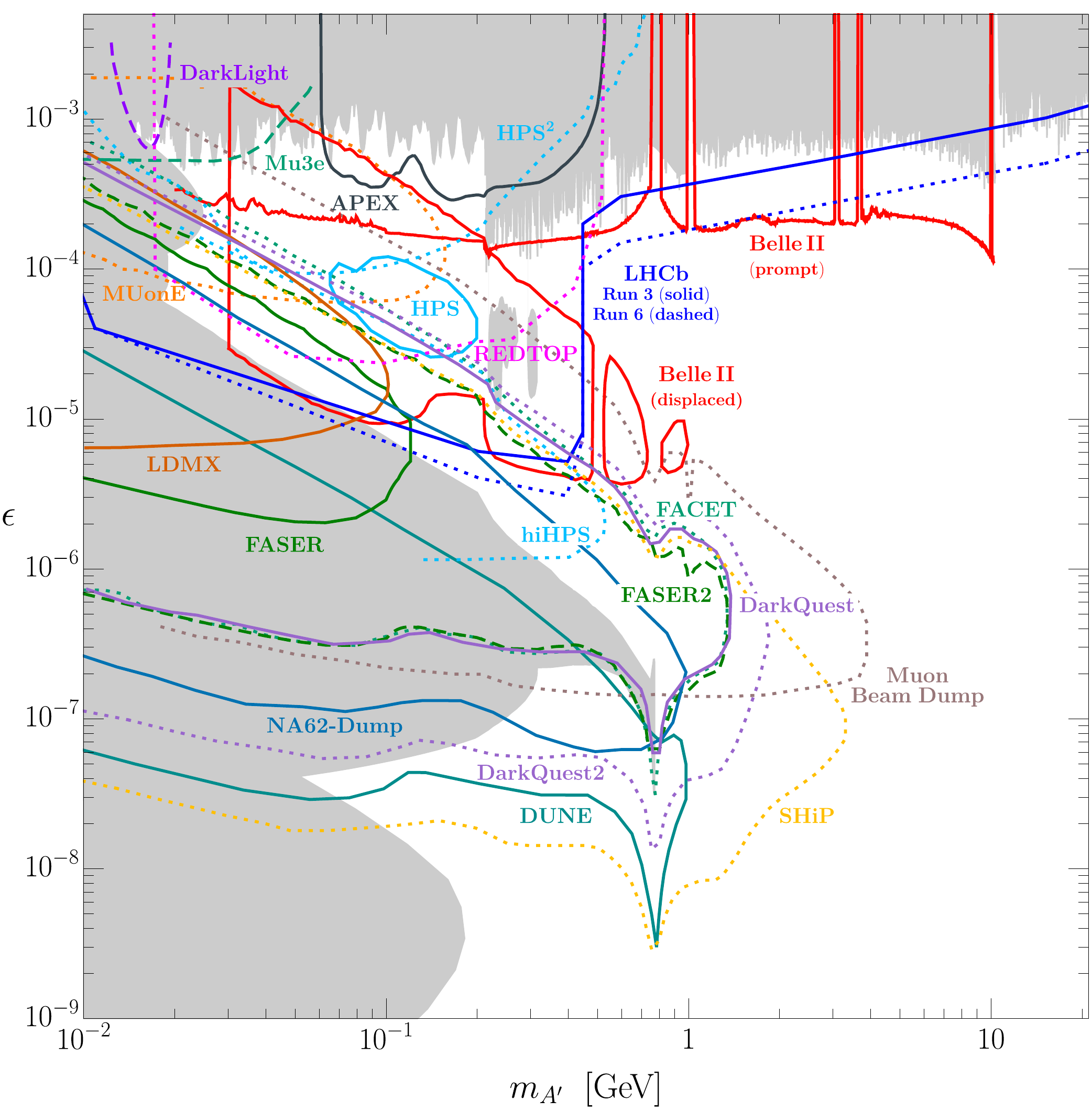}\vspace{10pt}
    \caption{Left panel: Current limits on visibly decaying minimal dark photon.  Most of the limits used to make this plot were drawn from darkcast data files \cite{Ilten:2018crw}, while the LSND decay limit is provided by Ref. \cite{Essig:2010gu}. 
  %
   %
Right panel: Projections from a number of existing and proposed future experiments, including 
DUNE~\cite{Berryman:2019dme},
Belle II~\cite{Belle-II:2018jsg,Ferber:2022ewf},
LHCb~\cite{Craik:2022riw},
FASER and FASER2~\cite{Feng:2017uoz,FASER:2018eoc}, 
HPS~\cite{Baltzell:2022rpd},
NA62-Dump~\cite{Beacham:2019nyx},
LDMX~\cite{Berlin:2018bsc},
DarkQuest~\cite{Berlin:2018pwi},
APEX~\cite{Essig:2010xa},
Mu3e~\cite{Echenard:2014lma},
DarkLight~\cite{DarkLight},
FACET~\cite{Cerci:2021nlb},
REDTOP~\cite{REDTOP:2022slw},
MUonE~\cite{Galon:2022xcl},
SHiP~\cite{SHiP:2020vbd}, and
a muon beam dump experiment~\cite{Cesarotti:2022ttv}.
  }
    \label{fig:dark_photon_limits}
\end{figure}

\begin{figure}[htb!]
    \centering
    \includegraphics[width=0.46\textwidth]{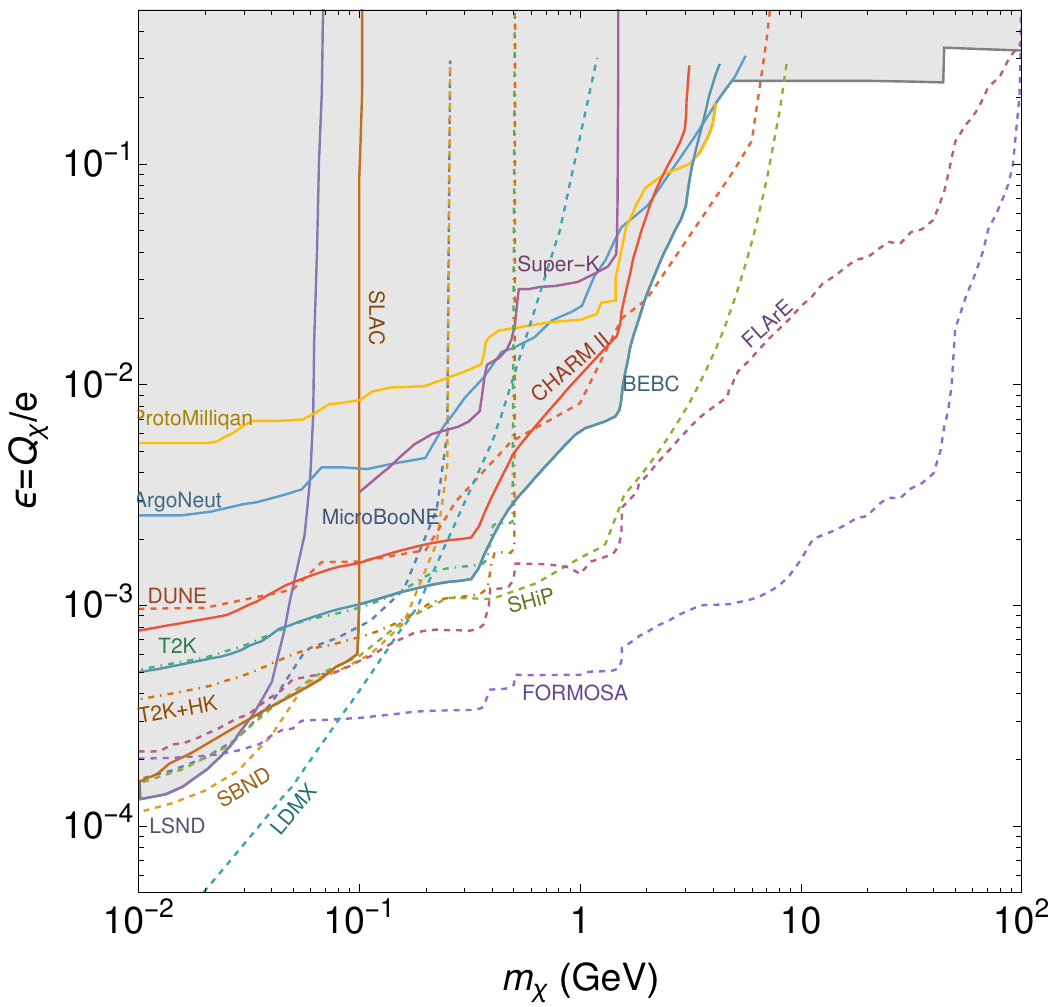}
    \caption{Limits and projections for mCP, with the $\chi$ mass on the x-axis and the electromagnetic charge of the $\chi$ as a multiple of the charge of an electron on the y-axis. Limits from BEBC and CHARM II come from \cite{Marocco:2020dqu}. Projections from MicroBooNE, SHiP, DUNE, SBND~\cite{Magill:2018tbb}, FORMOSA~\cite{Anchordoqui:2021ghd}, FLArE~\cite{Kling:2022ykt}, and T2K(+HK)~\cite{Gorbunov:2021jog} are shown as dotted lines.}
    \label{fig:mCP_Limits}
\end{figure}

\subsubsection{Experimental Prospects}
\label{ssec:vector_portal_experiments}

\begin{figure}[htb!]
    \centering
    \includegraphics[width=0.46\textwidth]{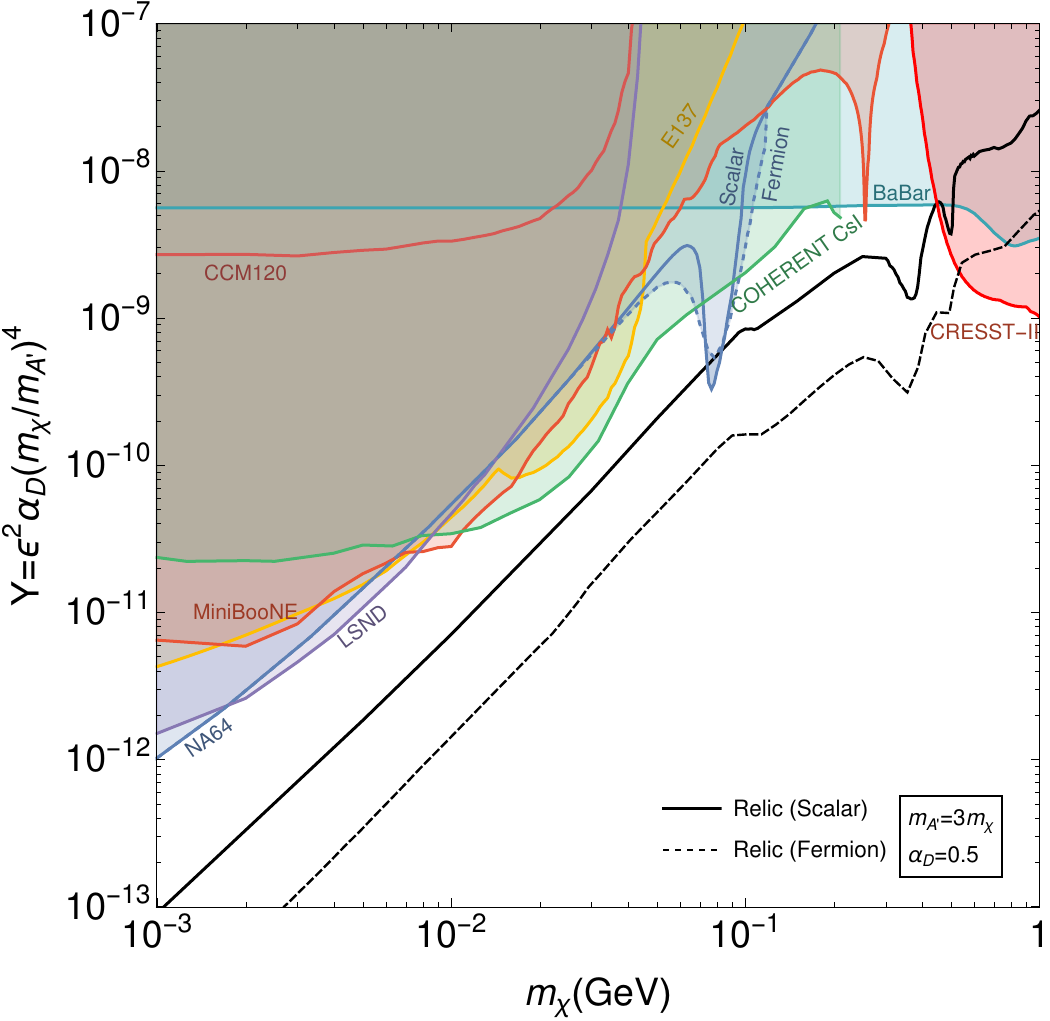}
    \centering
    \includegraphics[width=0.46\textwidth]{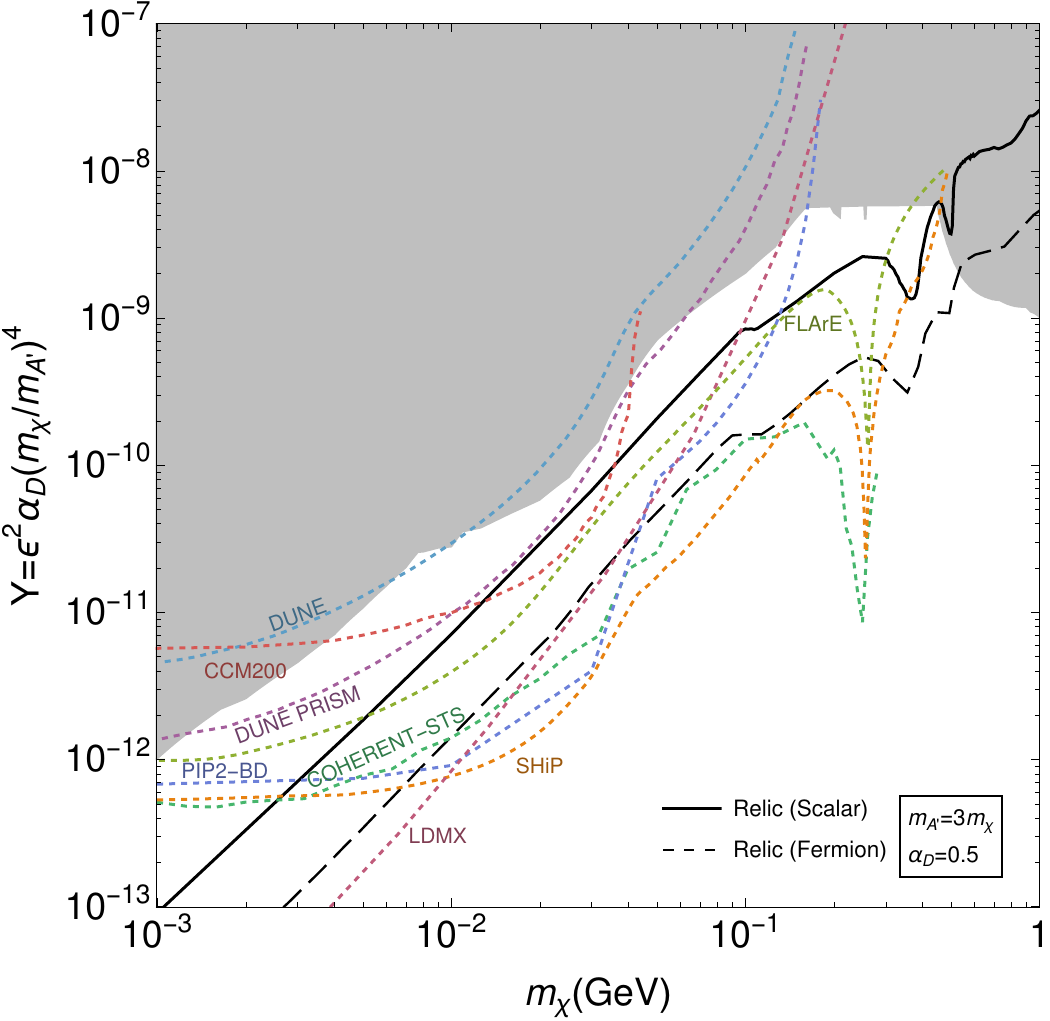}
    \caption{
        Left panel: Current limits on vector portal dark matter. The relic lines are the points in the model parameter space that reproduce the observed dark matter relic abundance, with the fermion relic line provided by \cite{Feng:2017drg}. The y-axis variable $Y$ was chosen such that the relic density line is approximately unchanged when model parameters are varied. $\alpha_D$ was set to be 0.5 and $m_{A^\prime} = 3 m_\chi$ to show a conventional two dimensional slice of the parameter space.
        Right Panel: Projections from future experiments: DUNE and DUNE-PRISM \cite{DeRomeri:2019kic,DUNE:2020ypp,DUNE:2020fgq}, COHERENT with the second target station, SHiP, CCM200 and PIP2-BD, FLArE \cite{Batell:2021blf}. Also shown for comparison is the first phase of the missing momentum experiment LDMX \cite{LDMX:2018cma}.}
        \label{fig:dark_matter_limits}
\end{figure}

\begin{figure}[htpb!]
    \centering
    \includegraphics[width=0.45\textwidth]{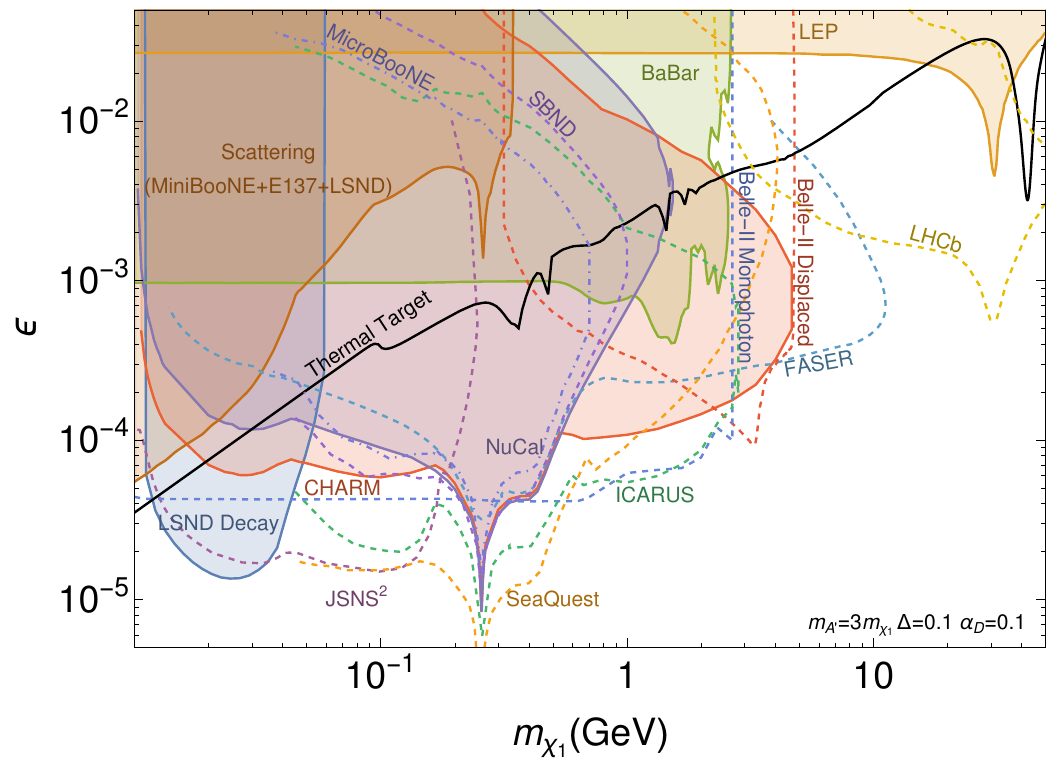}
    \centering
    \includegraphics[width=0.45\textwidth]{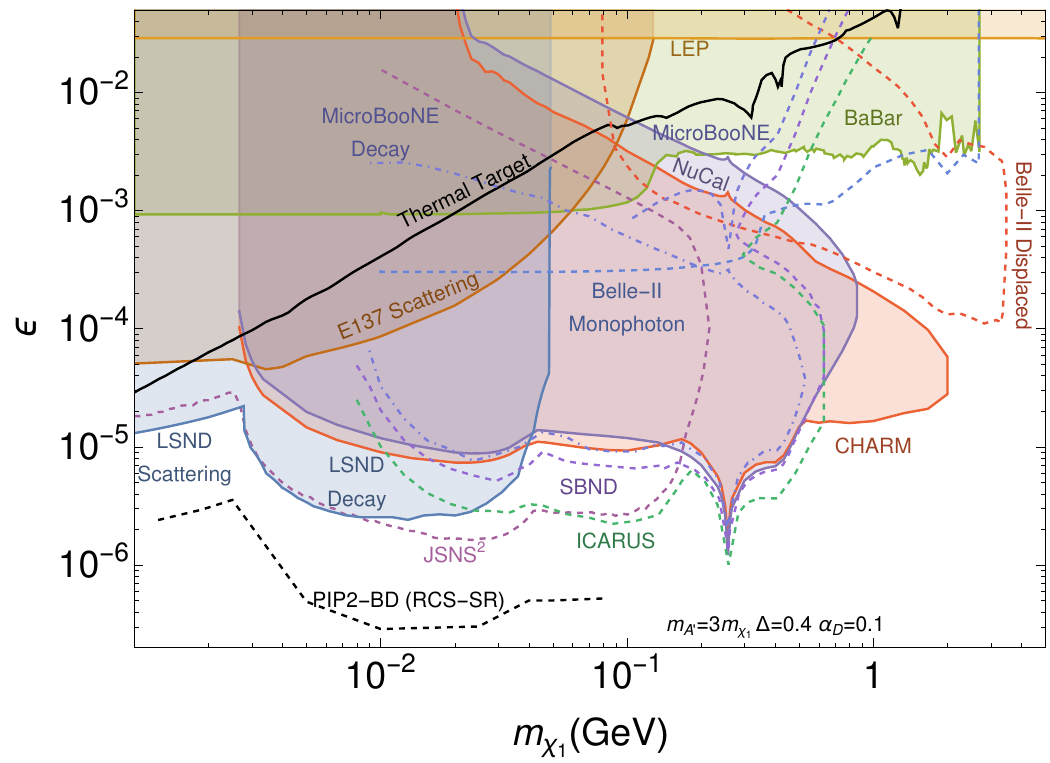}
    \caption{Two different slices of the inelastic dark matter parameter space. Left Panel: Projections of Short Baseline Neutrino experiments on the inelastic dark matter parameter space \cite{Batell:2021ooj}. Right Panel: Projections from FASER \cite{Berlin:2018jbm} and JSNS$^2$ \cite{Jordan:2018gcd}.}
    \label{fig:idm_limits}
\end{figure}

Many planned neutrino experiments provide complementary sensitivity to multiple vector portal models. The right panel of Fig. \ref{fig:dark_photon_limits} shows projections for sensitivity to minimal dark photon, Fig.~\ref{fig:mCP_Limits} shows projections for millicharged particle models, and Fig. \ref{fig:dark_matter_limits} shows projections for vector portal dark matter scenarios, and Fig. \ref{fig:idm_limits} for two different slices of the inelastic dark matter parameter space corresponding to different values of $\Delta$, $\alpha_D$ and mass ratios between the dark photon and dark matter candidate. 

\begin{figure}[htpb!]
    \centering
    \includegraphics[width=0.45\textwidth]{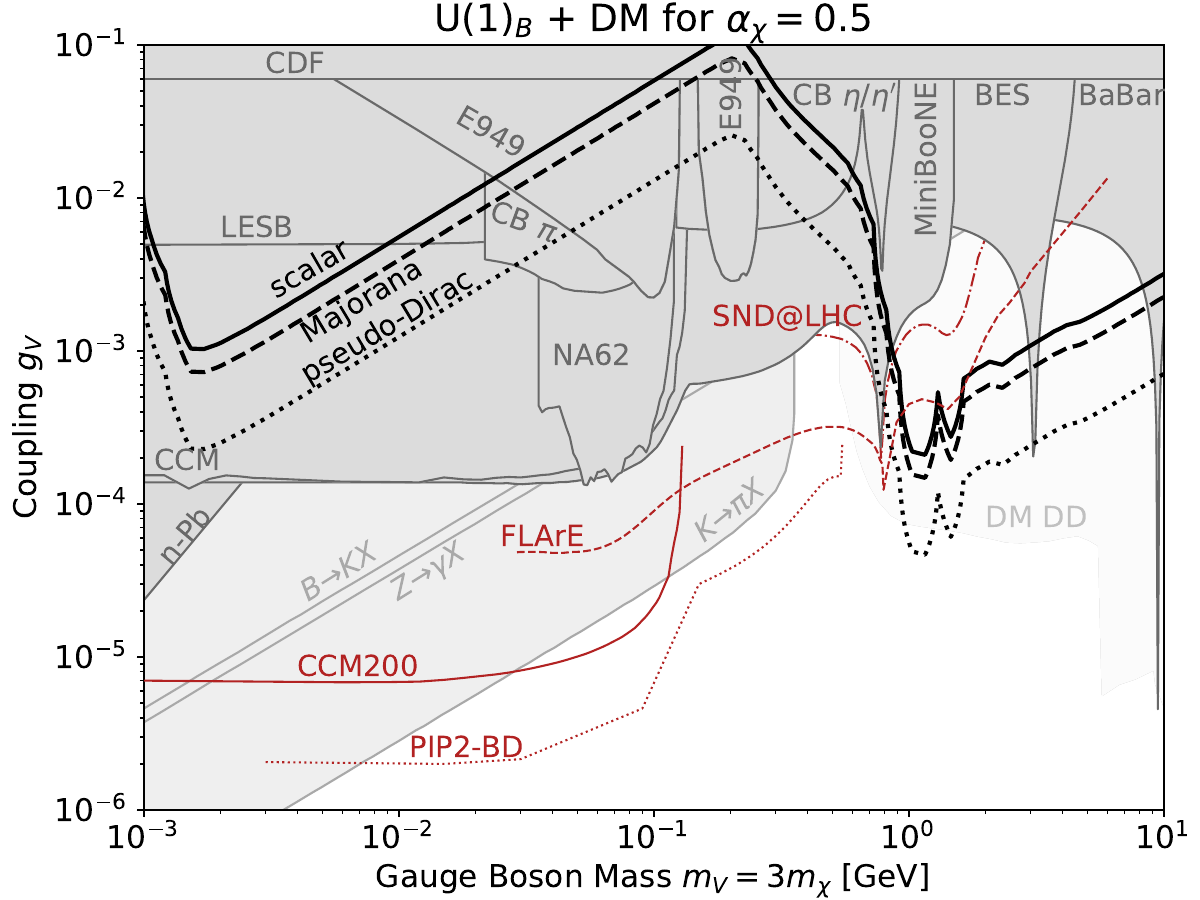}
    \centering
    \includegraphics[width=0.45\textwidth]{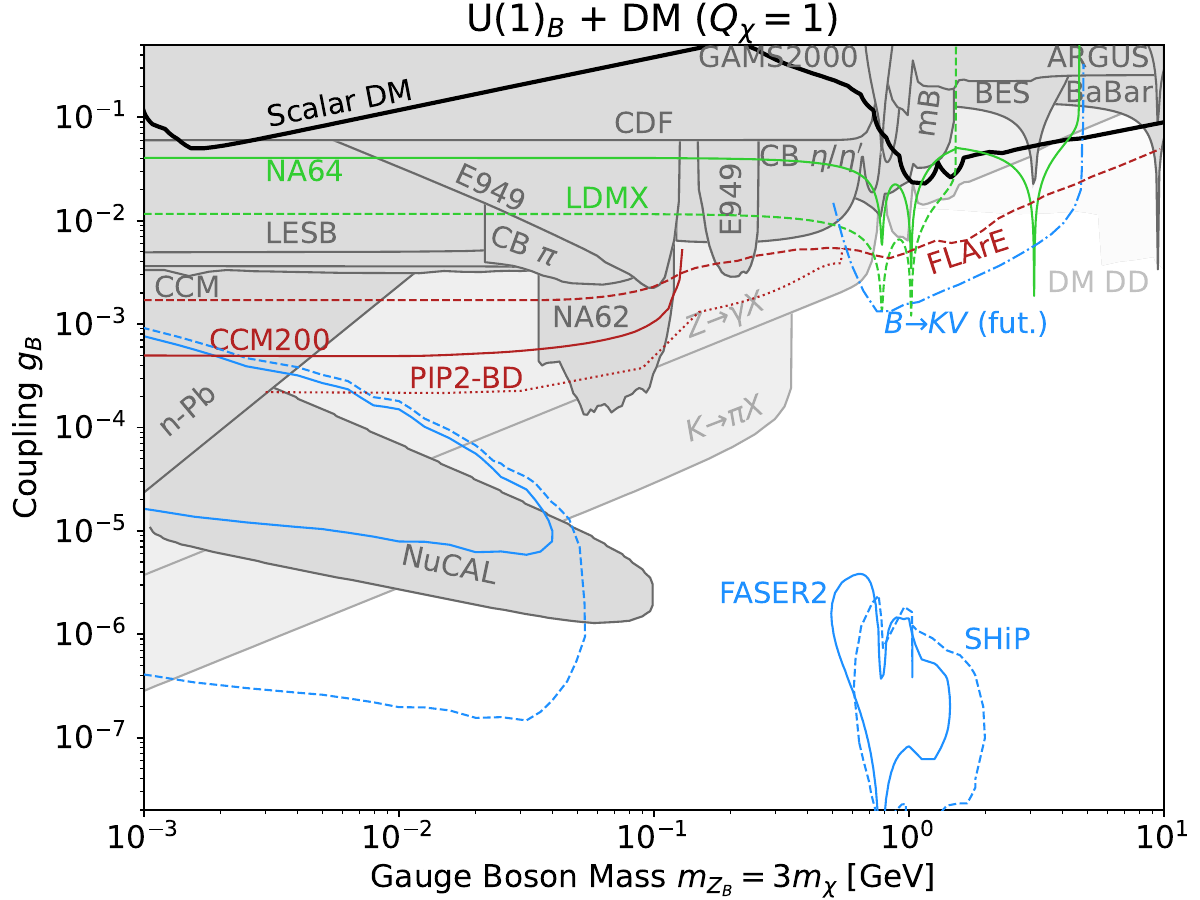}
    \caption{Slices of the parameter space of the hadrophilic model with the $U(1)_B$ mediator between DM and the SM species for $\alpha_\chi = Q_\chi^2g_\chi^2/4\pi= 0.5$ (left) and $Q_\chi = 1$ (right). Projections of future CCM-200~\cite{CCM:2021leg,Aguilar-Arevalo:2021sbh}, FLArE~\cite{Batell:2021snh}, and SND@LHC~\cite{Boyarsky:2021moj} searches for DM scattering signatures are shown with red, while blue lines correspond to visible dark vector decays in FASER 2~\cite{Batell:2021snh} and SHiP detectors, and in $B\to KV$ searches~\cite{Dror:2017ehi}. The expected senitivity lines of missing energy/momentum searches at LDMX-II and NA64 are shown with blue~\cite{Schuster:2021mlr}. Taken from Refs~\cite{Batell:2021snh,Krnjaic:2022ozp}.}
    \label{fig:u1b}
\end{figure}

\begin{description}
    \item[CCM, COHERENT and PIP2-BD] CCM \cite{CCM:2021leg} and COHERENT \cite{COHERENT:2021pvd} have both completed searches for light dark matter coherently scattering off of detector material, and projections of future work can draw on these analyses to estimate both the magnitude and scale of the backgrounds at future experiments. The projections from CCM are based on three years of running with CCM200, improved shielding and underground Argon, while the COHERENT projections are a combination of a 10t Argon and 700 kg Cesium Iodide detectors with beam from the Second Target Station (STS). PIP2-BD, a 100-ton-scale liquid argon detector discussed in section \ref{PIP2-BD}, could achieve similar sensitivity to COHERENT in dark matter searches. CCM, COHERENT and PIP2-BD are also uniquely positioned to probe $U(1)_{L_\mu-L_\tau}$ models due to their contributions to CE$\nu$NS processes \cite{Hapitas:2021ilr}.  
    
    \item[DUNE]  The future DUNE offers an interesting opportunity to probe vector portal light dark matter. Thanks to the high intensity of the beam, large signal statistics is expected thus allowing for considerable sensitivity despite the smallness of dark matter interactions with the SM particles. DUNE will search for the relativistic scattering of light-mass dark matter at the near detector \cite{DeRomeri:2019kic,DUNE:2020ypp,DUNE:2020fgq}. Light dark matter particles (either fermions of scalars) may be produced mainly via neutral pseudoscalar meson decays -- via an on-shell dark photon -- and subsequently scatter elastically off electrons (or nucleons, though with larger backgrounds) in the near detector, via a $t$-channel dark photon. We show in Fig.~\ref{fig:dark_matter_limits} the 90\% CL sensitivity reach of the DUNE near detector using $\chi e^- \to \chi e^-$ scattering. It assumes fermionic dark matter in a vector portal-type scenario, with $\alpha_D = 0.5$ and $m_{A^\prime} = 3 m_\chi$.
    For the analysis, 3.5 years of data collection each in neutrino and antineutrino modes have been assumed. Moreover, the cyan/dashed contour assumes the DUNE near detector to be all the time on axis, while the purple/dashed sensitivity is obtained allowing the near detector to move off-axis up to 24 m transverse to the beam direction (DUNE-PRISM option \cite{DUNE:2021tad}). This last configuration allows for a larger signal-to-background ratio and hence a better experimental sensitivity to the light dark matter scenario.

    \item[Short baseline neutrino experiments] SBN experiments ICARUS, SBND and MicroBooNE are all sensitive to inelastic dark matter through its inelastic decays \cite{Batell:2021ooj}. The SBN experiments can also place limits on milli-charged particle scattering \cite{Magill:2018tbb} (see Fig. \ref{fig:mCP_Limits}), and should be equally capable of searching for vector portal dark matter. JSNS$^2$ has similar capabilities, and is also sensitive to low mass inelastic dark matter \cite{Jordan:2018gcd}.
    
    \item[T2K/Hyper-Kamiokande] The T2K experiment \cite{T2K:2011qtm}, located in Japan, is sensitive to numerous models, taking benefit of the J-PARC neutrino facility, the composite near detector complex and the well established water Cherenkov technology for the far detector (Super-Kamiokande up to $\sim 2026$, and Hyper-Kamiokande -HK- from then). In particular, a recent study has estimated the future sensitivity of T2K and HK to millicharged particles \cite{Gorbunov:2021jog}. Pairs of mCP would be produced in meson decays alongside with neutrinos. They will propagate to the near detector ND280 and a clear (background-free) signature could be identified by considering mCP interacting twice in the Super-FGD scintillator (to be installed in 2022--2023). Within the coming years, this will allow to probe regions that have been up to now inaccessible to direct search experiments, with fractional charges $\varepsilon \sim 5 \cdot 10^{-4} - 10^{-2}$ and masses in the range $0.1 - 0.5$\,GeV, as illustrated on Fig.~\ref{fig:dark_matter_limits}. 
    
    \item[FASER(2)] is capable of searching for both the minimal dark photon \cite{FASER:2018eoc} (see right panel of Fig. \ref{fig:dark_photon_limits}) and inelastic dark matter \cite{Berlin:2018jbm} (see right panel of Fig. \ref{fig:idm_limits}) through their visible decay products. FASER2 is projected to attain far greater sensitivity to decay processes than FASER due to its larger decay volume and the increased LHC luminosity~\cite{FASER:2018eoc}.

    \item[FLArE] is proposed to study neutrino interactions and search for scattering signatures of new physics particles in the far-forward region of the LHC. In particular, light DM species can be studied via their scatterings off electrons~\cite{Batell:2021blf} and nuclei~\cite{Batell:2021aja}, both elastically and inelastically. The latter signatures are also well-suited to search for hadrophilic DM produced in $pp$ collisions at the LHC~\cite{Batell:2021snh}. In the left panel of Fig.~\ref{fig:u1b}
    , we illustrate the relevant projected exclusion bounds for FLArE in the model with the $U(1)_B$ vector mediator between the DM and SM species, assuming $m_V = 3m_\chi$ and the dark coupling constant $\alpha_\chi = g_\chi^2 Q_\chi^2/4\pi = 0.5$ which corresponds to large values of the dark charge $Q_\chi\gg 1$. We also show there the relic target lines for scalar, Majorana, and pseudo Dirac DM, as well as expected sensitivities of CCM-200, PIP2-BD, and SND@LHC~\cite{Boyarsky:2021moj} experiments. Further complementarities between different types of searches can be found for smaller values of $Q_\chi$ for which the dark vector can both decay invisibly to DM and visibly to the lighter SM particles. We show this for $Q_\chi = 1$ in the right panel of Fig.~\ref{fig:u1b}, where blue lines are obtained for visible decay products of $V$ in FASER 2 and SHiP, as well as $B\to KV$ searches~\cite{Dror:2017ehi}. We also present there the expected sensitivity of missing energy/momentum searches at NA64 and LDMX-II~\cite{Schuster:2021mlr}. 
\end{description}

\subsection{Neutrino Portal
}\label{subsec:Benchmark:NeutrinoPortal}

We now discuss the phenomenology of a heavy neutrino $N$ in the context of the minimal neutrino portal model introduced in Section~\ref{subsec:Theory:NeutrinoMass}. Since the interactions of $N$ with the SM take place exclusively through mass mixing with active-neutrinos, their couplings to ordinary matter are necessarily \emph{weaker-than-weak}, in the sense that at low energies all observables are suppressed by a small mixing parameter times $G_F$. Nevertheless, they can still lead to striking signatures at neutrino experiments, even for interaction strengths that are several orders of magnitude below $G_F$.

\subsubsection{Models}

A complete seesaw extension of the SM is expected to contain several heavy neutrinos that mix with all SM flavors. However, to discuss the low-energy phenomenology of these particles it pays of to work under a few simplifying assumptions. First, we consider the existence of a single heavy neutrino $N$, with a mass $m_N$. This mass may be of Majorana or Dirac. Instead of considering the Yukawa couplings $y_\nu^{\alpha i}$, we instead work in terms of the mixing that arises -- introducing $N$ will require extending the $3\times3$ Leptonic mixing matrix to $4\times4$, yielding new independent angles that we express in terms of the elements $|U_{e4}|^2$, $|U_{\mu4}|^2$, and $|U_{\tau4}|^2$. The final simplifying assumption we make is that only one of these new elements is nonzero. This simple picture is insufficient, if $N$ exists, to account for the observed spectrum of light neutrino masses -- at least two additional heavy neutrinos, and a less trivial mixing structure are required for this. Nevertheless, this simple, phenomenologically-driven approach is useful in comparing current and future experimental searches in a consistent way. If -- and hopefully, when -- signatures of heavy neutrinos are observed in experiments, they will surely be analyzed in a more complete approach.

Throughout this discussion, we will neglect any additional beyond-the-SM interactions that $N$ might have. Many well-motivated extensions of the SM exist where heavy neutrinos interact with new $B-L$ or secluded gauge bosons, as well as with new scalar and pseudo-scalar particles, such as the majoron, for example. Such additional interactions may allow for better or worse detection prospects at neutrino experiments, depending on how they modify the production and decay of these particles. For further discussion on this topic, see \ref{sec:Benchmark:DarkNus}.

\subsubsection{Experimental signatures}

Schematically, neutrino-beam environments are sensitive to the neutrino portal in the following way. The large flux of charged mesons (especially pions, kaons, and $D$ mesons) is capable of decaying into SM neutrinos and, if mixing $|U_{\alpha 4}|^2$ exists, heavy neutrinos $N$. The branching ratio into these new states is suppressed relative to the SM process by the small mixing $|U_{\alpha 4}|^2$. The $N$ produced in the beam can travel in the same direction as the neutrino beam, towards any detectors situated along the beamline. Because of the same mixing that allows for their production, the $N$ are metastable and have the potential of decaying in/near detectors into one or more SM particles. The signature from these decays is often distinct from backgrounds such as cosmic rays and neutrino interactions, and has features (such as reconstructing the invariant mass of the $N$) that allows for additional signal characterization.

This production/decay picture leads to a restriction on the range of $m_N$ that we can consider in the neutrino-beam context. In order for detectable signatures, we need SM decay products of $N$. For this reason, we focus on $m_N \gtrsim$ 1 MeV. In principle, $N$ can decay to a neutrino and a photon, but this is a difficult signature for all current/upcoming neutrino-beam experiments to disentangle from backgrounds. The upper limit of $m_N$ is determined by the various production mechanisms we can consider. The highest energy neutrino beams are produced with $\mathcal{O}(100)$ GeV proton beams, which produce mesons up to $D^\pm/D_{s}^\pm$ (a very small number of $B$ mesons could be produced as well). This allows us to consider heavy neutral leptons with masses up to ${\sim}1$ GeV.

For the remainder of this subsection, we will discuss current constraints on $m_N$ for the three different mixing scenarios: nonzero $|U_{e4}|^2$, $|U_{\mu 4}|^2$, and $|U_{\tau 4}|^2$. We will compare the existing constraints from neutrino-beam searches against those from other probes of these particles. Afterwards, we will demonstrate the sensitivity of several upcoming neutrino-beam searches for these particles and some possibilities for determining further properties of $N$ in the event of such a discovery. Finally, we will discuss some prospects for searches beyond this minimal scenario, where we consider additional new-physics particles and interactions involving $N$.

\subsubsection{Existing neutrino-beam searches} 

Older experiments have also provided limits, including CHARM~\cite{Orloff:2002de}, PS191~\cite{Bernardi:1985ny,Bernardi:1987ek}, and NOMAD~\cite{NOMAD:2001eyx}, although these latter limits have now been superseded by the new generation of neutrino experiments. Among the most stringent searches for the decays of heavy neutrinos in a neutrino-beam are the searches performed at the T2K~\cite{T2K:2019jwa,Arguelles:2021dqn}, MicroBooNE~\cite{MicroBooNE:2019izn,Kelly:2021xbv}, and ArgoNeuT~\cite{ArgoNeuT:2021clc} experiments. These have focused on a variety of model scenarios (e.g. different nonzero mixing angles) and different detectable signatures in their detector volumes. 
\begin{figure}
    \centering
    \includegraphics[width=\linewidth]{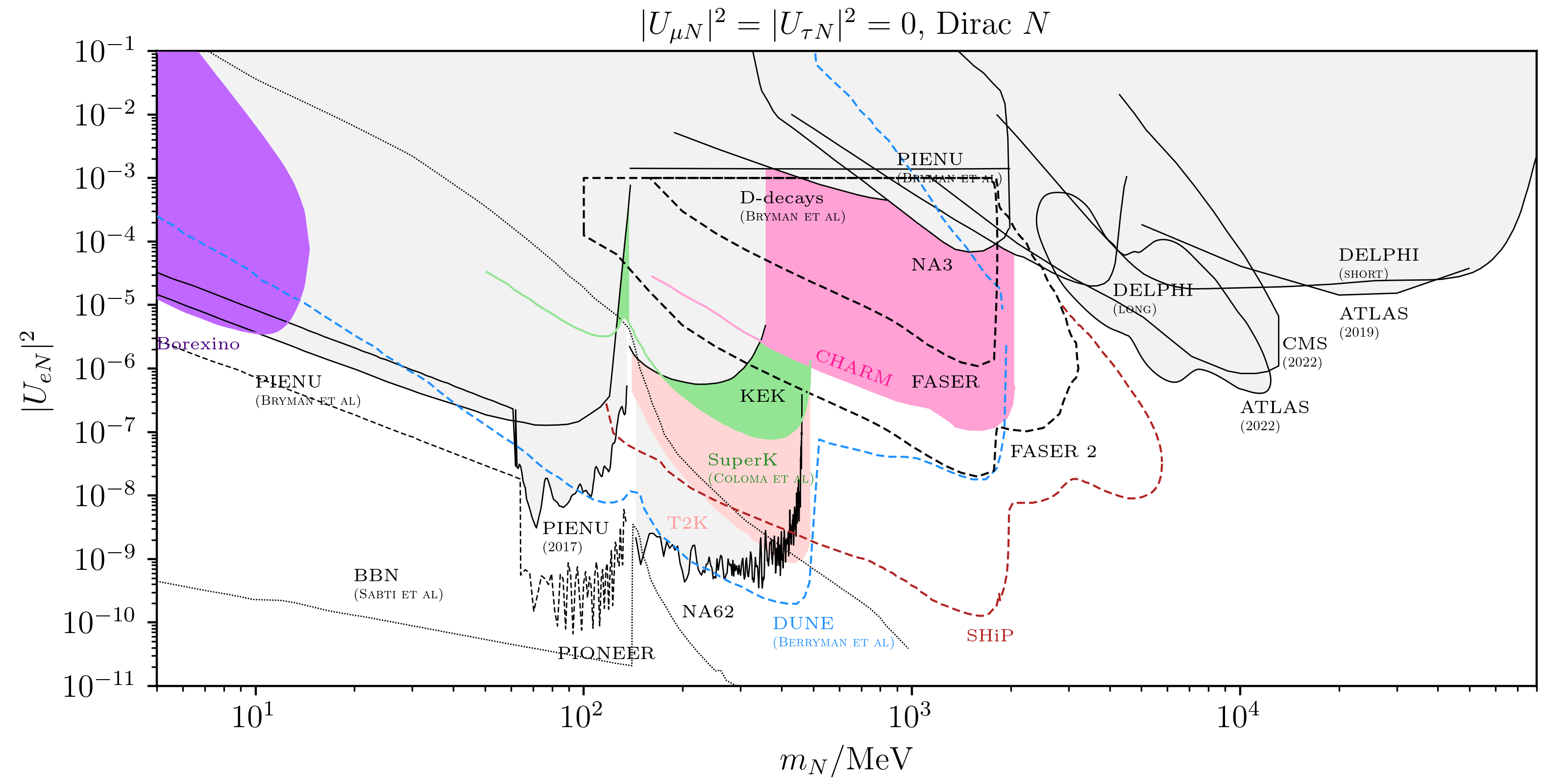}
    \caption{Constraints as a function of heavy neutrino mass $m_N$ for the electron-mixing scenario, where $|U_{e N}|^2$ is the only non-zero mixing angle. Neutrino experiments are shown in color, existing limits are shaded, and projected sensitivities are shown in dashed.}
    \label{fig:HNL:Electron}
\end{figure}

\begin{figure}
    \centering
    \includegraphics[width=\linewidth]{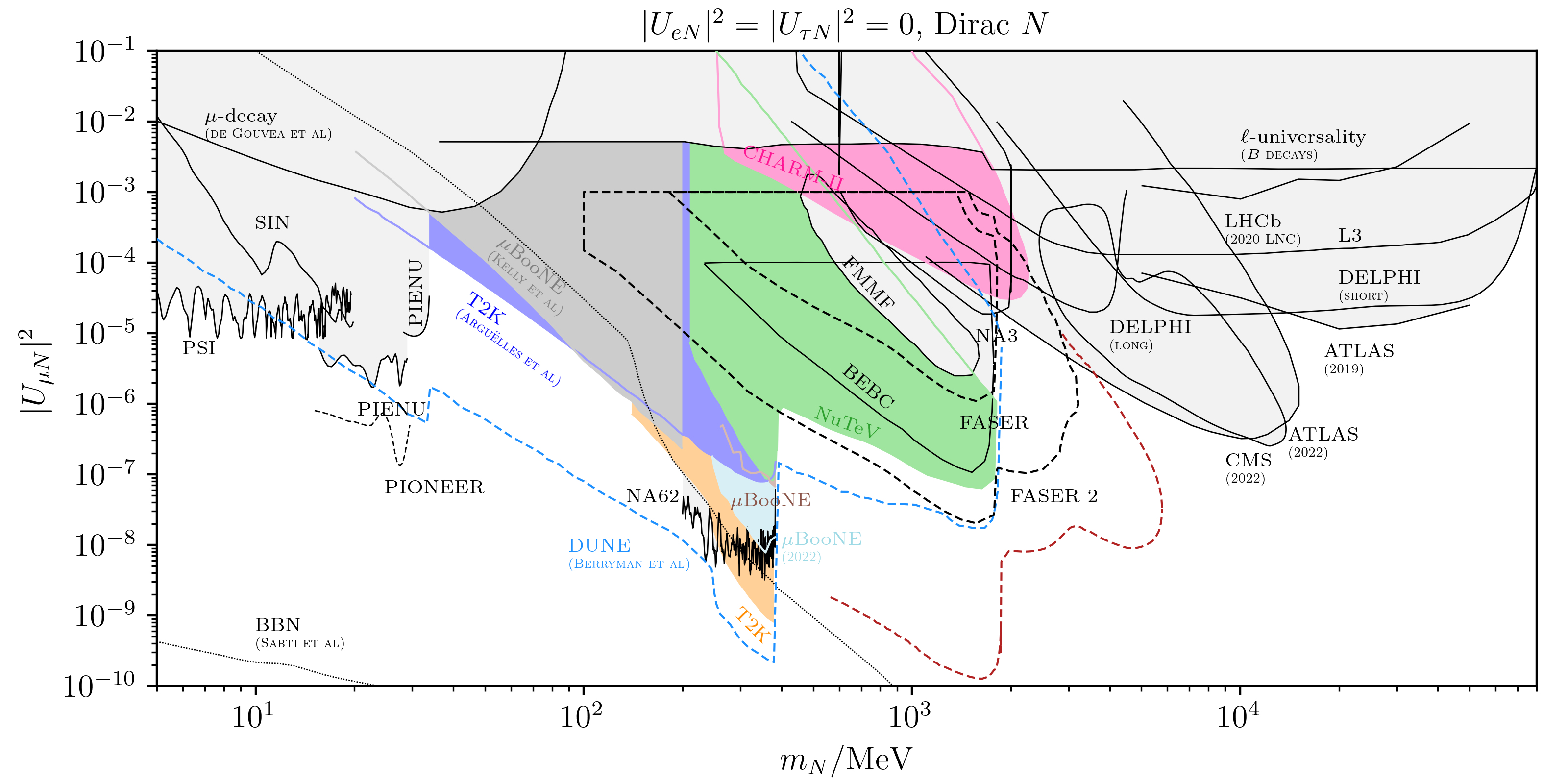}
    \caption{Constraints as a function of heavy neutrino mass $m_N$ for the muon-mixing scenario, where $|U_{\mu N}|^2$ is the only non-zero mixing angle. Neutrino experiments are shown in color, existing limits are shaded, and projected sensitivities are shown in dashed.}
    \label{fig:HNL:Muon}
\end{figure}

In its near detector ND280, T2K has performed a search for $N$ being produced from charged pion/kaon decays and subsequently decaying into the final states $N \to \ell_\alpha^\pm \pi^\mp$ and $N\to \ell_\alpha^+ \ell_\beta^- \nu$, where $\alpha,\, \beta = e,\, \mu$. In the absence of any observed excess, the results of Ref.~\cite{T2K:2019jwa} place the most stringent constraints on $|U_{e4}|^2$ and $|U_{\mu 4}|^2$ for a variety of masses -- these are presented in Figs.~\ref{fig:HNL:Electron} and~\ref{fig:HNL:Muon}. Other constraints that are competitive with T2K in these regions of parameter space include those from PS191~\cite{Bernardi:1985ny,Bernardi:1987ek,Kusenko:2004qc,Ruchayskiy:2011aa,Arguelles:2021dqn}, E949~\cite{Artamonov:2014urb}, and NA62~\cite{NA62:2020mcv,NA62:2021bji}. The constraints from Ref.~\cite{T2K:2019jwa} have been reinterpreted in Ref.~\cite{Arguelles:2021dqn} to extend to lower masses, demonstrating impressive results from T2K, compared against a variety of existing constraints in the $20$ MeV $\lesssim m_N \lesssim 200$ MeV window~\cite{Bryman:2019bjg,deGouvea:2015euy,Hayano:1982wu,Bernardi:1987ek,Artamonov:2014urb}.

The MicroBooNE collaboration published its first results from searching for kaon-decay production of $N$ and its subsequent decay into a charged pion/muon pair in Ref.~\cite{MicroBooNE:2019izn}. Due to large backgrounds from neutrino scattering (CC$1\mu1\pi$), the time structure of the neutrino beam was leveraged to reduce backgrounds and optimize signal searches. This (null) result leads to the constraint shown in Fig.~\ref{fig:HNL:Muon}, which is relatively weak compared to T2K and other existing constraints. MicroBooNE has also searched for Higgs-Portal Scalars (see Section~\ref{subsec:Benchmark:HiggsPortal} for more)~\cite{MicroBooNE:2021usw} where the new-physics particle is produced in the absorber of the NuMI neutrino beam. Ref.~\cite{Kelly:2021xbv} reinterpreted this result in the neutrino-portal context and found that it improves on existing constraints in the ${\sim}30-150$ MeV region, comparable to the results demonstrated for T2K in Ref.~\cite{Arguelles:2021dqn}.
\begin{figure}
    \centering
    \includegraphics[width=\linewidth]{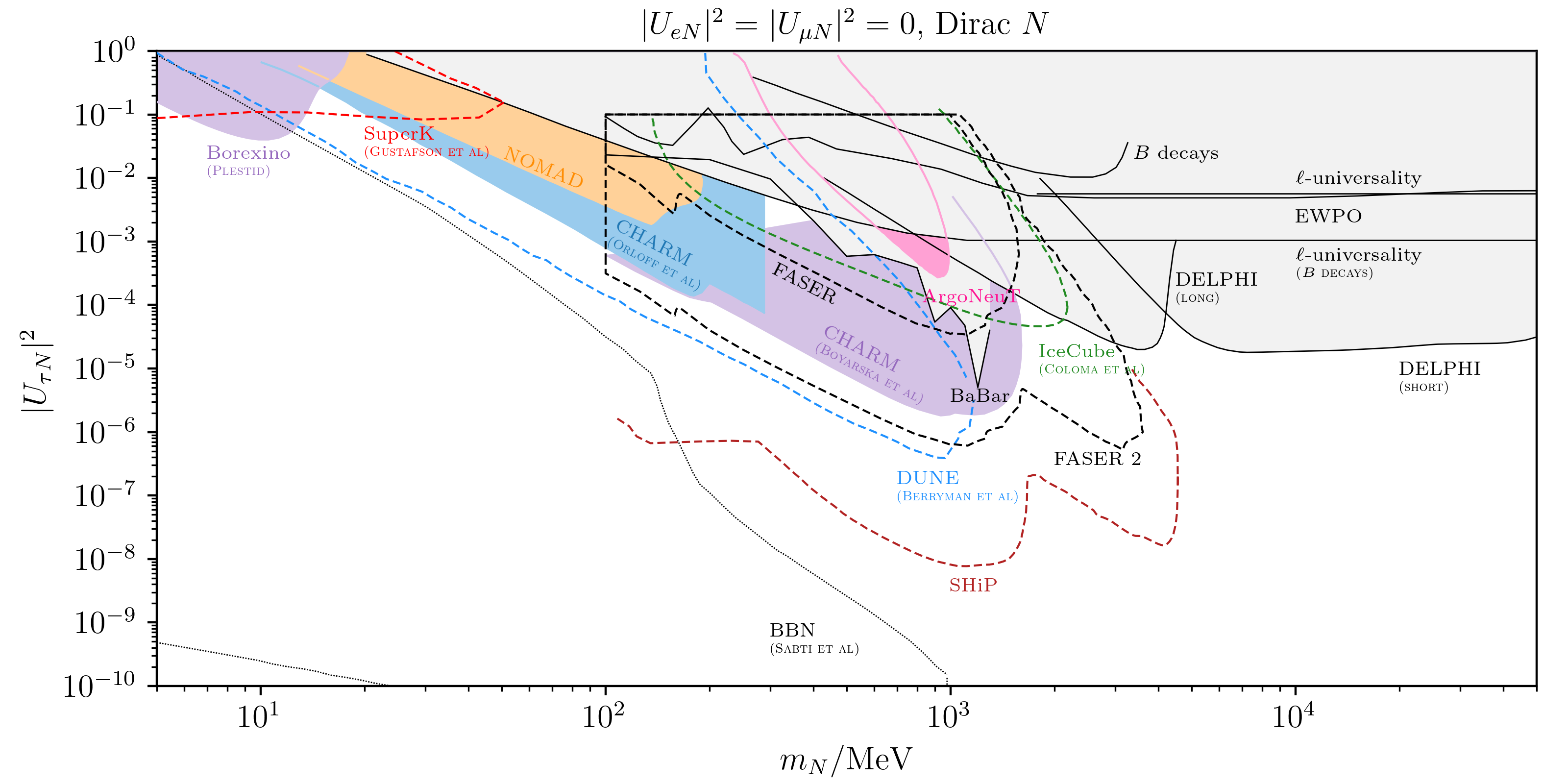}
    \caption{Constraints as a function of heavy neutrino mass $m_N$ for the tau-mixing scenario, where $|U_{\tau N}|^2$ is the only non-zero mixing angle. Neutrino experiments are shown in color, existing limits are shaded, and projected sensitivities are shown in dashed.}
    \label{fig:HNL:Tau}
\end{figure}

Finally, the ArgoNeuT collaboration~\cite{ArgoNeuT:2021clc} carried out a search for heavy neutral leptons with nonzero $|U_{\tau 4}|^2$ where the $N$ are produced in a chain of $D$ meson decays -- $D_{(s)}^\pm \to \tau^\pm \nu_\tau$, $\tau^\pm \to N X^\pm$ (where $X^\pm$ is one or more SM particles). These are produced in the NuMI beam and absorber, and then can decay either inside or in front of the ArgoNeuT detector via $N \to \nu_\tau \mu^+ \mu^-$. With no observed signal, Ref.~\cite{ArgoNeuT:2021clc} places a constraint demonstrated in Fig.~\ref{fig:HNL:Tau}, improving on existing searches in this mass range for tau-coupled $N$ from CHARM~\cite{Orloff:2002de} and DELPHI~\cite{Abreu:1996pa}. However, in the time since, Ref.~\cite{Boiarska:2021yho} demonstrated that CHARM has sensitivity beyond the reach of Ref.~\cite{Orloff:2002de}, yielding a more stringent constraint than ArgoNeuT. This is summarized in Fig.~\ref{fig:HNL:Tau}.

It is also worth noting that cosmological constraints set strong limits on heavy neutrinos below the kaon mass. These limits, however, are model dependent and may be avoided if $N$ has new secluded interactions. In Ref.~\cite{Arguelles:2021dqn} it was argued that heavy neutrinos may be as light as tens of MeV without violating constraints from, for instance, Big Bang Nucleosynthesis.

\subsubsection{Future prospects with neutrino beams} Upcoming experiments have great sensitivity in searches for neutrino-portal particles, much in the same way that the existing searches have operated. These next-generation experiments additionally offer excellent particle identification capabilities, allowing for searches of a variety of final states of $N$ decays simultaneously. Existing sensitivity studies have been carried out for the Fermilab SBN detectors~\cite{Ballett:2016opr}, the DUNE Near Detector complex~\cite{Ballett:2019bgd,Berryman:2019dme,Coloma:2020lgy}, and FASER~\cite{Kling:2018wct}.

The SBN detectors (with sensitivity driven mostly by the nearest of the three, SBND, to the production target) offer moderate improvement over current capabilities of detectors like what MicroBooNE has demonstrated to date~\cite{MicroBooNE:2019izn}. In contrast, the DUNE Near Detector, if equipped with a gaseous argon option, will enable a nearly background-free search due to the relatively low detector density (signal rates in this context scale like the detector volume, whereas neutrino-induced backgrounds scale like mass). Refs.~\cite{Ballett:2019bgd,Berryman:2019dme,Coloma:2020lgy} have demonstrated the capability of DUNE in this context, showing great improvement on current constraints in all mixing scenarios.

Finally, the upcoming FASER and FASER$\nu$ experiments at the LHC offer unique sensitivity to many new-physics models, including the one under consideration here. The $N$ can be produced in a variety of meson decays, up to the $B$ meson mass, and decay inside the spectrometer. This leads to very powerful sensitivity, especially if future extensions such as FASER-2 are funded and operate.

\subsubsection{Post-Discovery potential} In the event of discovery of a new neutrino-portal particle, the next step will be to understand its properties and address a large number of questions -- does it have interactions beyond those of mixing with the SM neutrinos, is it connected to the origin of light neutrino masses, do its interactions preserve or violate Lepton Number? In neutrino-beam environments, it is possible to study the possibility of Lepton-Number Violation (LNV) and whether $N$ is a Dirac or Majorana fermion. This is attainable by measuring the relative rate of fully-visible decays in a detector, e.g., $N \to \mu^+ \pi^-$ and $N \to \mu^- \pi^+$. If $N$ is a Majorana fermion, then these will occur with equal probability, whereas if $N$ is a Dirac fermion, only one process can occur (depending on the polarity of the beam, which must be pure to perform this separation). Ref.~\cite{Berryman:2019dme} demonstrated that for large parts of parameter space, DUNE can not only discover a new particle $N$ but also determine, via its decay rates, whether it is a Dirac or Majorana fermion. If $N$ decays in a partially-invisible channel, then decay kinematics can be used to perform this separation~\cite{BahaBalantekin:2018ppj,Balantekin:2018ukw,deGouvea:2021ual,deGouvea:2021rpa}.

When considering non-minimal scenarios, predictive phenomenology can connect constraints/observations in neutrino-beam experiments with those from other sectors, allowing for additional handles on discovering or constraining a particular model scenario~\cite{Ballett:2019pyw,Abdullahi:2020nyr}.

\subsubsection{Neutrinophilic Mediators}\label{subsubsec:Neutrinophilic}
Neutrino facilities are also capable of testing a different nature of neutrino-portal interactions that may be connected to dark matter. Instead of the Lagrangian considered in Section~\ref{subsec:Theory:NeutrinoMass} that leads to additional heavy, neutral fermions beyond those of the SM, we consider the following SM extension:
\begin{equation}
    \mathcal{L} \supset \frac{\left(L_\alpha H\right) \left(L_\beta H\right) \phi}{\Lambda_{\alpha\beta}^2},
\end{equation}
where $\alpha$ and $\beta$ indicate lepton flavor indices. This dimension-six operator, after electroweak symmetry breaking, leads to interaction terms of the type $g_\phi^{\alpha\beta} \nu_\alpha \nu_\beta \phi$, sourcing additional interactions between neutrinos and this new particle $\phi$.

There is also the possibility that additional BSM particles exist that interact with $\phi$. Through fairly simple model realizations, these particles may be responsible for the observed abundance of dark matter in the universe. This includes scenarios in which the particle thermalizes with the SM in the early universe and then freezes out its relic abundance through its interaction with $\phi$ and neutrinos~\cite{Kelly:2019wow}, as well as scenarios in which the abundance of a sterile neutrino dark matter particle freezes in through these weak interactions with $\phi$ as well as mixing with the light neutrinos~\cite{deGouvea:2019wpf}. These dark matter scenarios provide interesting targets in the $m_\phi$ vs. $g_\phi$ parameter space that are worth testing.

\begin{figure}
    \centering
    \includegraphics[width=\linewidth]{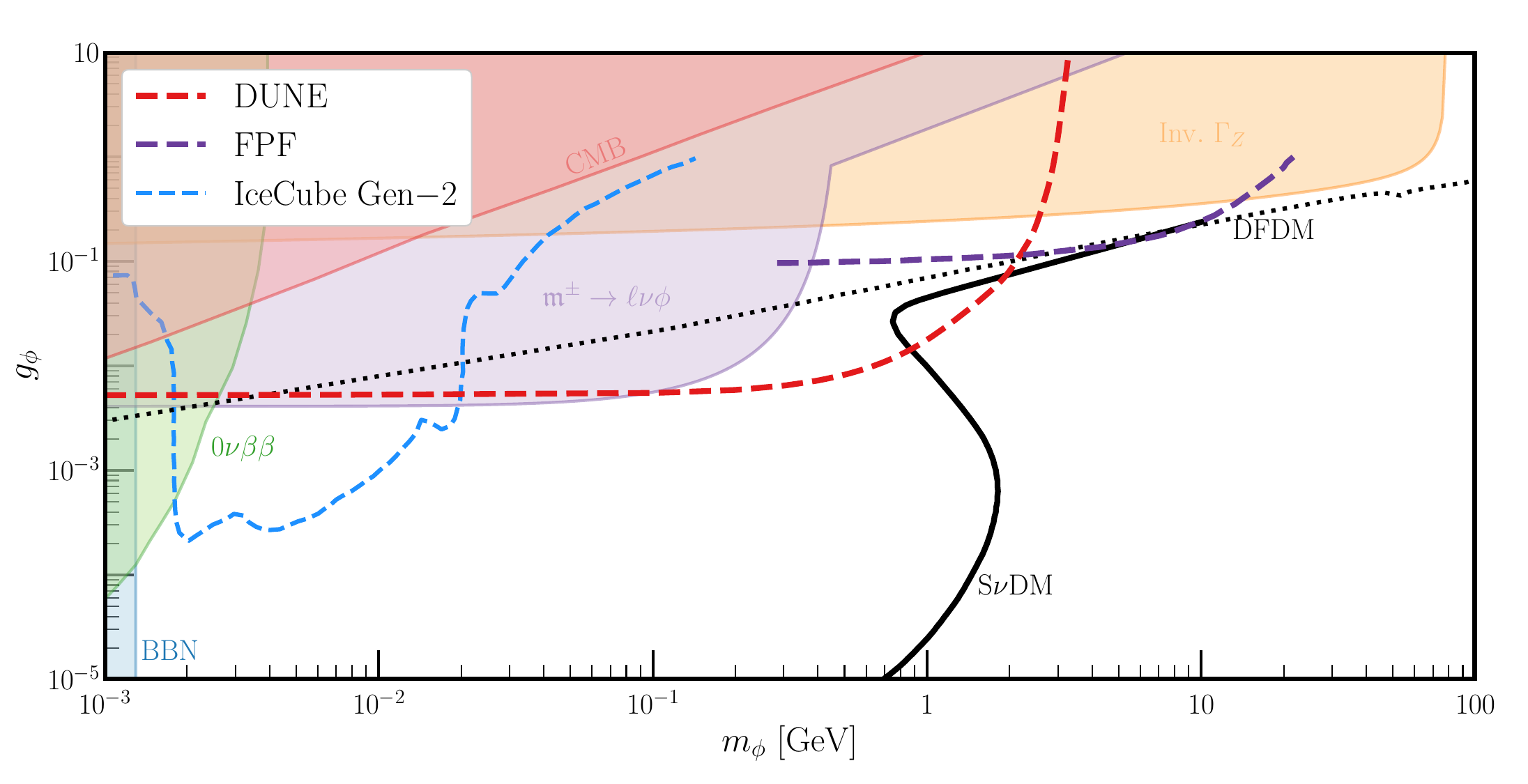}
    \caption{A summary of searches for neutrinophilic mediators $\phi$ as a function of their mass and their (assumed flavor-universal) coupling to neutrinos $g_\phi$. Future neutrino-facility searches are highlighed: DUNE~\cite{Kelly:2019wow}, the FPF~\cite{Kelly:2021mcd}, and IceCube Gen-2~\cite{Esteban:2021tub}.\label{fig:SInuTargets}}
\end{figure}

Neutrino experiments offer a wide variety of complementary ways to search for these new particles $\phi$ -- Ref.~\cite{Berryman:2022hds} explored these and other probes in great detail. Here, we highlight the types of searches that can be performed in the coming decades at these experiments, summarized in Fig.~\ref{fig:SInuTargets}.

The IceCube-Gen2 upgrade will allow for even more powerful studies of atmospheric and astrophysical neutrinos. While propagating from their source to the Earth, these high-energy neutrinos have the possibility of interacting with the at-rest cosmic neutrino background neutrinos along the way. For TeV-PeV energy neutrinos, this process can occur resonantly if $m_\phi \approx$ 10 MeV. IceCube may search for evidence of these new interactions by searching for features in the astrophysical neutrino spectrum -- the projected capabilities~\cite{Esteban:2021tub} are shown as a dashed blue line in Fig.~\ref{fig:SInuTargets}.

In neutrino-beam environments, it is possible to search for the presence of $\phi$ in neutrino scattering events. This can occur for scenarios in which $E_\nu \gtrsim m_\phi$, in processes such as $\nu+X \rightarrow \ell^{+} \phi X'$, where $\phi$ is emitted off the initial-state neutrino (and $X$ and $X'$ are hadronic initial/final states). The DUNE near detector~\cite{Kelly:2019wow} (red dashed line) and a potential liquid argon detector at the Forward Physics Facility~\cite{Kelly:2021mcd,Feng:2022inv} (purple dashed line) offer great sensitivity for these searches, as shown in Fig.~\ref{fig:SInuTargets} as well.

\subsubsection{Neutrino Portal Dark Matter}\label{subsubsec:NeutrinophilicDS}
Scenarios where sterile neutrinos interact with a dark sector containing the DM have also been studied recently~\cite{Bertoni:2014mva,Batell:2017rol,Batell:2017cmf}. A simple example involves augmenting the sterile neutrino's coupling to active leptons in~(\ref{eq:NeutrinoMassLagrangian}) with a Yukawa interaction with SM singlets, ${\cal L}\supset-\lambda \phi\chi N$. Because $N$ can decay invisibly into the dark sector, the allowed active-sterile mixing can be relatively large, particularly in the case of mixing with $\nu_\tau$. Despite this, it can be particularly difficult to test this simple setup, although it could potentially be probed at neutrino facilities through $\nu_\tau$ production and PMNS unitarity~\cite{DeGouvea:2019kea}.

\subsection{ALP Portal
}
\subsubsection{Introduction}
The axion is a well-motivated solution to the strong CP problem. The experimental searches for axions are nowadays extended toward probing more general pseudo-scalar axion-like particles (ALPs), which are ubiquitous in the string theory (e.g., Refs.~\cite{Arvanitaki:2009fg,Cicoli:2012sz}). 
Due to their rich structure, ALPs often behave as mediators and allow dark-sector particles to communicate with the SM particles in various dark-sector scenarios~\cite{Park:2015ysf}.   
We here focus on the search for ALP mediators whose production is related to photons. 
Among various ALP interactions, many of the phenomenological considerations are predicated upon their couplings described in the following interaction Lagrangian: 
\begin{equation}
    \mathcal{L_{\rm ALP}}\supset \frac{1}{4} g_{a\gamma\gamma}aF_{\mu\nu}\tilde{F}^{\mu\nu}+ig_{aee}a \bar{e}\gamma^5 e +i  a\bar{\psi}_N \gamma^5(g_{ann}^{(0)}+g_{ann}^{(1)}) \psi_N, \label{eq:alpint}
\end{equation}
where $g_{a\gamma\gamma}$, $g_{aee}$, $g_{ann}^{(0)}$, and $g_{ann}^{(1)}$ parametrize the coupling strengths of the ALP field $a$ to the SM photon, electron, nucleon isoscalar, and nucleon isovector, respectively. Here, again $\tilde{F}_{\mu\nu}$ is the dual field strength tensor of the SM photon, as defined earlier, and $\psi_N=\left( \begin{matrix} p\\ n \end{matrix} \right)$ denotes the nucleon field.
Neutrino experiments provide a great platform to explore the ALP parameter space associated with the interactions defined in Eq.~\eqref{eq:alpint}, as they generate copiously the fluxes of not only neutrinos but photons. 

Depending on the underlying model details, ALPs can couple to other SM particles such as gluon and quarks and other production and/or detection channels may become important. 
For example, in models of ALPs coupling to SM gluon through the $ a G_{\mu\nu}^a \tilde{G}^{a,\mu\nu}$ operator, the ALPs can be produced via the mixing with SM pseudoscalar mesons or gluon-gluon fusion in the process of the collision between an incoming beam proton and a proton inside the target system. Therefore, neutrino facilities carry great potential of testing these models~\cite{Kelly:2020dda}. 

\subsubsection{Production}

Depending on the ALP interaction under consideration, ALPs can be produced in various ways in the neutrino facilities. In particular, wherever photons emerge during the sequence of interactions between the proton beam and the target material, production of ALPs can be induced via the couplings in Eq.~\eqref{eq:alpint}.
Photons can be produced not only by standard mechanisms including proton-beam bremsstrahlung and neutral meson decays but by the cascade showering processes of charged particles (e.g., $e^\pm$) out of the primary interaction between the beam proton and the target~\cite{Dent:2019ueq,Brdar:2020dpr}. 
The main ALP production mechanisms include:
\begin{description}
\item[Primakoff process] For the ALPs interacting with photons, a photon can turn into an ALP through the Primakoff-like scattering process with a nucleus $\mathcal{N}$, i.e., $\gamma +\mathcal{N} \to a +\mathcal{N}$. As it takes place via a $t$-channel exchange of a photon, the typical momentum transfer to the target nucleus is very small, and as a result, i) the whole nucleus participates in the process coherently and the production scattering cross-section benefits from a $Z^2$ enhancement with $Z$ being the atomic number of the nucleus of interest and ii) the momentum of the outgoing ALP inherits from the incoming photon. 

\item[Compton-like process] In the presence of the ALP-electron coupling $g_{aee}$, an ALP can be produced by a Compton-like scattering process which is initiated by a photon, i.e., $\gamma +e^- \to a +e^-$. 
Unlike the Primakoff process above, the outgoing electron can carry a sizable fraction of the incoming photon energy and the ALP can be emitted at a large angle with respect to the initial-state photon. 

\item[Nuclear reactions] If the ALP-to-nucleon coupling $g_{ann}^{(0,1)}$ is non-zero, an ALP can be produced by nuclear reactions; for example, neutron captures, i.e., $n +\mathcal{N}(Z,A) \to \mathcal{N}(Z, A+1)+a$ with $n$ and $A$ respectively denoting neutron and atomic mass number, and nuclear de-excitations (radioactive decays), i.e., $\mathcal{N}^*(Z,A) \to \mathcal{N}(Z,A)+ a$. 
In the SM, both neutron captures and nuclear de-excitations accompany a photon, so any photons originating from these nuclear processes can be connected to the production of ALPs with a non-zero value of $g_{ann}^{(0,1)}$. 
\end{description}

Inside the target material, the processes described in the above three mechanisms compete with ordinary SM processes; for example, the photons in the first two mechanisms can go through a pair conversion or an ordinary Compton process, and the nuclear reactions create a photon instead of an ALP in the final state. 
For a given source particle, therefore, the likelihood of ALP production is given by the competition between the ALP production rate and other SM events rate.  

\subsubsection{Detection}
In the presence of the couplings in Eq.~\eqref{eq:alpint}, ALP can leave various signatures in the experiments. Broadly speaking, three channels are available in the neutrino experiments: ALPs can be detected through their i) decays, ii) scattering, and iii) conversion.
The three channels show best signal sensitivity to ALPs in different mass regions, providing complementary information in the exploration of ALP parameter space. 

\begin{description}
\item[Decays] Once the ALP is massive enough and/or its coupling to SM particles is sizable enough, a large fraction of the produced ALPs can decay within the detector fiducial volume. 
With a non-zero value of $g_{a\gamma\gamma}$, an ALP can leave a diphotonic decay signal (i.e., $a \to \gamma+\gamma$), whereas in the presence of non-vanishing $g_{aee}$, an ALP can decay to a pair of electrons (i.e., $a\to e^++e^-$) as long as its mass is greater than twice the electron mass.  
\item[Scattering] If an ALP produced at the target does not decay and reaches the detector, it can scatter off the nucleus or the electrons inside the detector material. If the coupling to photon $g_{a\gamma\gamma}$ is non-zero, the ALP can be converted to a photon through the inverse Primakoff scattering process, i.e., $a+ \mathcal{N} \to \gamma +\mathcal{N}$. The scattering process involves the whole nucleus, resulting in $Z^2$ enhancement as in the ALP production by the Primakoff process. 
Since typical nuclear recoils are soft, the detectable experimental signature is a single photon signal unless the detector features sensitivity to small energy deposits. 
By contrast, if the coupling to electron $g_{aee}$ is non-vanishing, the ALP can undergo a Compton-like scattering process, giving rise to an electron and a photon in the final state, i.e., $a +e^- \to \gamma+ e^-$. 
Both the electron and the photon can carry sizable energy fractions of the incoming ALP, and therefore, detectors can record and identify both electron and photon, depending on their capability, which provides an additional handle to reduce potential backgrounds.  
\item[Conversion] For the ALPs interacting with photons, there exists an additional detection channel if a detector accompanies a magnetic field region. 
The detection principle is the same as that in helioscopes such as CAST; an incoming ALP can convert to a photon by the external magnetic field. 
In the beam-dump type experiments, this idea was proposed as Particle Accelerator helioScopes for Slim Axion-like-particle deTection, PASSAT~\cite{Bonivento:2019sri,Dev:2021ofc}.
In the limits of $m_a\to 0$ and $g_{a\gamma\gamma}\to 0$, the ALP-to-photon conversion probability is proportional to $(g_{a\gamma\gamma}BL)^2$ with $B$ and $L$ being the strength of magnetic field and the length of magnetic field region, respectively. Once $m_a$ becomes large, the probability is quickly suppressed by the decoherency factor (see e.g., \cite{Irastorza:2018dyq,Fortin:2021cog}) and the conversion channel starts losing its sensitivity. 
The conversion channel, therefore, shows the best sensitivity to the ALP of a very small mass value. 
\end{description}

\subsubsection{Experimental Prospects}

Most of the experimental prospects are based on the simplified approach in which only the coupling of interest is turned on while the others are suppressed enough to be irrelevant.
Various types of neutrino facilities including the reactor-based, the stopped-pion, and the beam-focusing neutrino experiments have sensitivity to ALP signals. They are sensitive to ALPs in different mass regimes, hence providing complementary information in constraining ALP parameter space. 

\begin{figure}[t]
    \centering
    \includegraphics[width=8.4cm]{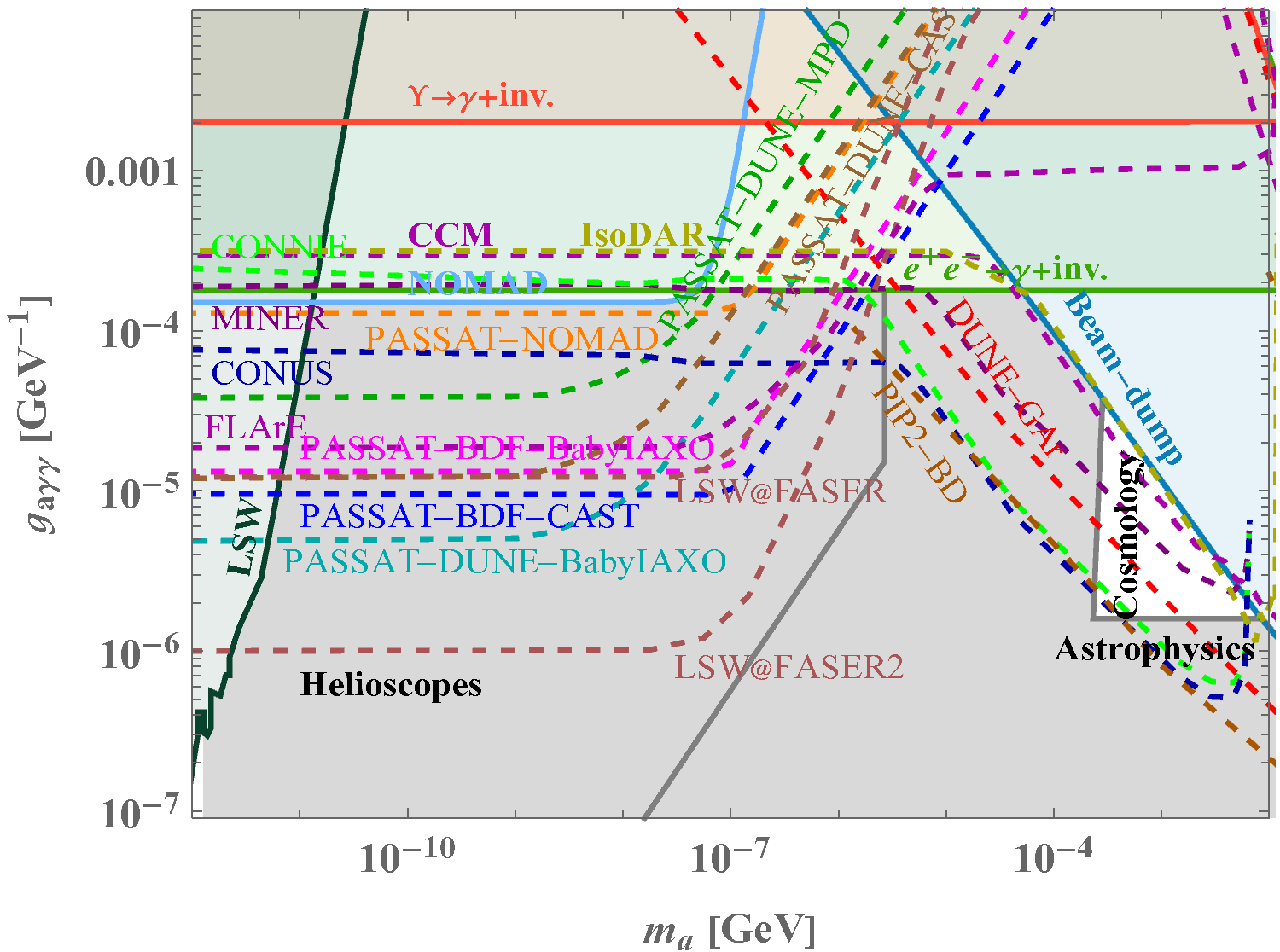}
    \includegraphics[width=8.4cm]{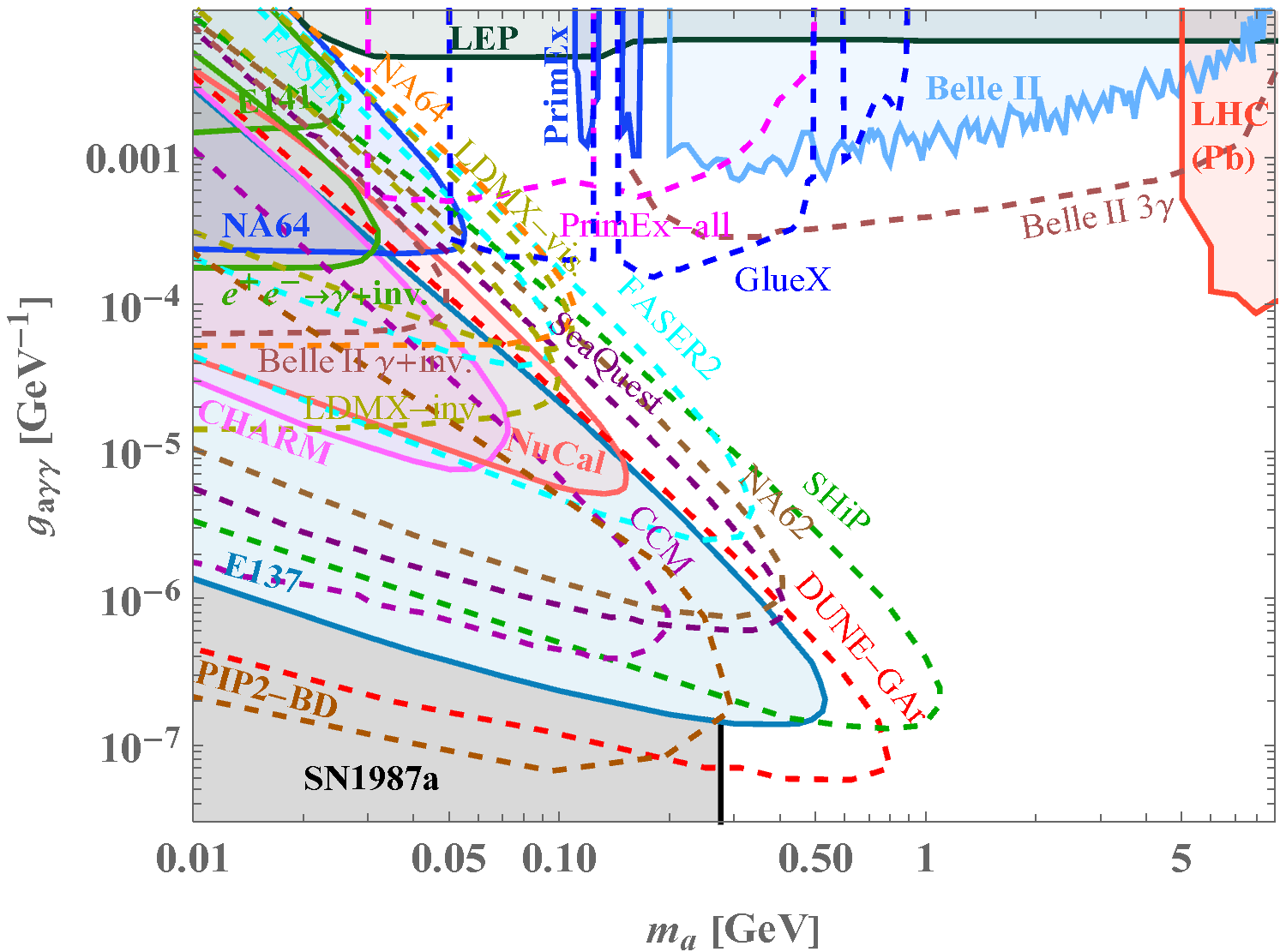}
    \includegraphics[width=8.4cm]{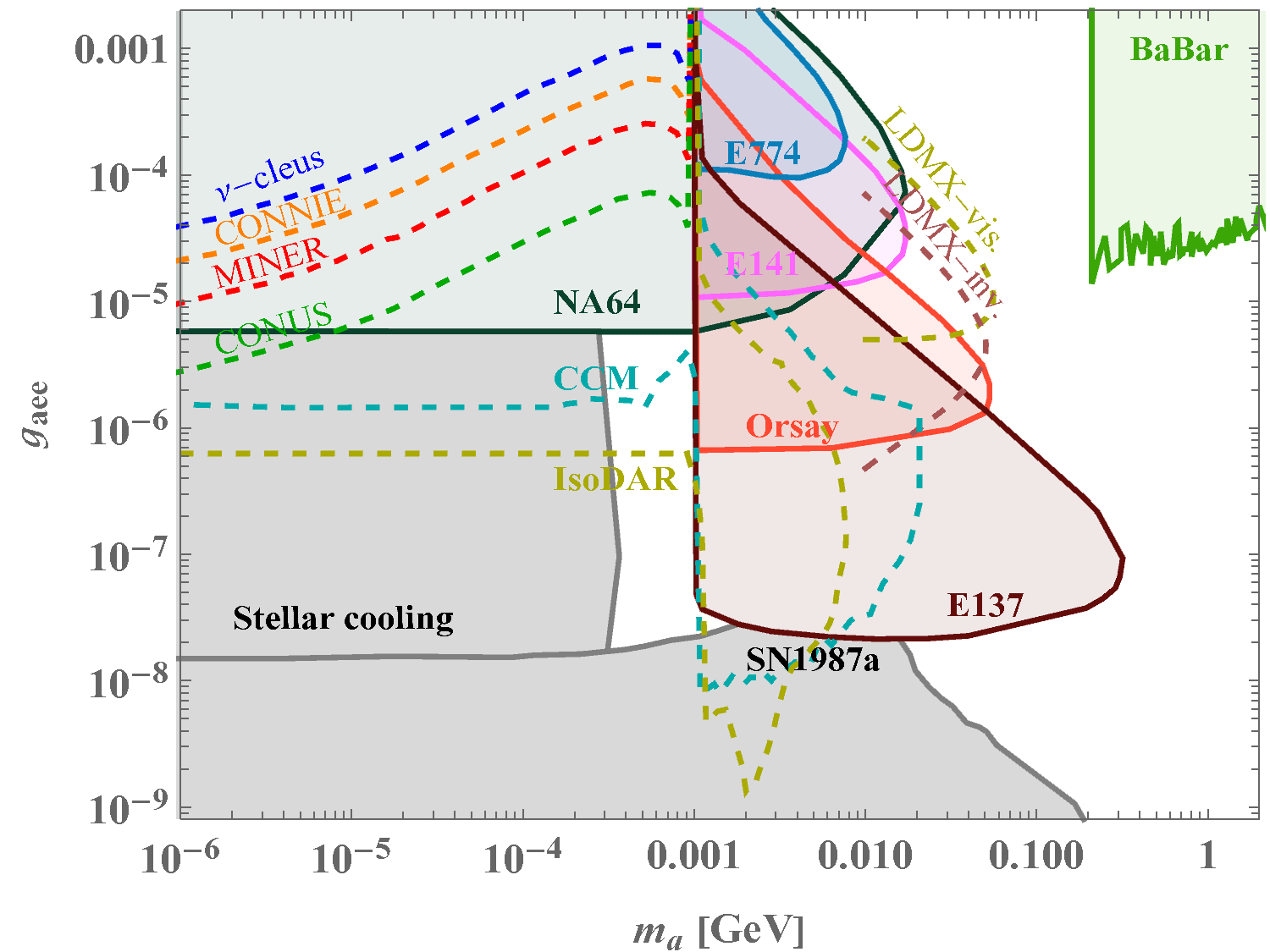}
    \includegraphics[width=8.4cm]{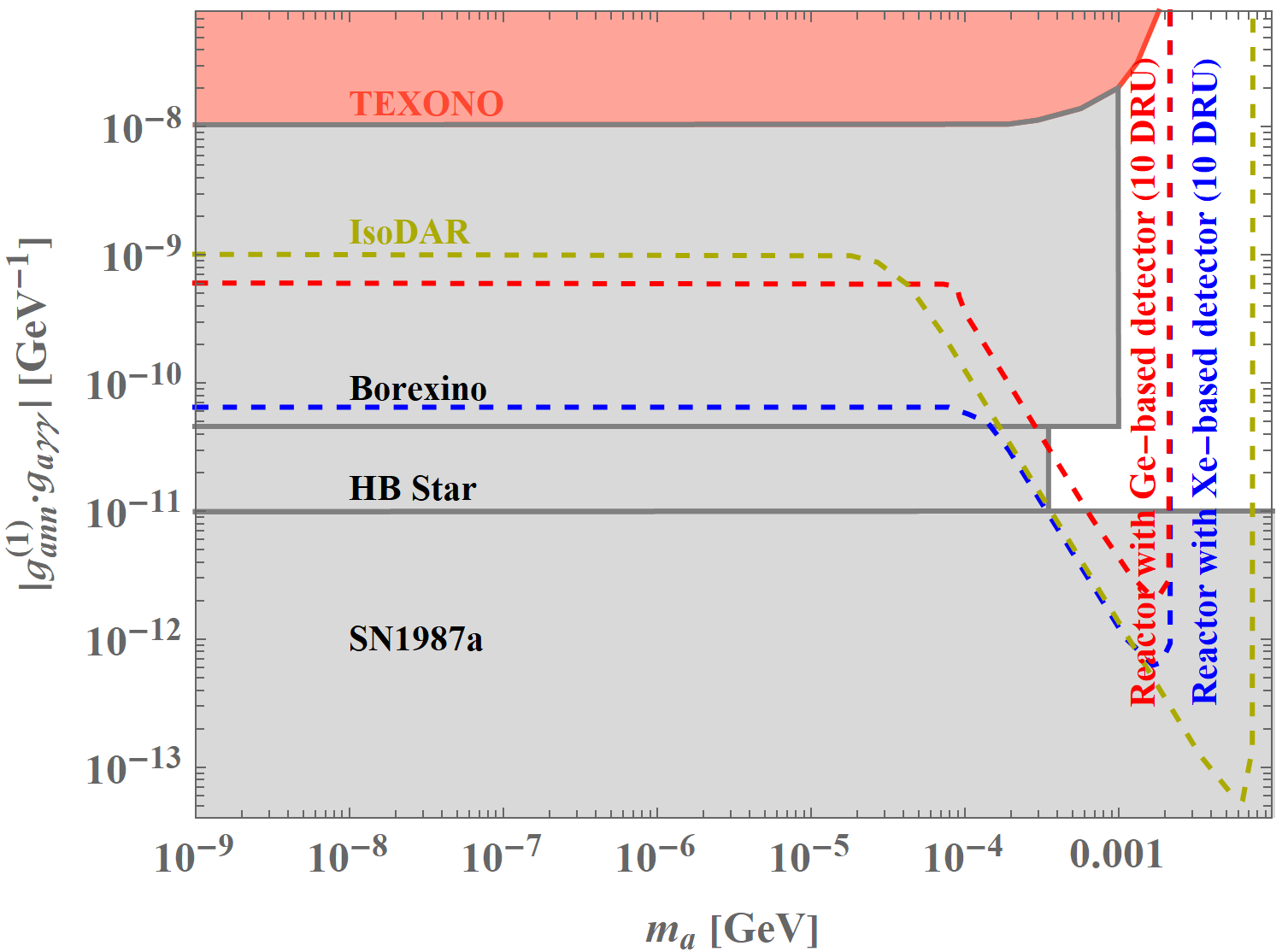}
    \caption{Existing constraints and future prospects of $g_{a\gamma\gamma}$ (upper panels), $g_{aee}$ (lower-left panel), and $g_{ann}^{(1)}$ (lower-right panel) at various experiments including neutrino facilities. The limits for $g_{a\gamma\gamma}$ are divided into the lower-mass regime (upper-left panel) and the higher-mass regime (upper-right panel). The plots for $g_{a\gamma\gamma}$ and $g_{aee}$ are adapted from \cite{Fortin:2021cog} while that for $g_{ann}^{(1)}$ is adapted from \cite{AristizabalSierra:2020rom}. See the main text for more details. }
    \label{fig:alpprospects}
\end{figure}

\begin{description}
\item[ALP-photon coupling $g_{a\gamma\gamma}$] The ALP-photon coupling is most extensively studied, not only at neutrino facilities but at colliders. 
The upper-right panel of Fig.~\ref{fig:alpprospects} shows existing (color-shaded regions) and future expected (dashed lines) limits in the $m_a$-$g_{a\gamma\gamma}$ plane for $m_a\gtrsim 10$~MeV to which the decay channels are relevant. 
The current constraints include $e^+e^-\to\gamma+{\rm inv.}$~\cite{OPAL:2000puu,ALEPH:2003ycd,L3:2003yon,DELPHI:2003dlq,Proceedings:2012ulb}, Belle-II~\cite{Belle-II:2020jti}, CHARM~\cite{CHARM:1985anb}, E137~\cite{Bjorken:1988as}, E141~\cite{Riordan:1987aw}, LEP~\cite{OPAL:2002vhf}, LHC (Pb)~\cite{CMS:2018erd,ATLAS:2020hii}, NA64~\cite{NA64:2020qwq}, NuCaI~\cite{Blumlein:1990ay}, and PrimEx~\cite{Aloni:2019ruo}. The limits from astrophysical and cosmological considerations compiled in Ref.~\cite{Bauer:2018uxu} are also displayed in gray-shaded regions.
Future sensitivity reaches at neutrino facilities include DUNE-GAr with a 1-year exposure~\cite{Brdar:2020dpr}, PIP-II with a 5-year exposure~\cite{pip2}, and SHiP with $2\times10^{20}$ POTs~\cite{Alekhin:2015byh}. For purposes of complementarity, we include the sensitivity reaches expected in non-neutrino experiments, e.g., Belle-II~\cite{Dolan:2017osp}, FASER/FASER2~\cite{Beacham:2019nyx}, Glue-X~\cite{Aloni:2019ruo}, LDMX~\cite{Berlin:2018bsc}, LHC (Pb)~\cite{Knapen:2016moh}, NA62~\cite{Dobrich:2019dxc}, NA64~\cite{Dusaev:2020gxi}, PrimEx~\cite{Aloni:2019ruo}, SeaQuest~\cite{Berlin:2018pwi}. 
The limits in the lower mass regime where the scattering and conversion channels are better suited are shown in the upper-left panel of Fig.~\ref{fig:alpprospects}. 
The current constraints (color-shaded regions) include $e^+e^-\to \gamma+{\rm inv.}$~\cite{OPAL:2000puu,ALEPH:2003ycd,L3:2003yon,DELPHI:2003dlq,Proceedings:2012ulb},
$\Upsilon\to\gamma+{\rm inv.}$~\cite{CLEO:1994hzy,BaBar:2010eww}, light-shining-through-wall type experiments~\cite{Redondo:2010dp}, NOMAD~\cite{NOMAD:2000usb}, and E137~\cite{Bjorken:1988as}, together with the limits from helioscopes and astrophysical/cosmological considerations (gray-shaded regions) compiled in Ref.~\cite{Bauer:2018uxu}.
Future sensitivity reaches (dashed lines) at neutrino experiments include a PASSAT interpretation~\cite{Bonivento:2019sri} of NOMAD~\cite{NOMAD:2000usb}, PASSAT implementations at the BDF facility with the CAST or BabyIAXO magnets with $2\times 10^{20}$ POTs~\cite{Bonivento:2019sri}, a PASSAT implementation at the DUNE MPD with a 7-year exposure~\cite{Dev:2021ofc}, PASSAT implementations at DUNE with the CAST or BabyIAXO magnets with a 7-year exposure~\cite{Dev:2021ofc}, IsoDAR with a 5-year exposure~\cite{we_isodar}, and reactor searches at CONNIE, CONUS, and MINER~\cite{Dent:2019ueq}.
Similar analyses for reactor neutrino experiments also appear in Ref.~\cite{AristizabalSierra:2020rom} by taking sets of benchmark ALP production and detection parameters, not specifying concrete experiments.
We also include the expected sensitivity reaches of LSW@FASER/FASER2~\cite{Kling:2022ehv} and FLArE~\cite{Kling:2022ehv} for comparison. 

\item[ALP-electron coupling $g_{aee}$] The ALP-electron coupling space can be also readily explored in many of the neutrino experiments, in both decay and scattering channels. The lower-left panel of Fig.~\ref{fig:alpprospects} displays the existing (color-shaded regions) and future expected (dashed lines) limits in the $m_a$-$g_{aee}$ plane. The current constraints include BaBar~\cite{BaBar:2016sci,Bauer:2017ris},  E137~\cite{Bjorken:1988as,Andreas:2010ms}, E141~\cite{Riordan:1987aw}, E774~\cite{Bross:1989mp}, NA64~\cite{NA64:2021aiq,Andreev:2021fzd,Gninenko:2017yus}, Orsay~\cite{Bechis:1979kp}, and astrophysical considerations (stellar cooling~\cite{Hardy:2016kme} and SN1987a~\cite{Lucente:2021hbp}). 
Future prospects include CCM (a stopped-pion neutrino experiment) with beam-on backgrounds~\cite{CCM:2021lhc}, IsoDAR with a 5-year exposure~\cite{we_isodar}, and reactor neutrino experiments (CONNIE, CONUS, MINER, and $\nu$-cleus)~\cite{Dent:2019ueq}.
For comparison, the expected LDMX are included~\cite{Berlin:2018bsc}. 
The reactor searches are sensitive to the ALP signals in the decay channel only within a narrow mass range since most of the photons created in the reactor cores are not energetic enough to overcome the production threshold $s \geq (m_a + m_e)^2$~\cite{Dent:2019ueq}. By contrast, the scattering channel is more promising in the search for light ALPs (i.e., $m_a < 2m_e$) and it has been investigated in reactor neutrino experiments~\cite{Dent:2019ueq,AristizabalSierra:2020rom} and neutrino beam experiments~\cite{Brdar:2020dpr}.
 
For the reactor neutrino experiments, the dark boson interaction through various portals can be produced in the reactor core and detected by the neutrino detector through the Compton-like scattering and its inversion scattering. The precision of the light-dark boson measurement in the reactor neutrino experiments is limited by the accidental background uncertainty. The further improvement of the sensitivity to an ALP signal can be achieved with a better background control, which is expected in the next generation of reactor neutrino experiments such as JUNO-TAO~\cite{Smirnov:2021alp}.

\item[ALP-nucleon couplings $g_{ann}$] The ALP-nucleon couplings were explored in reactor neutrino experiments, e.g., TEXONO~\cite{TEXONO:2006spf}. The TEXONO Collaboration assumes that ALPs are produced via the $pn\to d\gamma$ isovector M1 transition inside the reactor core [i.e., proportional to $(g_{ann}^{(1)})^2$] while ALP detection is done either via the inverse Primakoff scattering process (with vanishing $g_{aee}$) or the Compton-like scattering process (with vanishing $g_{a\gamma\gamma}$). As an example, the lower-right panel of Fig.~\ref{fig:alpprospects} shows the existing (color-shaded regions) and future expected (dashed lines) limits in the $(m_a,g_{ann}^{(1)}\cdot g_{a\gamma\gamma})$ plane. 
TEXONO constrains models of ALP in this parameter space~\cite{TEXONO:2006spf}, while more stringent limits from various astrophysical considerations (e.g., Borexino~\cite{Borexino:2008wiu,Borexino:2012guz}, HB stars~\cite{Lee:2018lcj}, and SN1987a~\cite{Lee:2018lcj}) are available. The reactor neutrino experiments have great potential to probe this parameter space like TEXONO; for example, the blue (red) dashed line represents the reactor-experiment limit expected from a 140 (50) cm long $10^3$ (10) kg Xe-(Ge-)based detector located 10 m away from the reactor core with a 10 DRU (/kg/keV/day) background. 
Similar limits in other parameter spaces, e.g., $(m_a, g_{ann}^{(1)}\cdot g_{aee})$ plane, $(g_{ann}^{(1)},g_{a\gamma\gamma})$ plane, $(g_{ann}^{(1)},g_{aee})$ plane, etc, appear in literature~\cite{TEXONO:2006spf,AristizabalSierra:2020rom}. We also add the limits expected at IsoDAR with a 5-year exposure~\cite{we_isodar}.

\end{description}

\subsection{Dark Neutrinos and Dipole Portal
}\label{sec:Benchmark:DarkNus}

In this section, we consider two types of BSM heavy neutral lepton models that yield different signatures compared to the traditional heavy sterile neutrino that mixes with SM flavors, discussed in Section~\ref{subsec:Benchmark:NeutrinoPortal}.
They are generally described as Heavy Neutral Leptons with a Dipole Portal, and dark neutrinos, described in the following subsections.

\subsubsection{Heavy Neutral Leptons with a Dipole Portal}

The dipole portal \textit{d} to the heavy neutral lepton (HNL) \textit{N} is contained within the effective Lagrangian:
\begin{equation}
    \mathcal{L} \supset \bar{N}(i\slashed{\partial}-m_{N})N + (d\bar{\nu}_{L}\sigma_{\mu\nu}F^{\mu\nu}N + h.c.)
\end{equation}
The traditional neutrino portal coupling $NLH$ \cite{Asaka:2005pn,Gorbunov:2007ak} is assumed to be absent, or subdominant, which allows the dipole portal operator to offer novel signatures and features in the production and decay of $N$. The energy, intensity, and astrophysical frontiers afford ideal probes for the sensitivity of \textit{d} and each case is studied within \cite{Magill:2018jla}. The main signal under consideration is the decay $N$ $\rightarrow \gamma \nu$. At high energy colliders, such as the LHC and LEP, the sensitivity for \textit{d} is $\sim$ 10 TeV$^{-1}$ and is largely independent of the $N$ mass, provided it is within the kinematic reach of the collider. In the high intensity regime where beam dump experiments dominate, such high masses are unattainable. Instead, beam dump experiments explore low coupling for sub-GeV masses. Experiments such as LSND \cite{Athanassopoulos:1996ds,Pospelov:2017kep}, MiniBooNE \cite{AguilarArevalo:2007it}, and MINER$\nu$A~\cite{Kamp:2022bpt} give sensitivity for \textit{d} at $\sim$ (10$^{-6}$--10$^{-7}$) GeV$^{-1}$. The upcoming experiments, including, e.g., SBND \cite{Antonello:2015lea}, DUNE \cite{DUNECollaboration2015}, SHiP \cite{Alekhin:2015byh}, and CCM \cite{CCM:2021leg} will provide increased sensitivity in this regime. At the astrophysical frontier supernovae explosions restrict the parameter space, but allow for probes for $m_{N}$ on the order of a few hundred MeV. This allows for astrophysical constraints to probe \textit{d} in the range of (10$^{-7}$--10$^{-10}$) GeV$^{-1}$.
The dipole portal HNL has diverse phenomenological consequences~\cite{Gninenko:2009ks, Gninenko:2010pr, McKeen:2010rx, Gninenko:2012rw, Magill:2018jla,Jodlowski:2020vhr,Ismail:2021dyp}, and could potentially explain an excess from the Xenon 1T experiment \cite{Shoemaker:2020kji,Brdar:2020quo}.

\begin{figure}
    \centering
    \includegraphics[width=0.5\textwidth]{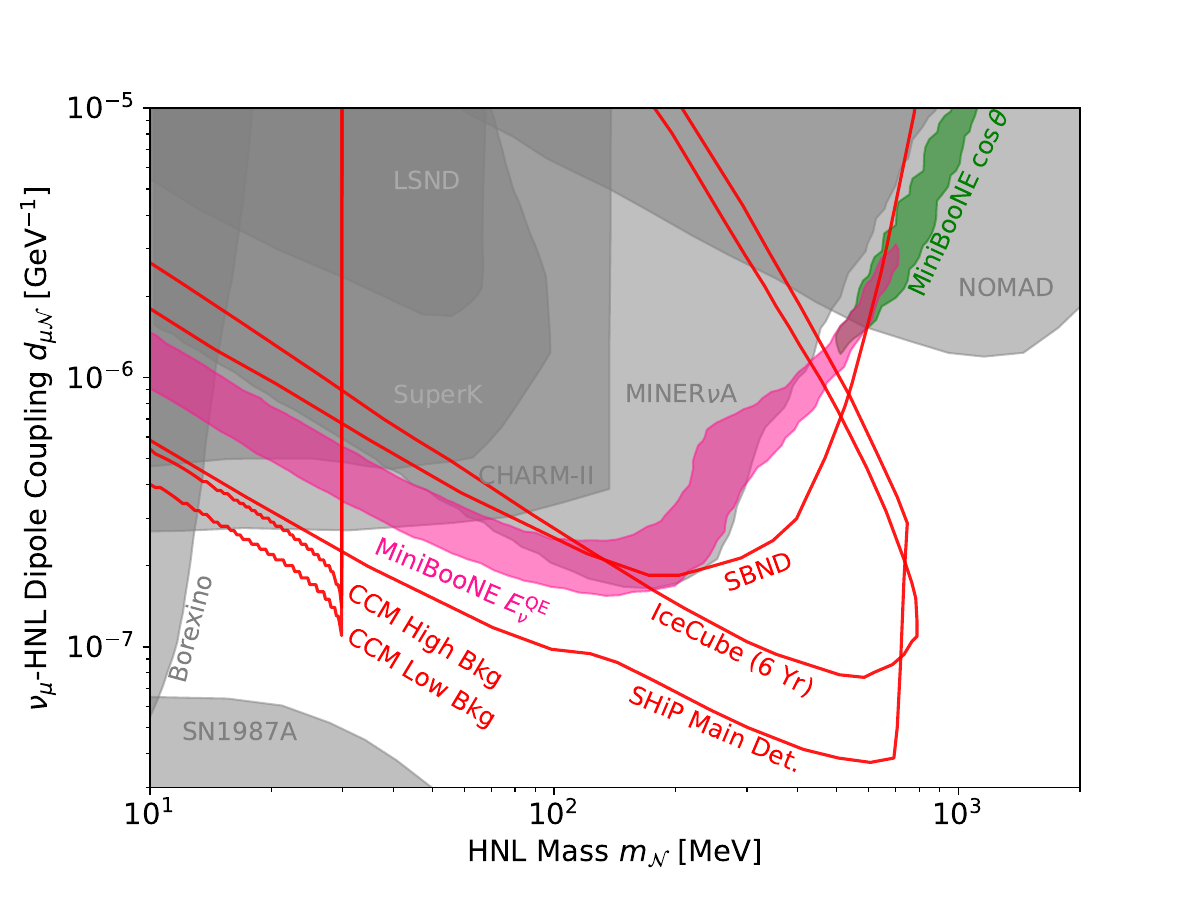}
    \caption{Global constraints on the dipole portal HNL model. Existing constraints are shown in grey, while projected constraints are shown in red. All constraints are at the 95\% CL with the exception of IceCube, for which the contour indicates one event in six years of data. The preferred regions to explain the energy and angular distributions of the MiniBooNE excess are shown in pink and green, respectively. See Refs.~\cite{Magill:2018jla,Kamp:2022bpt,Coloma:2017ppo} and references therein for a description of the features of this figure. The reported CCM sensitivity is expected to be conservative: additional sensitivity may be gained by considering production channels beyond neutrino upscattering and detection channels beyond HNL decay.}
    \label{fig:dipole_global}
\end{figure}

The dipole portal HNL has often been invoked as at least a piece of the solution to the MiniBooNE excess of electron-like events~\cite{Gninenko:2009ks,Magill:2018jla,Kamp:2022bpt,Vergani:2021tgc}.
Recent studies indicate that constraints from MINER$\nu$A neutrino-electron scattering analyses come close, but do not rule out the preferred region of dipole portal parameter space which can explain the MiniBooNE excess~\cite{Kamp:2022bpt}; a dedicated MINER$\nu$A analysis would likely be sensitive to the MiniBooNE solution.
A global picture of existing and projected constraints in dipole portal parameter space (considering the effective coupling to the muon neutrino) is shown in Fig.~\ref{fig:dipole_global}.

\subsubsection{Heavy Neutral Leptons With Dark Forces}

\begin{figure}
    \centering
    \includegraphics[width=\textwidth]{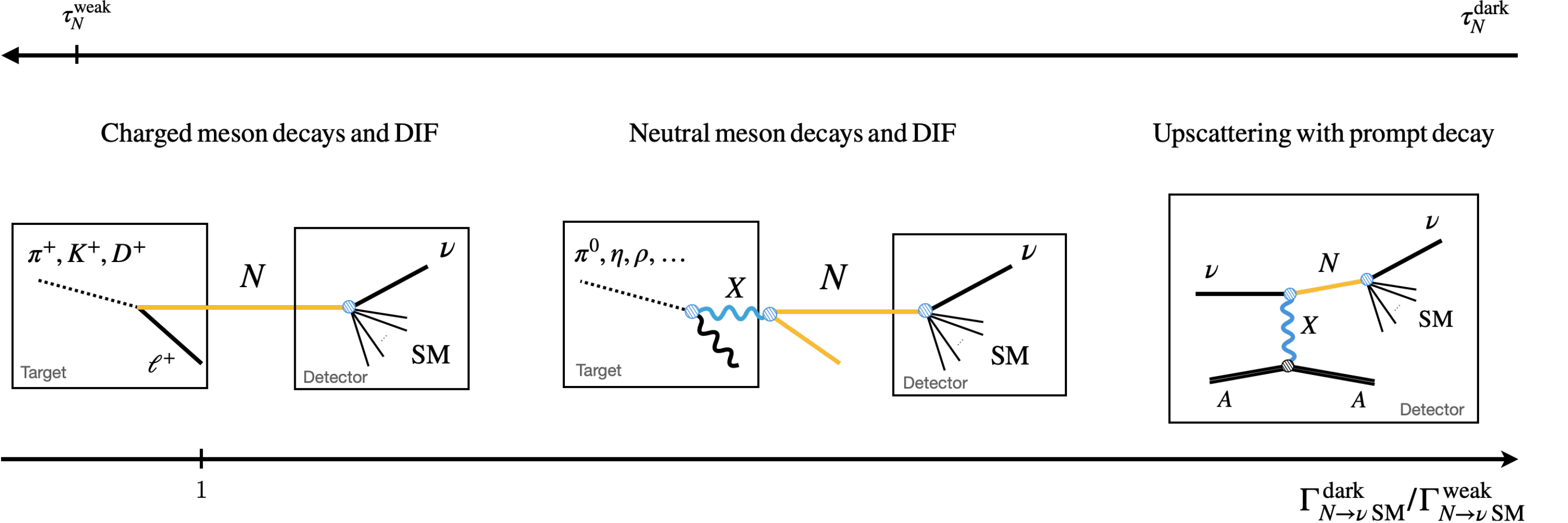}
        \caption{Signatures of heavy neutrinos that interact with additional dark forces as a function of the new interaction strength. Stronger interactions lead to shorter $N$ lifetimes and additional production modes. \label{fig:HNL:dark_forces}}
\end{figure}

Heavy neutral leptons may not be completely sterile and could engage in beyond-the-SM gauge interactions. These interactions may be shared by the SM fields, such as in $B-L$ or $L_\alpha-L_\beta$ gauge symmetries, or not, such as in a secluded $U(1)$ gauge symmetry wherein SM particles only interact with the dark photon via kinetic mixing. In the secluded case, the new particles can often evade strong experimental constraints, as they only communicate with the SM via the renormalizable portals: scalars, neutrino, and vector portals. Nevertheless, the phenomenology of the heavy neutrinos can be substantially different from that of the minimal models where one considers each portal interaction individually.

In neutrino experiments, the new forces can lead to new production as well as decay channels. If the new force is much weaker than the weak interactions (further suppressed by mixing), then the phenomenology will be very similar to that already discussed in \ref{subsec:Benchmark:NeutrinoPortal}. However, when it is stronger, it shortens the lifetime of the heavy neutrino, enhancing the rate of decay-in-flight signatures up until the new states become too short-lived to propagate between the beam target and the detector. Still, in the latter scenario, heavy neutrinos can still be searched for when they are produced by neutrino scattering inside the detector. A schematic of the phenomenology in that case is show in Fig.~\ref{fig:HNL:dark_forces}, where the new signatures are shown as a function of the new interaction strengths.

Dark neutrinos have been investigated in various contexts. For instance, dark neutrinos can provide an explanation for the long-standing excess of electron-like events at MiniBooNE~\cite{Bertuzzo:2018ftf,Bertuzzo:2018itn,Ballett:2018ynz,Arguelles:2018mtc,Ballett:2019cqp,Ballett:2019pyw,Abdullahi:2020nyr,Abdallah:2020vgg,Abdallah:2020biq}, long-range vector-mediated neutrino self-interactions \cite{Berbig:2020wve, Schoneberg:2021qvd, Archidiacono:2020yey}, and as an explanation to the XENON1T electron recoil excess \cite{Bally:2020yid}. 
Dark neutrinos also lead to active to sterile neutrino transition magnetic moments (dipole portal HNL), which is further discussed in the previous section.

In general, this dark neutrino model may, in principle, also give contributions to the muon $g-2$, to atomic parity violation, polarized electron scattering, neutrinoless double $\beta$ decay, rare meson decays as well as to other low energy observables such as the running of the weak mixing angle $\sin^2\theta_{W}$. There may also be consequences to neutrino experiments. It can, for instance, modify neutrino scatterings, such as coherent neutrino-nucleus scattering, or impact neutrino oscillations experimental results as this model may give rise to non-standard neutrino interactions in matter.

\subsection{$\nu$-philic Interactions (CE$\nu$NS, I$\nu$NS, Neutrino Tridents) 
}

The study of neutrino interactions can be used to shed light on new physics communicating to neutrinos, in particular in scenarios with $\nu-$philic interactions, where the dark sector only communicates with neutrinos. 
New $\nu-$philic physics can manifest itself through many distinctive signatures. Focusing on neutrino beams,  scattering measurements can probe the existence of neutrino non-standard interactions or of new light dark sector states, whose existence can also enhance rare SM processes
such as trident events.
 
\subsubsection{Coherent elastic neutrino-nucleus scattering (CE$\nu$NS)}
 CE$\nu$NS is a neutral-current process in which the neutrino interacts with the nucleus as a whole~\cite{Freedman:1973yd,Kopeliovich:1974mv}. One basic requirement for the coherence of this interaction is the smallness of the momentum transfer, which needs to be smaller than the inverse size of the nucleus. The fact that all nucleons recoil coherently gives rise to an increase in the scattering amplitude of the process, in turn leading to an enhancement of the cross section proportional to the number of nucleons squared.
 The tiny recoil energy transferred to the nucleus is then observable, although it requires using low-threshold detectors capable of observing $\sim\mathcal{O}($keV) recoil energies. Despite the magnitude of the CE$\nu$NS cross section, the largest among neutrino scattering channels for energies below $\sim$ 100 MeV and for most nuclei, its detection has proven to be demanding because of the required low thresholds and reduction of backgrounds. However, the use of intense, pulsed sources of low-energy neutrinos in combination with low-threshold detectors and a low-background environment allow for a successful measurement of the CE$\nu$NS process.
 The first detection of the CE$\nu$NS process in 2017 in a sodium-doped cesium-iodide (CsI[Na]) by the COHERENT experiment \cite{COHERENT:2017ipa,COHERENT:2018imc}, has fuelled the interest in new physics opportunities which can be explored with this channel. Then, the recent observation of CE$\nu$NS also in argon~\cite{COHERENT:2020iec}, together with the new data collected by the COHERENT experiment using the CsI[Na] detector \cite{Akimov:2021dab} have further demonstrated that we are entering an era of CE$\nu$NS precision measurements, allowing not only to perform accuracy tests of the SM but also to probe physics beyond the SM \cite{Barranco:2005yy,Papoulias:2019xaw}.
 
Accelerator-based experiments using low-energy neutrinos from pion decays at rest ($\pi-$DAR), designed for the detection of the CE$\nu$NS process, have the further advantage of offering several opportunities to probe dark sectors. These may include light, weakly-coupled states such as sub-GeV dark matter candidates, axion-like particles (ALPs), sterile neutrinos, new heavy or light mediators in the neutrino sector and other new particles which can appear in several extensions of the SM.

\begin{figure}[!hbt]
\centering
\includegraphics[width=0.33\textwidth]{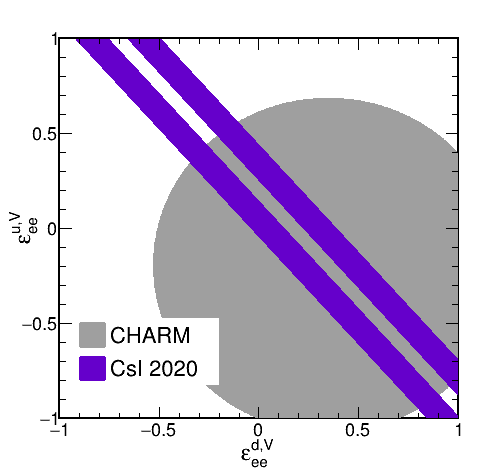}
\includegraphics[width=0.35\textwidth]{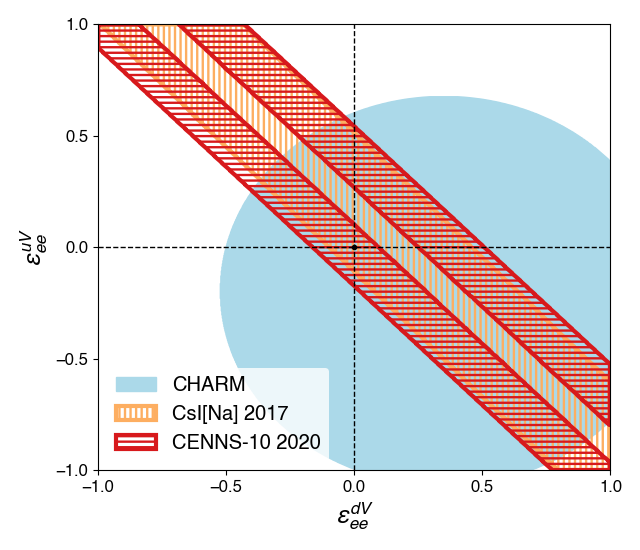}
\caption{COHERENT constraints on non-zero vector-like neutrino-quark NSI couplings $\epsilon_{\rm ee}^{\rm d, V}$ and $\epsilon_{\rm ee}^{\rm u, V}$, compared to the CHARM~\cite{CHARM:1986vuz} constraint. Constraints in the left panel have been obtained using data collected by the COHERENT experiment with the CsI[Na] detector \cite{Akimov:2021dab}. The right panel shows the bounds extracted from the argon measurement \cite{COHERENT:2020iec} and plotted together with
the previous COHERENT CsI[Na] measurement \cite{COHERENT:2017ipa}.} 
\label{fig:COH_NSI}
\end{figure}

\paragraph{Non-standard neutrino interactions}
The spectrum of coherent elastic neutrino–nucleus scattering
events can be used to probe new subdominant neutrino Non-Standard Interactions (NSI) with matter fields mediated by a new heavy vector particle \cite{Farzan:2017xzy,Proceedings:2019qno}. These are commonly parameterized 
by a matrix of parameters $\epsilon_{ij}^{q,P}$, with $q=u,d$, $i,j = e, \mu, \tau$ and $P= V (\rm vector), A (\rm axial)$, describing  either  ``non-universal''  ($i=j$)  or  flavor-changing
($i \neq j$) interactions  of  neutrinos  with  quarks. (A scattering experiment using neutrinos from  $\pi-$DAR  will  have  sensitivity  only  to NSI couplings  with ($i=e, \mu$).)
Neglecting the axial contribution, the existence of non-zero NSI would result in an overall scaling of the event rate, either enhancement or suppression. While a positive signal would undoubtedly hint towards new physics, existing data already constrain the effective NSI parameters \cite{Papoulias:2017qdn,Coloma:2017ncl,Liao:2017uzy,Coloma:2017egw,Denton:2018xmq,Dent:2017mpr,Abdullah:2018ykz,Dutta:2019eml,Coloma:2019mbs,Akimov:2021dab,Miranda:2020tif} -- encoding the magnitude of the new interaction in terms of $G_F$ -- and consequently disfavour models of new physics, as for instance those that were motivated by the Dark-LMA solution \cite{Miranda:2004nb,Escrihuela:2009up,Gonzalez-Garcia:2013usa}. 
Fig.~\ref{fig:COH_NSI} shows current constraints on non-zero vector-like neutrino-quark NSI couplings $\epsilon_{\rm ee}^{\rm d, V}$ and $\epsilon_{\rm ee}^{\rm u, V}$, obtained with COHERENT CsI~\cite{Akimov:2021dab} and LAr~\cite{COHERENT:2020iec} data.
In full generality, not only vector operators but a larger set of interactions can be explored with CE$\nu$NS experiments, including all Lorentz invariant non-derivative interactions of neutrinos with first generation quarks \cite{Lindner:2016wff,AristizabalSierra:2018eqm,Flores:2021kzl}.

\paragraph{CE$\nu$NS and reactor experiments} CE$\nu$NS can be investigated at the reactor experiments using electron type anti-neutrino emerging from the reactors, e.g.,~\cite{Dent:2017mpr}. Various reactor based experiments, e.g., TEXONO~\cite{Wong:2004ru},
$\nu$GeN~\cite{Belov:2015ufh}, CONUS~\cite{CONUS:2020skt}, MINER~\cite{MINER:2016igy}, CONNIE~\cite{CONNIE:2016ggr},RED-100~\cite{RED-100:2019rpf}, Ricochet~\cite{Billard:2016giu},NEON~\cite{Choi:2020gkm}, $\nu$-cleus~\cite{Strauss:2017cuu},  NCC-1701 at Dresden-II~\cite{Colaresi:2022obx} etc (see also Table~\ref{tab:listreactor}). Among these experiments, recently, 
NCC-1701  at  Dresden-II nuclear reactor has  observed an excess of events  which is claimed to be compatible with the expectations from the SM CE$\nu$NS signal. This observation will implement constraints on non-standard neutrino interactions, neutrino electromagnetic properties and new light mediators~\cite{Coloma:2022avw,AristizabalSierra:2022axl,Liao:2022hno,AtzoriCorona:2022qrf}.  
\paragraph{Light mediators}

CE$\nu$NS experiments are also ideal laboratories for searches of new physics in the form of light mediators, since in these scenarios the scattering is proportional to the inverse of the momentum transferred squared, which we recall is small for the CE$\nu$NS process. For the case of vector mediators, there is an interference with the SM vector couplings that can be either constructive or destructive, thus leading also to a possible depletion of the event rate \cite{Billard:2018jnl,AristizabalSierra:2019ufd}. In scenarios with a scalar boson mediating the CE$\nu$NS process, the new contribution has no interference with the SM one, and it simply adds to it. Low-energy high-intensity measurements at CE$\nu$NS experiments can provide a valuable probe of light mediator scenarios, complementary to high-energy collider searches and electroweak precision measurements \cite{Dutta:2015vwa,Dent:2016wcr,Billard:2018jnl,Denton:2018xmq,Cadeddu:2020nbr,Dutta:2020enk,Miranda:2020tif,Miranda:2020zji,Banerjee:2021laz,AtzoriCorona:2022moj}.
CP-violating effects may also be investigated in the context of
light vector mediator scenarios with CE$\nu$NS \cite{AristizabalSierra:2019ufd}. It is possible to
 utilize detector with directional sensitivities to CE$\nu$NS to various kinds of light mediators and their interactions with neutrinos. Both stopped pion and reactor neutrino sources can be used with gaseous helium and fluorine as possible detector materials. It is  shown that directional detectors show distinguishing spectral features in the angular and the recoil energy spectrum for light mediators  even for nuclear recoil thresholds as high as 50 keV. Directional information combined with energy and timing information, can help extracting new physics from CE$\nu$NS experiments~\cite{Abdullah:2020iiv}.

Apart from the CE$\nu$NS measurements, the light mediators can be constrained at the reactor experiments utilizing electron recoil with low threshold detectors.

\paragraph{Sterile neutrinos}
Many SM extensions capable of accommodating neutrino masses and mixings call upon the introduction of new neutral sterile leptons. From a theoretical point of view, there is no limit on the number of these sterile states and their masses can span over several orders of magnitude. Because they are flavor blind, CE$\nu$NS experiments, are sensitive to the total flux of active neutrinos and they can be used to probe neutrino oscillations to the sterile states \cite{Anderson:2012pn,Blanco:2019vyp,Miranda:2020syh}, in
complementarity to neutrino-electron scattering experiments. The sterile neutrino oscillation also can be probed at the reactor based CE$\nu$NS experiments utilizing low threshold detectors~\cite{Dutta:2015nlo}. 
Future, more ambitious CE$\nu$NS experiments can provide complementary information to the conventional neutrino oscillation data, offering an alternative way to search for light sterile neutrinos.

\paragraph{Low-mass dark matter}
Utilizing the high intensity  neutrino fluxes originated from pulsed proton beams, detectors sensitive to CE$\nu$NS processes can search for light dark matter \cite{deNiverville:2015mwa,Dutta:2019nbn,Dutta:2020vop}. Probing light dark matter with masses below $\sim 1$ GeV this analysis  can provide complementary information to direct detection searches. The characteristic timing composition of $\pi-$DAR beams provide useful opportunities to constrain systematic uncertainties, which, together with the coherent enhancement of the CE$\nu$NS-like dark matter cross section on nuclei allow for competitive sensitivities. Currently, both a scenario
with a vector portal that mixes kinetically with the photon (see also section \ref{subsec:Benchmark:VectorPortal}) \cite{COHERENT:2019kwz} as well another scenario with a lepto-phobic portal coupling to any SM baryon \cite{COHERENT:2019kwz,Aguilar-Arevalo:2021sbh} have been investigated with CE$\nu$NS experiments.

The production of dark matter $\chi$ can occur from the decay of a dark gauge boson $X$ via a  relevant Lagrangian  given by
\begin{equation}
    \mathcal{L}_{\rm prod} \supset \sum_f  \kappa_f x_f X_\mu \bar{f} \gamma^\mu f + \kappa_D X_\mu \bar{\chi} \gamma^\mu \chi\,, \label{eq:lagprod}
\end{equation}
where $x_f$ is the gauge charge of SM fermion species $f$ and $\kappa_f$ denotes the coupling constant associated with the dark gauge boson $X$.  
By contrast, $\kappa_D$ parameterizes the dark-sector coupling of $X$ to $\chi$. The dark matter can produce nuclear and electron recoils at the detector. Using the energy and momentum, $E_\chi$ and $p_\chi$, of incoming dark matter, the differential scattering cross section in recoil energy $E_{r,N}$ of the target nucleus is expressed as~\cite{Dutta:2019nbn}
\begin{equation}
    \frac{d\sigma}{dE_{r,N}}=\frac{(x_f\kappa_f \kappa_D)^2 Z^2 \cdot |F_X|^2}{4\pi p_\chi^2(2m_NE_{r,N}+m_X^2)^2} \left\{2E_\chi^2 m_N\left( 1-\frac{E_{r,N}}{E_\chi}-\frac{m_NE_{r,N}}{2E_\chi^2}\right) +m_N E_{r,N}^2\right\},
\end{equation}
where $Z$ and $m_N$ are the atomic number and the mass of the target nucleus and where $F_X$, which is a function over $2m_NE_{r,N}$, denotes the form factor associated with the dark gauge boson $X$. 

It has been shown that the timing information available in neutrino experiments with pulsed beam from  COHERENT, CCM and JSNS$^2$ is a powerful probe of new physics~\cite{Dutta:2019nbn, Dutta:2020vop}.
A combination of energy and timing cuts can eliminate SM neutrino events which emerge as prompt decays of pion with energy $\sim$ 30 MeV and delayed from the decays of the muons, thereby allowing the possibility of isolating DM-induced events which are in the prompt window with larger energy.

\begin{table}[t]
    \centering
    \resizebox{\columnwidth}{!}{
    \begin{tabular}{c|c c c}
    \hline \hline
         &  Channel & $E_r$ cut & $t$ cut\\
    \hline
    COHERENT-CsI & Nucleus scattering & 14~keV$<E_r<$26~keV & $t<1.5~\mu$s \\
    COHERENT-LAr & Nucleus scattering & $E_r>21$~keV & $t<1.5~\mu$s \\
    \multirow{2}{*}{CCM} & \multirow{2}{*}{Nucleus scattering} & \multirow{2}{*}{$E_r>50$~keV} & $t<0.1~\mu$s (Tight WP) \\
     & & & $t<0.4~\mu$s (Loose WP) \\
    JSNS$^2$ & Electron scattering & $E_r>30$~MeV & $t<0.25~\mu$s \\
    \hline \hline
    \end{tabular}
    }
    \caption{A summary of the recoil energy and timing cuts that we use for our data analysis.}
    \label{tab:cutsummary}
\end{table}

\begin{figure}[t]
    \centering
    \includegraphics[width=9cm]{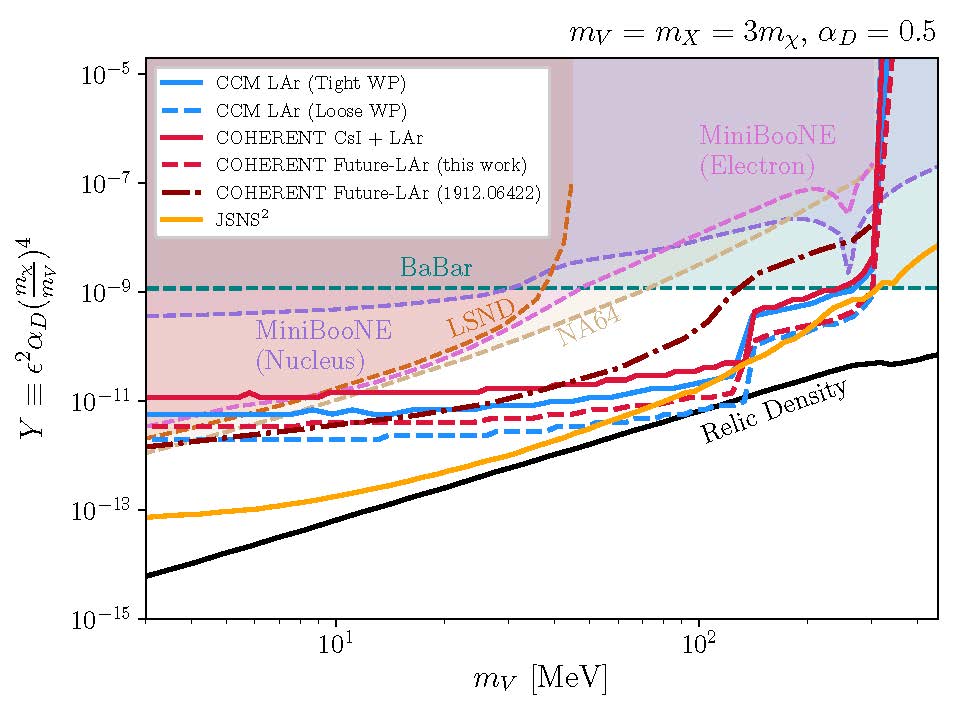}
    \caption{90\% C.L. projected experimental sensitivity to the model couplings and mediator masses in the single-mediator scenario ($X = V$) for our benchmark detectors.
    We take a dark photon $A'$ as our mediator, and conventionally plot the sensitivities on $Y\equiv \epsilon^2 \alpha_D (\frac{m_\chi}{m_V})^4$ where $\alpha_D = (\kappa^V_D)^2 / (4\pi)=0.5$ and $m_V/m_\chi=3$.
    The relevant existing limits from BaBar~\cite{Lees:2017lec}, LSND~\cite{deNiverville:2011it}, MiniBooNE~\cite{Aguilar-Arevalo:2018wea}, and NA64~\cite{NA64:2019imj} are shown by the shaded regions. 
    The parameter sets that are consistent with the observed dark matter relic abundance are shown by the black solid line.
    } 
    \label{fig:limits_single_mediator}
\end{figure} 

For Fig. \ref{fig:limits_single_mediator}, the energy and timing cuts are utilized as it allows for eliminating a large portion of backgrounds.
The choices for the cuts are summarized in Table~\ref{tab:cutsummary}.
In order to evaluate future sensitivities at COHERENT, CCM, and JSNS$^2$ to dark matter signals, a likelihood analysis is performed using simulation data at each experiment for nominal choices of the expected exposure.
For COHERENT, we have utilized the CsI and LAr data as released by the collaboration.

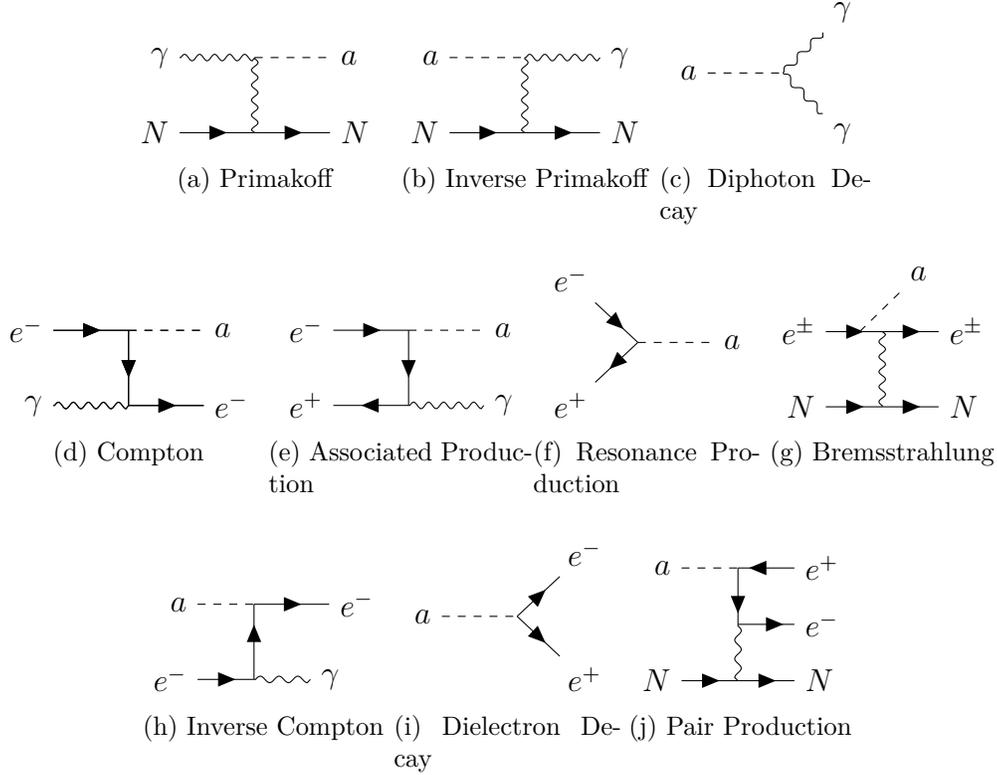
\begin{figure}[tbh]
 \centering
\subfloat[Primakoff]{
     \begin{tikzpicture}
              \begin{feynman}
         \vertex (o1);
         \vertex [right=1cm of o1] (f1) {\(a\)};
         \vertex [left=1cm of o1] (i1){\(\gamma\)} ;
         \vertex [below=1cm of o1] (o2);
         \vertex [right=1cm of o2] (f2) {\(N\)};
         \vertex [left=1cm of o2] (i2) {\(N\)};

         \diagram* {
           (i1) -- [boson] (o1) -- [scalar] (f1),
           (o1) -- [boson] (o2),
           (i2) -- [fermion] (o2),
           (o2) -- [ fermion] (f2),
         };
        \end{feynman} 
       \end{tikzpicture}
       \label{fig:axionPrimakoff}
}
\subfloat[Inverse Primakoff]{  
       \begin{tikzpicture}
              \begin{feynman}
         \vertex (o1);
         \vertex [right=1cm of o1] (f1) {\(\gamma\)};
         \vertex [left=1cm of o1] (i1){\(a\)} ;
         \vertex [below=1cm of o1] (o2);
         \vertex [right=1cm of o2] (f2) {\(N\)};
         \vertex [left=1cm of o2] (i2) {\(N\)};

         \diagram* {
           (i1) -- [scalar] (o1) -- [boson] (f1),
           (o1) -- [boson] (o2),
           (i2) -- [fermion] (o2),
           (o2) -- [fermion] (f2),
         };
        \end{feynman}
       \end{tikzpicture}
       \label{fig:axionInvPrimakoff}
} 
\subfloat[Diphoton Decay]{
       \begin{tikzpicture}
       \begin{feynman}
         \vertex (o1);
         \vertex [left=1cm of o1] (i) {\(a\)};
         \vertex [above right=0.75cm of o1] (f1) {\(\gamma\)};
         \vertex [below right=0.75cm of o1] (f2) {\(\gamma\)};

         \diagram* {
           (i) -- [scalar] (o1),
           (o1) -- [boson] (f1),
           (o1) -- [boson] (f2),
         };
        \end{feynman}
       \end{tikzpicture}
        \label{fig:axionDecayDiphoton}
} 
\\
\subfloat[Compton]{
    \begin{tikzpicture}
        \begin{feynman}
         
        \vertex (o1);
         \vertex [right=1cm of o1] (f1) {\(a\)};
         \vertex [left=1cm of o1] (i1){\(e^-\)} ;
         \vertex [below=1cm of o1] (o2);
         \vertex [right=1cm of o2] (f2) {\(e^-\)};
         \vertex [left=1cm of o2] (i2) {\(\gamma\)};

         \diagram* {
           (i1) -- [fermion] (o1) -- [scalar] (f1),
           (o1) -- [fermion] (o2),
           (i2) -- [boson] (o2),
           (o2) -- [ fermion] (f2),
         };
         
          \diagram* {
           (i1) -- [fermion] (o1) -- [scalar] (f1),
           (o1) -- [fermion] (o2),
           (i2) -- [boson] (o2),
           (o2) -- [ fermion] (f2),
         };
        \end{feynman}
    \end{tikzpicture}
    \label{fig:axionCompton}
}
\subfloat[Associated Production]{
    \begin{tikzpicture}
              \begin{feynman}
         \vertex (o1);
         \vertex [left=1.0cm of o1] (i1) {\(e^-\)};
         \vertex [right=1.0cm of o1] (f1){\(a\)};
         \vertex [below=1.0cm of o1] (o2);
         \vertex [right=1.0cm of o2] (f2){\(\gamma\)};
         \vertex [left=1.0cm of o2] (i2) {\(e^+\)};

         \diagram* {
           (i1) -- [fermion] (o1) -- [fermion] (o2) -- [fermion] (i2),
           (o2) -- [boson] (f2),
           (o1) -- [scalar] (f1),
         };
        \end{feynman}
       \end{tikzpicture}
    \label{fig:axionAssociated}
}
\subfloat[Resonance Production]{
    \begin{tikzpicture}
              \begin{feynman}
         \vertex (o1);
         \vertex [above left=0.75cm of o1] (i1) {\(e^-\)};
         \vertex [below left=0.75cm of o1] (i2) {\(e^+\)};
         \vertex [right=1.0cm of o1] (f1) {\(a\)};
         \diagram* {
           (i1) -- [fermion] (o1) -- [fermion] (i2),
           (o1) -- [scalar] (f1)
         };
        \end{feynman}
       \end{tikzpicture}
    \label{fig:axionProductionResonance}
}
\subfloat[Bremsstrahlung]{
    \begin{tikzpicture}
              \begin{feynman}
         \vertex (o1);
         \vertex [left=0.75cm of o1] (i1) {\(e^\pm\)};
         \vertex [right=0.75cm of o1] (f1) {\(e^\pm\)};
         \vertex [below=1cm of o1] (o2);
         \vertex [right=0.75cm of o2] (f2) {\(N\)};
         \vertex [left=0.75cm of o2] (i2) {\(N\)};
         \vertex [left=0.3cm of o1] (b1);
         \vertex [above right=0.75cm of b1] (f3) {\(a\)};

         \diagram* {
           (i1) -- [fermion] (o1) -- [fermion] (f1),
           (i2) -- [fermion] (o2) -- [fermion] (f2),
           (o2) -- [boson] (o1),
           (b1) -- [scalar] (f3),
         };
        \end{feynman}
       \end{tikzpicture}
    \label{fig:axionBrem}
}
\\
\subfloat[Inverse Compton]{
    \begin{tikzpicture}
        \begin{feynman}
         \vertex (o1);
         \vertex [right=1.0cm of o1] (f1) {\(e^-\)};
         \vertex [left=0.75 of o1] (i1){\(a\)} ;
         \vertex [below=1cm of o1] (o2);
         \vertex [right=0.75 of o2] (f2) {\(\gamma\)};
         \vertex [left=0.75 of o2] (i2) {\(e^-\)};

         \diagram* {
           (i1) -- [scalar] (o1) -- [fermion] (f1),
           (o1) -- [anti fermion] (o2),
           (i2) -- [fermion] (o2),
           (o2) -- [ boson] (f2),
         };
        \end{feynman}
    \end{tikzpicture}
    \label{fig:axionInvCompton}
} 
\subfloat[Dielectron Decay]{
    \begin{tikzpicture}\begin{feynman}
         \vertex (o1);
         \vertex [left=1cm of o1] (i) {\(a\)};
         \vertex [above right=0.75cm of o1] (f1) {\(e^-\)};
         \vertex [below right=0.75cm of o1] (f2) {\(e^+\)};

         \diagram* {
           (i) -- [scalar] (o1),
           (o1) -- [fermion] (f1),
           (o1) -- [fermion] (f2),
         };
        \end{feynman}
    \end{tikzpicture}
    \label{fig:axionDecayDielectron}
} 
\subfloat[Pair Production]{
    \begin{tikzpicture}
              \begin{feynman}
         \vertex (o1);
         \vertex [left=0.75cm of o1] (i1) {\(a\)};
         \vertex [right=0.75cm of o1] (f1){\(e^+\)};
         \vertex [below=0.75cm of o1] (o2);
         \vertex [right=0.75cm of o2] (f2){\(e^-\)};
         \vertex [below=0.75cm of o2] (o3);
         \vertex [left=0.75cm of o3] (i2) {\(N\)};
         \vertex [right=0.75cm of o3] (f3) {\(N\)};

         \diagram* {
           (i1) -- [scalar] (o1),
           (f2) -- [anti fermion] (o2) -- [anti fermion] (o1) -- [anti fermion] (f1),
           (o2) -- [boson] (o3),
           (i2) -- [fermion] (o3),
           (o3) -- [fermion] (f3),
         };
        \end{feynman}
       \end{tikzpicture}
    \label{fig:axionPairProduction}
}
    \caption{Contributions to ALP production  and detection considered in this analysis~\cite{CCM:2021lhc}.}
    \label{fig:axion}
\end{figure}

\paragraph{ALPs}Recently, it has been realized that the high intensity sources of photons, electron-positrons at neutrino experiments (reactor and beam based) can  provide an excellent opportunity to probe ALP ~\cite{Dent:2019ueq,AristizabalSierra:2020rom,Brdar:2020dpr}. In this regard the beam based CE$\nu$NS experiments, e.g., CCM, COHERENT, JSNS$^2$ with proton beam energies ranging from 800 MeV to  3 GeV can play important roles by producing  high intensity photon and electron-positron flux from bremsstrahlung, cascades, meson decays etc. at the target.  The photons can be utilized via 
Primakoff processes (enhanced by a factor of $Z^2$ where $Z$ is  the atomic number of the target nucleus)  for ALP production and similarly, inverse Primakoff can be utilized for detection  as shown in Figs. \ref{fig:axionPrimakoff}, \ref{fig:axionInvPrimakoff}. Further, axions can also decay into two photons in the detector as shown in Fig. \ref{fig:axionDecayDiphoton}).  ALPs, with couplings to electrons, can be produced via Compton-like scattering (Fig. \ref{fig:axionCompton}), bremsstrahlung, and $e^+e^-$ annihilation and can be detected via inverse-Compton scattering (Fig. \ref{fig:axionInvCompton}), decay to $e^+e^-$  (Fig. \ref{fig:axionDecayDielectron}).

\begin{figure}[h]
\centering
    \includegraphics[width=9cm]{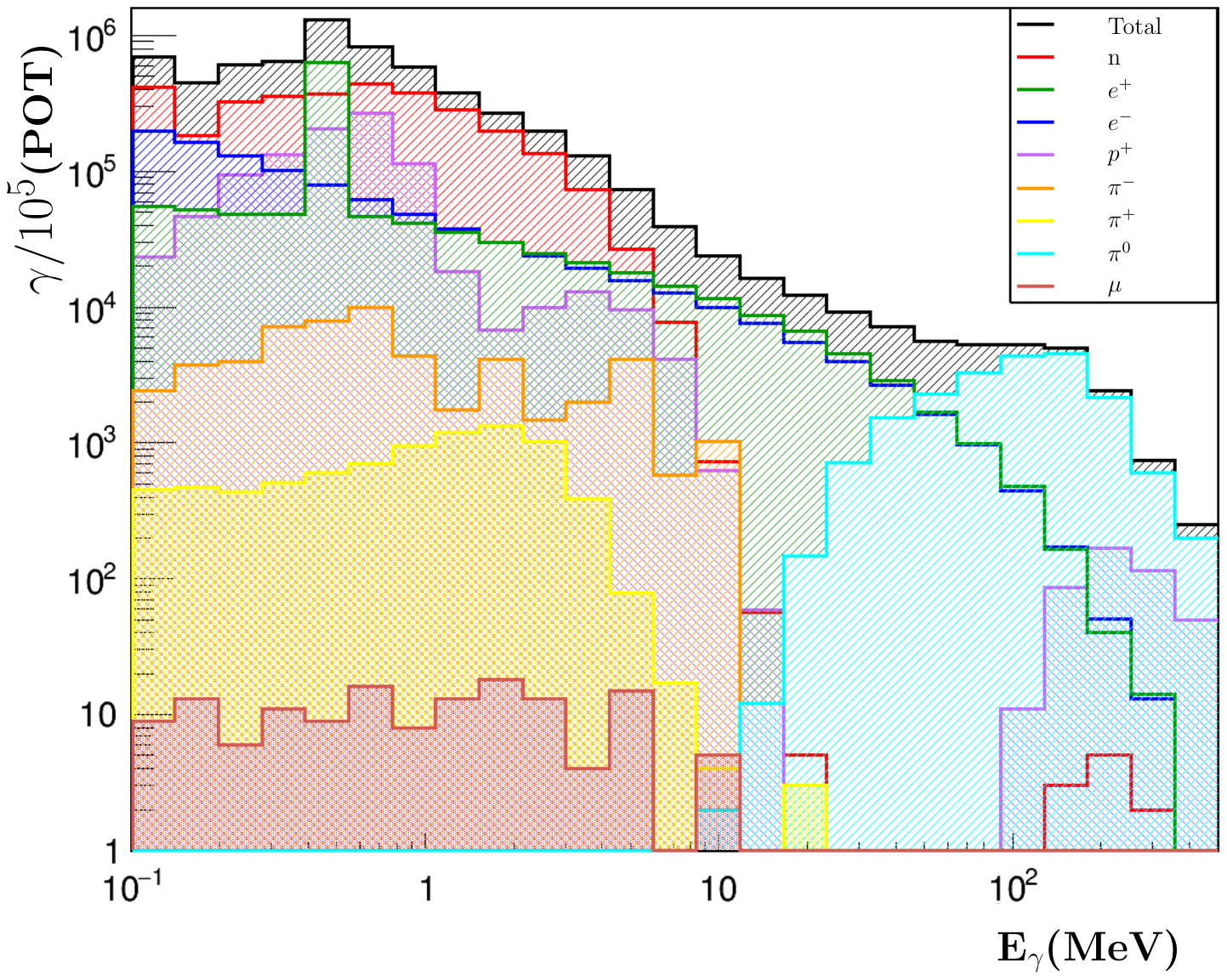}
    \includegraphics[width=9cm]{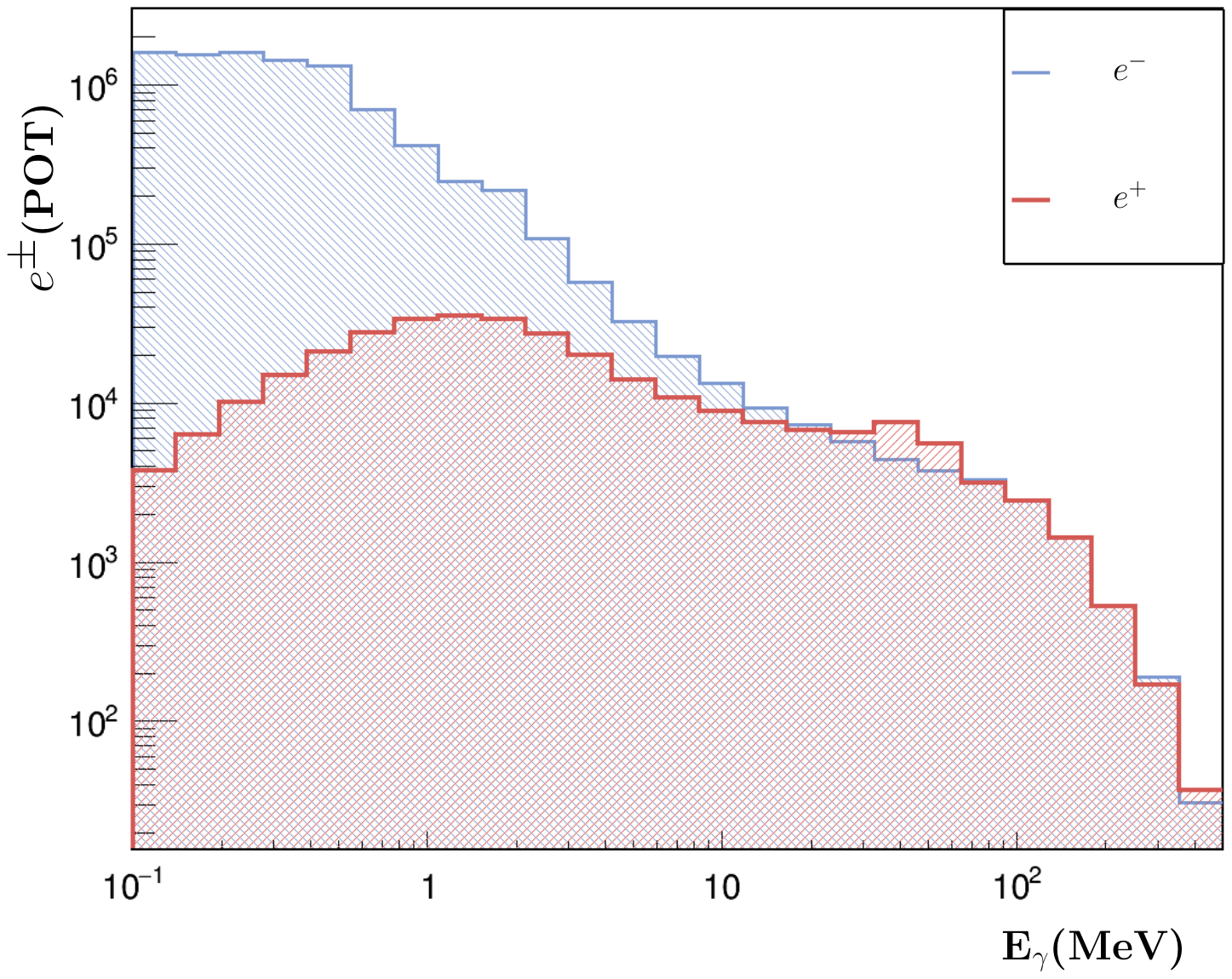}
    \caption{$\gamma$ and $e^\pm$ energy spectrum at the Lujan source. \texttt{GEANT4 10.5} is used with the \texttt{QGSP\_BIC\_HP} library~\cite{Agostinelli:2002hh} by generating $10^5$ protons incident on a tungsten target~\cite{CCM:2021lhc}.}
    \label{fig:photonspec}
\end{figure}

In order to search for ALPs, we use the Coherent CAPTAIN-Mills (CCM) experiment, as an example where an 800 MeV proton beam hits a tungsten target~\cite{CCMcollaboration-ALP}. The \texttt{GEANT4 10.7} along with the \texttt{QGSP\_BIC\_HP} library~\cite{Agostinelli:2002hh} is used to model the photon and electron-positron production processes. The resulting photon, electron/positron spectra are  shown in Fig.\,\ref{fig:photonspec} for the energy range $E_\gamma = 1\, \text{MeV}- 1\, \text{GeV}$.  

The ALP interactions  are parametrized by the following Lagrangian terms for this analysis:
\begin{equation}
\mathcal{L}_{\rm ALP} ~\supset~ -\frac{g_{a\gamma}}{4}\,aF_{\mu\nu}\tilde{F}^{\mu\nu}\,-\,g_{ae}\,a\,\bar e \,i \gamma_5\, e\end{equation}
where $F_{\mu\nu}\tilde{F}^{\mu\nu}$ is the electromagnetic dual field-strength operator.

\begin{figure}
    \centering
    \includegraphics[width=0.48\textwidth]{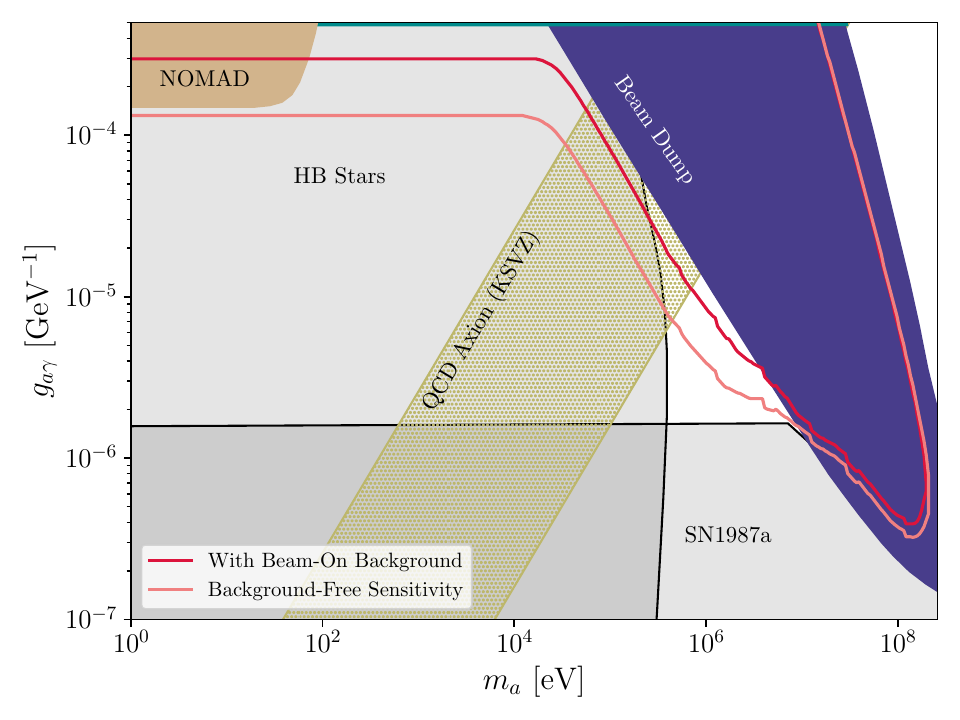}
    \includegraphics[width=0.49\textwidth]{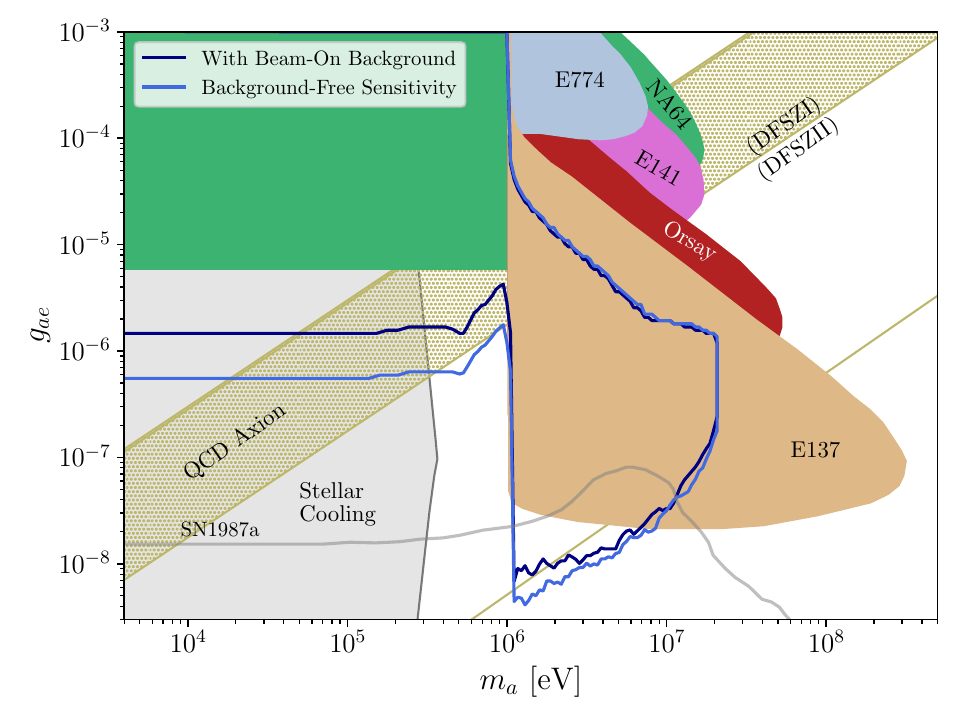}
    \caption{CCM sensitivities over the axion-photon coupling  and electron coupling parameter spaces  are shown for 3 year ($2.25 \cdot 10^{22}$ POT)~\cite{CCM:2021lhc}. }
    \label{fig:reach}
\end{figure}

In Fig.~\ref{fig:reach}, the CCM reach for the ALP parameter space for the coupling $g_{a\gamma}$ is shown where all the existing constraints from various experiments and astrophysical bounds are included. CCM's reach is shown in dark red (with projected background based on the current measurement) and pink (no backgrounds). However, the sensitivity would be as good as the zero background limit once the purity of the LAr is improved. The ``cosmological triangle"~\cite{Carenza:2020zil,Depta:2020wmr}, can be explored at CCM via ALP decaying into two photons which is not excluded by any lab based experiment or astrophysical constraint at present. The upper region of the triangle overlaps some of the parameter space allowed within various QCD axion models~\cite{DiLuzio:2020wdo}. The parameter space bounded by the horizontal line portion of the sensitivity graph is due to the inverse Primakoff scattering and extends to a very low axion mass and in this region of the parameter space, CCM would also be able to put the best laboratory bound. The astrophysical bounds are  generally dependent on the underlying model assumed~\cite{Jaeckel:2006xm,Khoury:2003aq,Masso:2005ym,Masso:2006gc,Dupays:2006dp, Mohapatra:2006pv,Brax:2007ak,DeRocco:2020xdt}. 

In Fig.~\ref{fig:reach},  we show the CCM reach for the ALP parameter space for the coupling $g_{ae}$. CCM's reach here is shown in navy blue (with projected background based on the current measurement) and light blue (no background scenario). The dotted line shows the projected limit.  The sensitivity reaches the parameter space where there exists no direct astrophysical or lab-based limits. The sensitivity reach also covers some unexplored regions of the QCD axion models.

The Primakoff and Compton channels can be important to probe ALPs at reactors ~\cite{Dent:2019ueq, AristizabalSierra:2020rom} both for production and detection.  Experiments like MINER, CONUS, CONNIE, $\nu$-cleus etc. dedicated to CE$\nu$NS observation can be utilized to search for ALPs.
The high intensity photon flux from the reactors can used 
to produce  ALP via Primakoff and Compton channels. Low-threshold detectors in close proximity to the core will be sensitive to ALP decays and inverse Primakoff and Compton scattering, providing sensitivity to the ALP-photon and ALP-electron couplings.

\begin{figure}
    \centering
    \includegraphics[width=0.48\textwidth]{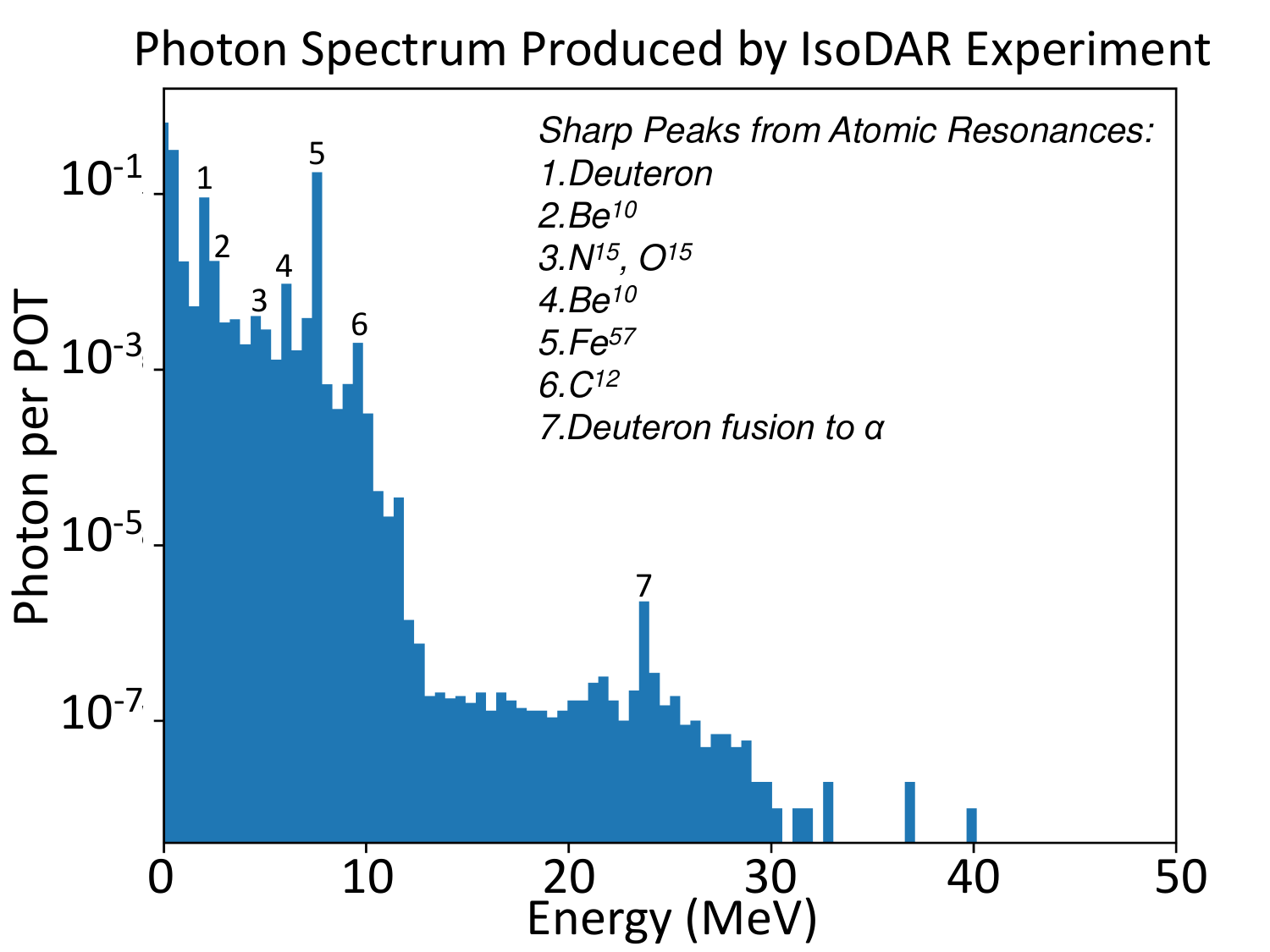}\includegraphics[width=0.48\textwidth]{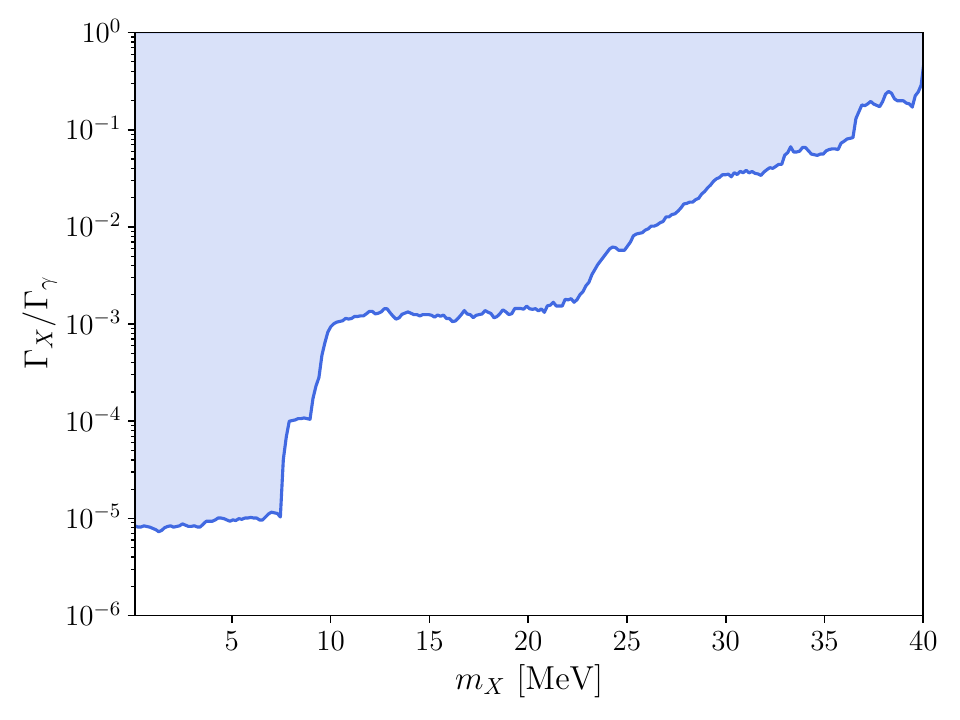}
     \includegraphics[width=0.58\textwidth]{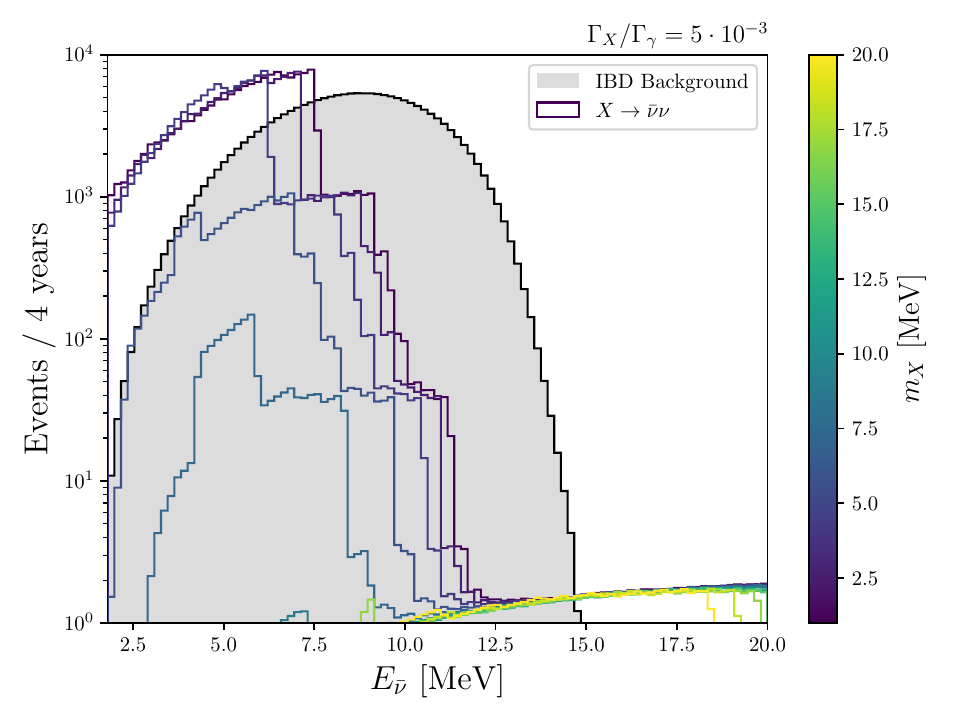}
    \caption{Top-left: The photon spectrum  from the IsoDAR target using  Geant 4-\texttt{QGSP\_BIC\_ALLHP} library. Top-right: The IsoDAR 4-year livetime sensitivity exclusion for the $N^* \to N X(\to \bar{\nu}\nu)$ branching ratio as a function of the $m_X$ at 90\% CL. The flat limit for $m_X \lesssim 5$ MeV can be  extended to arbitrarily small masses.~\cite{Alonso:2021kyu}.Bottom: The IBD ($\bar{\nu}_e + p \to e^+ + n$) rates from the decaying $X\to\bar{\nu}_e\nu_e$ arriving at the Yemilab detector  plotted along with the expected IBD background (gray). The spectral shape is due to the convolution of the boosted 2-body decay spectrum with the IBD cross section which is summed over all kinematically accessible nuclear transitions to produce the $X$ states }
    \label{fig:photon_isodar}
\end{figure}

\paragraph{``Bump" hunting using low energy proton beam based IsoDAR experiment}
The photon flux from a low energy  proton beam experiment, e.g., IsoDAR  can provide interesting BSM search prospects~\cite{Alonso:2021kyu}. In this experiment the photon flux exhibits interesting line-structures which emerge from the transitions of excited nuclei Fig.~\ref{fig:photon_isodar}(top-left). The transition lines can produce  a light mediator, $X$(scalar, pseudoscalar, vector etc), coupled to the nucleons and the types of mediators  determine the  magnetic ($M_i$) and electric ($E_i$) moments of these transitions~\cite{Avignone:1988bv,Feng:2016ysn,Dent:2021jnf}. 
The production rate of the new mediator  can be expressed as a branching ratio for a given transition which depends on the coupling and   the mass. One can assume that $X$ has couplings  to both quarks and neutrinos, e.g., ~\cite{Datta:2018xty,AristizabalSierra:2019ykk, Boehm:2018sux}.  

In Fig.~\ref{fig:photon_isodar}(top-right), the 90\%~CL sensitivity to the mediator  $X$  is shown. Here $X$   decays to $\bar{\nu}\nu$ pairs  where the antineutrino is detected via IBD. 
Fig.~\ref{fig:photon_isodar}(bottom)  shows $\bar{\nu}_e$ energy spectra  for several masses which are  compared with the IBD rate from $^{8}$Li ( expected to be the only significant background for this search). The edges in the spectrum emerge due to  the photon spectrum  and IBD cross section convolution. From this analysis, $X$-boson coupling can be constrained to be $\leq 10^{-3}$ for mass $O(10)$~MeV when the $X$ boson decays promptly into neutrinos with coupling values $\geq 10^{-7}$. The existing constraints from COHERENT, CCM have not ruled out this parameter space completely.  

The ``bump search'' utilizing the transition lines at the low energy beam dump experiments  have  many interesting consequences:\\
(i) We can probe low mass mediators which  can  provide positive and negative contributions to $\Delta N_{eff}$ depending on its decay branching ratio into neutrino-anti-neutrinos and electron-positrons, respectively~\cite{Escudero:2018mvt}. Further, the interactions involving the decay into neutrino final states have impact on the neutrino floor for dark matter direct detection experiments~\cite{AristizabalSierra:2019ykk, Boehm:2018sux,AristizabalSierra:2021kht}.\\
(ii) We can achieve sensitivity to the light mediator claimed to explain the Atomki anomaly~\cite{Krasznahorkay:2015iga, Gulyas:2015mia, Aleksejevs:2021zjw, DelleRose:2018pgm}. This is a reported excess of $e^+e^-$ pairs observed in the decay of the 18 MeV excited state of beryllium produced through $^7$Li(p,n)$^8$Be$^*$, and the set of 20 MeV excited states of helium produced through $^3$H(p,$\gamma$)$^4$He.\\
(iii) We can probe the origin of the  5~MeV reactor bump, which is observed in the event distribution of most modern reactor experiments, e.g., PROSPECT~\cite{prospect2} STEREO~\cite{stereo},  NEOS~\cite{NEOS:2016wee}, RENO~\cite{reno}, Daya Bay~\cite{dayabay} and Double Chooz~\cite{doublechooz}.

\begin{figure}
    \centering
    \includegraphics[width=0.48\textwidth]{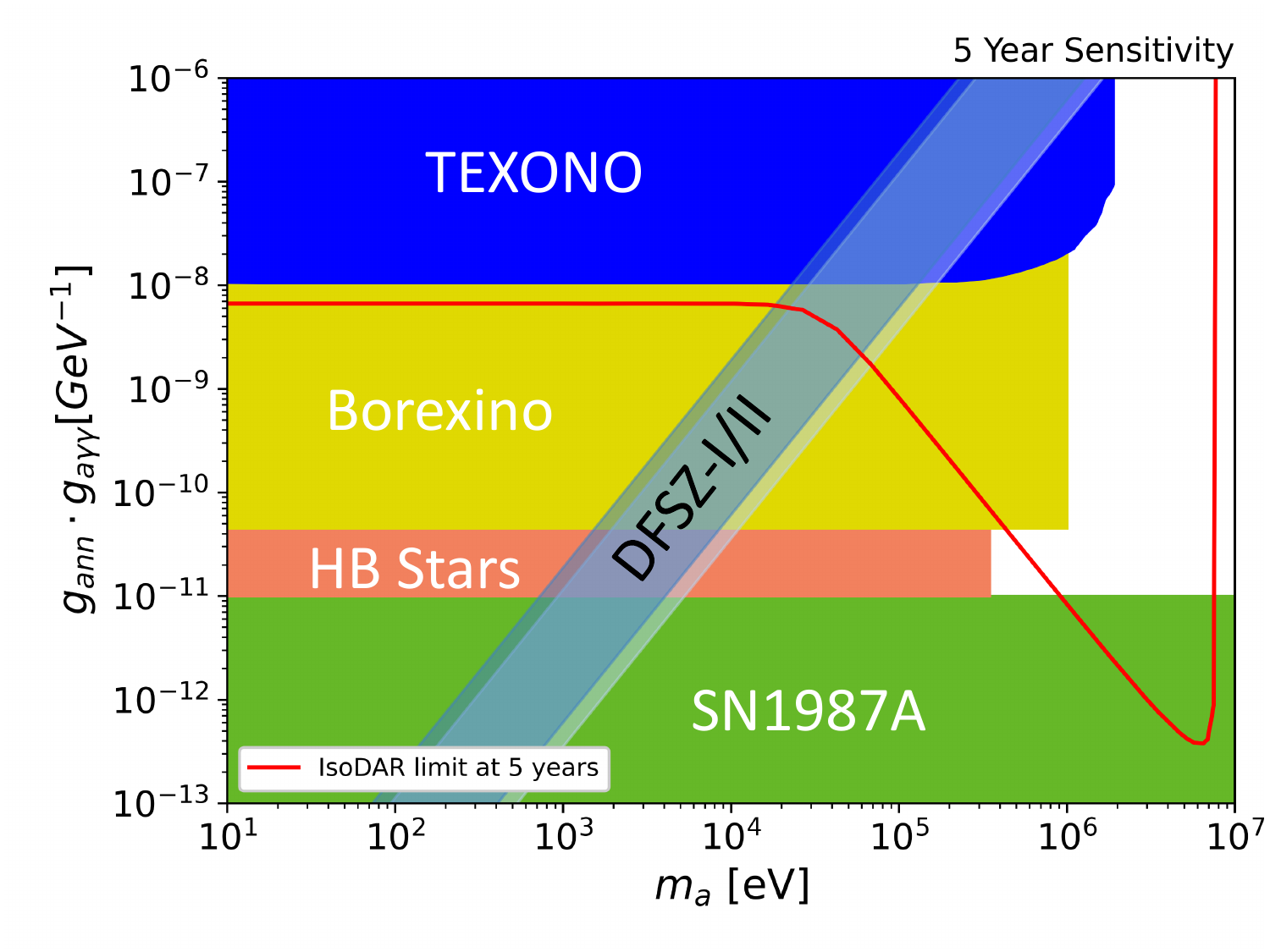}
    \caption{[Preliminary]The IsoDAR  sensitivity exclusion $g_{aNN}\times g_{a\gamma\gamma}$ is shown as function of ALP mass $m_a$.~\cite{we_isodar}. }
    \label{ALP_sensitivity}
\end{figure}

{\bf{ALP at IsoDAR}}
The M-type transition lines will be very useful to investigate ALP at IsoDAR. We will be able to get world leading constraint in some of the channels. In figure~\ref{ALP_sensitivity}, the 5 year sensitivity exclusion plot is shown for the product of couplings $g_{aNN}\times g_{a\gamma\gamma}$. We find that IsoDAR will provide the best constraint in the parameter space which has constraints from TEXONO, Borexino, SN1987A, HB stars etc. Similar exclusion plots can be obtained for $g_{aNN}\times g_{aee}$ as a function of $m_a$. 

ALPs at IsoDAR can be also obtained from the Primakoff process using the bremsstrahlung and cascade photons and from the cascade electrons which would provide constraints on $g_{a\gamma\gamma}$ and $g_{aee}$ separately. A detailed study is underway~\cite{we_isodar} 
\subsubsection{Incoherent neutrino-nucleus scattering (I$\nu$NS)}
Neutral-current neutrino-nucleus
scattering also presents an inelastic (incoherent) channel (I$\nu$NS), in which the quantum state of the nucleus in the final state is changed. Contrary to the elastic channel, this cross section has a linear dependence on the number of nucleons~\cite{Bednyakov:2018mjd,Pirinen:2018gsd,Bednyakov:2019dbl}. Going to higher neutrino energies, there is a smooth transition between the coherent and incoherent neutrino-nucleus scattering regimes. The  I$\nu$NS channel can affect the high-energy tail of the recoil spectrum of neutrino experiments like COHERENT~\cite{Sahu:2020kwh}.
A correct treatment of both channels requires an accurate evaluation of the transition matrix elements describing the various interaction channels between the initial and a final nuclear states. 

\begin{figure}[tbh]
    \centering
    \includegraphics[width=7cm]{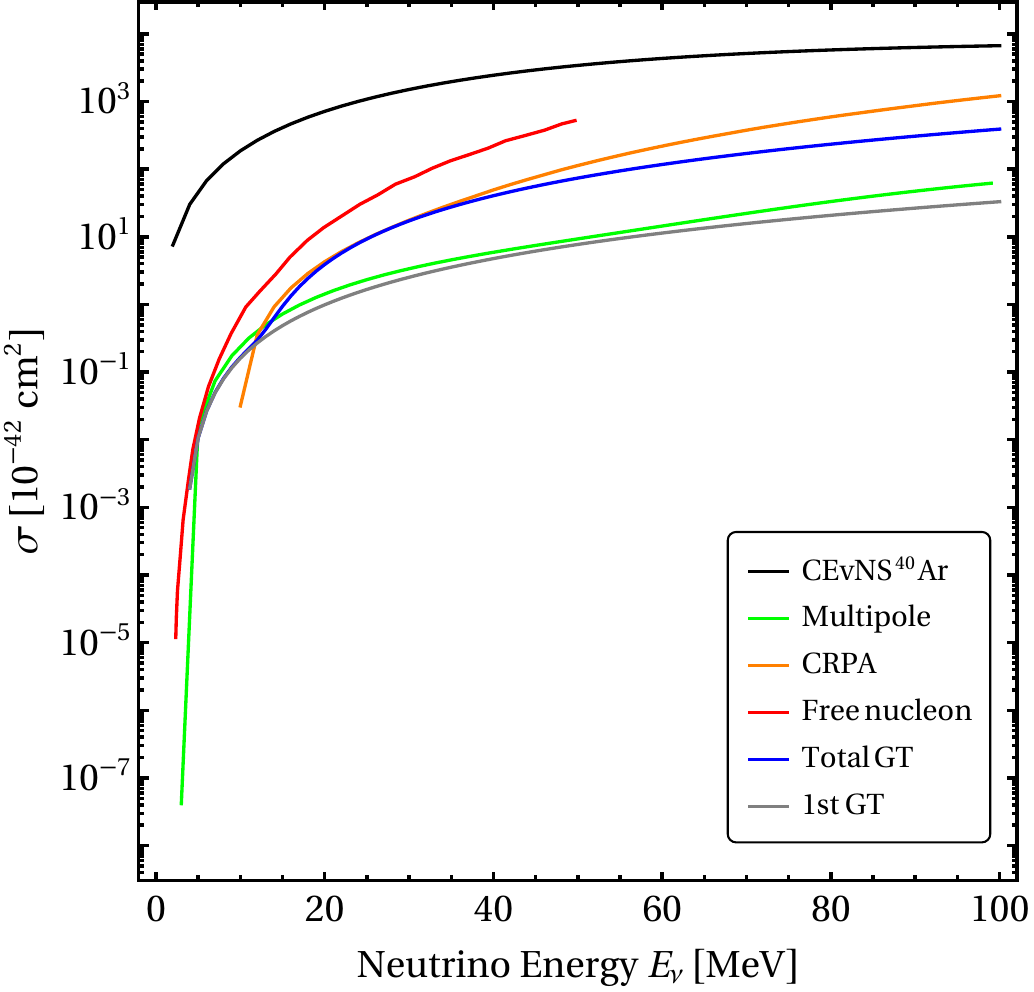}
    \includegraphics[width=7cm]{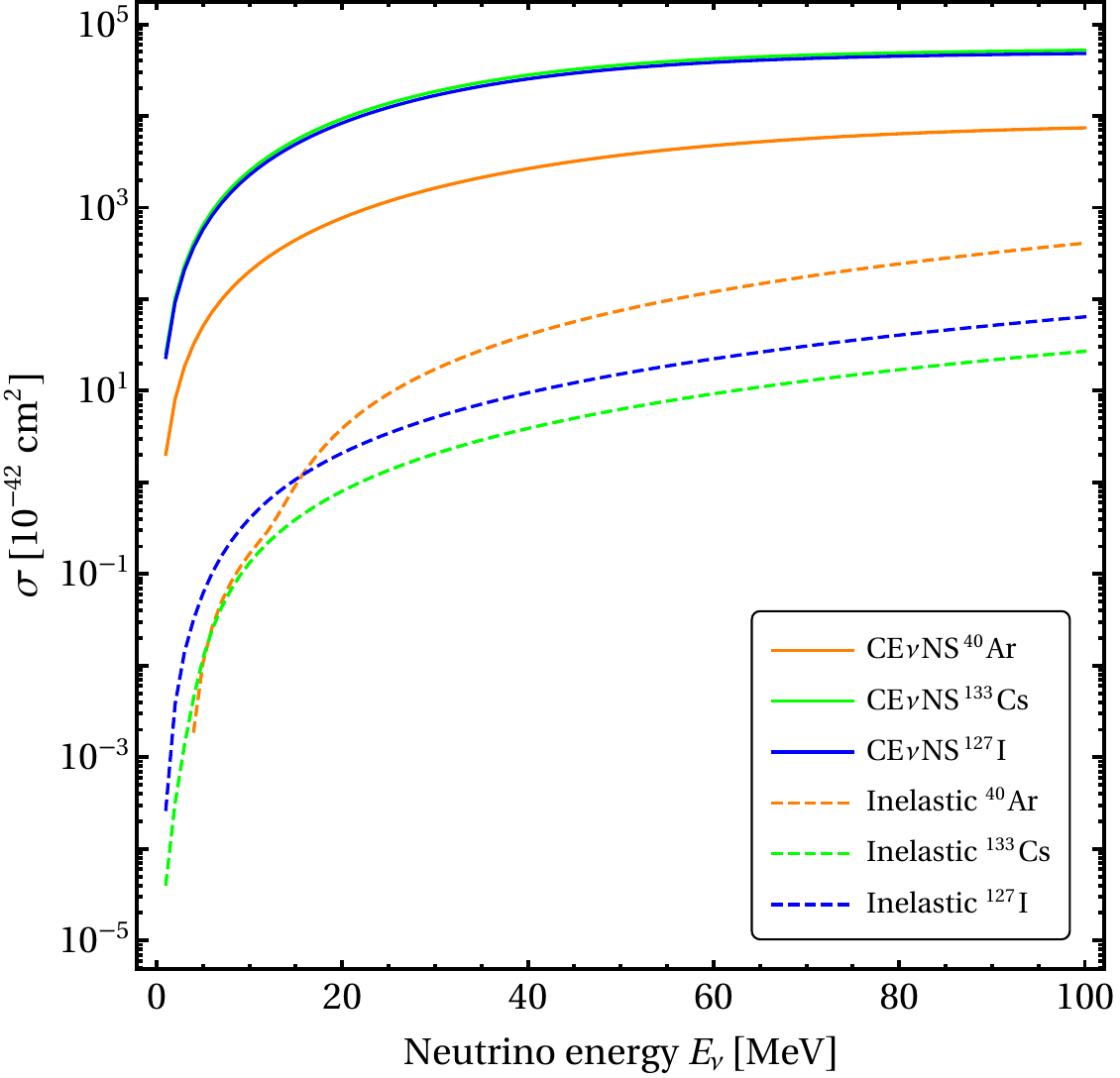}
    \includegraphics[width=7cm]{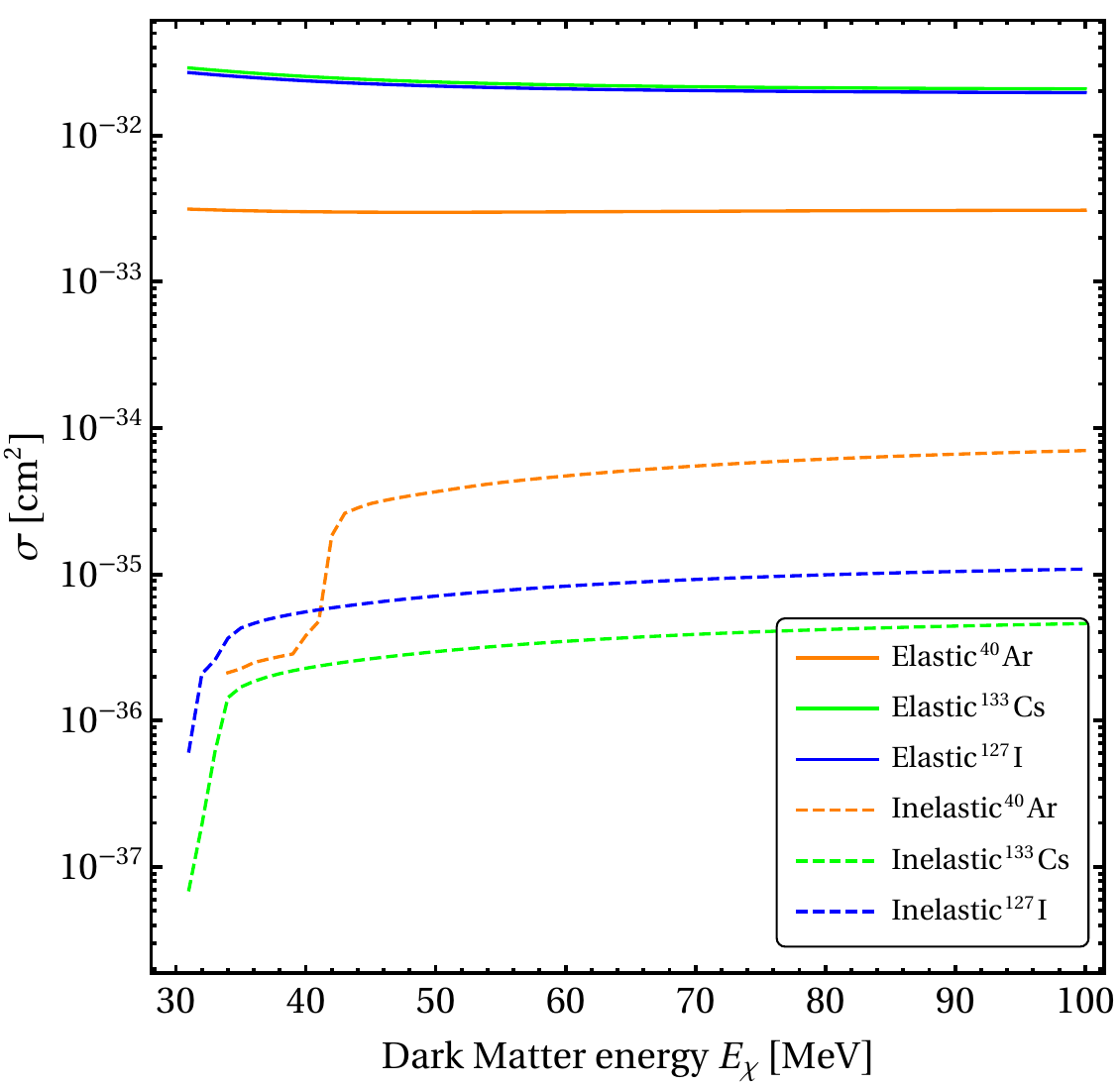}
    \caption{Comparison of multipole operator analysis, total GT and 1st GT are compared with the CRPA \cite{VanDessel:2020epd} and free nucleon~\cite{Bednyakov:2018PRD} predictions (top-left). Total elastic and inelastic $\nu$-nucleus (top-right) DM-nucleus (bottom) scattering cross section for $^{40}$Ar, $^{133}$Cs, and $^{127}$I nuclei are shown~\cite{Dutta:2022tav}.}\label{ar40}
\end{figure}

In Fig.\ref{ar40}(top-left) we show various neutrino-nucleus inelastic calculations. The top left plot shows a comparison of multipole operator analysis, total Gammow-Teller (GT) and 1st GT are compared with the CRPA \cite{VanDessel:2020epd} and free nucleon~\cite{Bednyakov:2018PRD} predictions for $^{40}Ar$~\cite{Dutta:2022tav}. The full multipole operator analysis and the GT transitions are calculated using the nuclear shell model code BIGSTICK\cite{Johnson:2018hrx,Johnson:2013bna}. For small momentum transfers, when a GT transition is kinematically accessible, the inelastic cross section will be dominated by it and in this limit the GT operator alone  can then provide an efficient way to approximate the scattering cross section.
The multipole results  only include transitions to the first 15 excited states and thus it is not capturing all accessible states. However it includes the first GT transition and one can see the GT dominates the inelastic cross-section. Total GT includes all possible GT transitions.
Computing the complete multipole results would be the most precise results, however,  the total GT strength is sufficiently precise. GT approximation makes large number of final states calculation a tractable more computational task. The total elastic and inelastic cross-sections for the $^{133}$Cs and $^{127}$I targets are shown in Fig.~\ref{ar40} (top-right).

The elastic and inelastic cross section for dark matter scattering on the target nuclei $^{40}$Ar, $^{133}$Cs and $^{127}$I are shown in Fig.~\ref{ar40}(bottom). We  assume $m_\chi=30$MeV, $m_{A'}=90$ MeV, $\epsilon=10^{-4}$  and $g_D=\sqrt{2\pi}$ in the allowed parameter space. $^{133}$Cs and $^{127}$I have a higher elastic cross section than $^{40}$Ar since $\sigma_{\rm el}$ has explicit $Z^2$ (atomic number) dependence, but have a lower inelastic cross section since the inelastic  cross section is the largest for $^{40}$Ar and is the lowest for $^{133}$Cs. 

\subsubsection{Neutrino tridents}
Neutrino trident production is a rare SM process in which a pair of leptons with opposite charge is produced from the scattering of a neutrino off the Coulomb field of a heavy nucleus \cite{Czyz:1964zz,Fujikawa:1971nx,Lovseth:1971vv,Koike:1971tu,Koike:1971vg,Brown:1972vne}.
Neutrino trident production has been searched for in several neutrino beam experiments and dimuon tridents have been observed at the CHARM-II \cite{CHARM-II:1990dvf} and CCFR \cite{CCFR:1991lpl} experiments. NuTeV found no conclusive signal \cite{NuTeV:1998khj,NuTeV:1999wlw}. While these measurements are consistent with SM predictions, this process offers a sensitive probe to constrain new physics, in particular new neutral currents among leptons. For instance, it has been shown to be capable of imposing stringent constraints on well-motivated extensions of the SM introducing a $Z'$ boson by gauging the difference of the lepton number between the muon and tau flavor, $L_\mu - L_\tau$ \cite{Altmannshofer:2014pba,Kaneta:2016uyt}. Sizeable number of trident events are expected at future near detector experiments receiving intense neutrino beams from accelerators \cite{Ballett:2018uuc,Altmannshofer:2019zhy,Ballett:2019xoj}, thus allowing to improve current measurements, including measuring dielectron and mixed flavor trident channels and to search for new physics effects \cite{Magill:2016hgc,Magill:2017mps}.


\section{Tools}
\label{sec:tools}

\begin{table}[!tbh]
\centering
	\begin{tabular}{c|c c c c|c|c c c c|c|c}
		\hline
		& \multicolumn{4}{c|}{Production}  &  & \multicolumn{4}{c|}{Dark $\rightarrow$ Standard}  & & \\
		Process & Brem.& Direct & Prompt & LL & Flux & Decay & $e$ & $N$ El. & $N$ Inel. & Det. & Reco. \\
		\hline
		\texttt{MadDump} &&  \textcolor{green!50!black}{$\checkmark$} & \textcolor{green!50!black}{$\checkmark$} &  & \textcolor{green!50!black}{$\checkmark$} & & \textcolor{green!50!black}{$\checkmark$} & & \textcolor{yellow!50!black}{$\checkmark$} & & \\
		\texttt{BdNMC} & \textcolor{green!50!black}{$\checkmark$} & \textcolor{green!50!black}{$\checkmark$} & \textcolor{green!50!black}{$\checkmark$} & & \textcolor{green!50!black}{$\checkmark$} & \textcolor{green!50!black}{$\checkmark$} & \textcolor{green!50!black}{$\checkmark$} & \textcolor{yellow!50!black}{$\checkmark$} & \textcolor{yellow!50!black}{$\checkmark$} &  &  \\
		\texttt{GENIE} & & & & & & & \textcolor{green!50!black}{$\checkmark$} & \textcolor{green!50!black}{$\checkmark$} & \textcolor{yellow!50!black}{$\checkmark$} &  & \\
		\texttt{Geant4} & & & 
		\textcolor{yellow!50!black}{$\checkmark$} & \textcolor{yellow!50!black}{$\checkmark$} & \textcolor{yellow!50!black}{$\checkmark$} & \textcolor{yellow!50!black}{$\checkmark$} & & & & \textcolor{yellow!50!black}{$\checkmark$} &  \\
		\texttt{ACHILLES} & & & & & & \textcolor{yellow!50!black}{$\checkmark$} & \textcolor{green!50!black}{$\checkmark$} & \textcolor{green!50!black}{$\checkmark$} & \textcolor{yellow!50!black}{$\checkmark$} & & \\
		\texttt{FORESEE} &
		\textcolor{green!50!black}{$\checkmark$} & \textcolor{green!50!black}{$\checkmark$} & \textcolor{green!50!black}{$\checkmark$} & \textcolor{green!50!black}{$\checkmark$} & \textcolor{green!50!black}{$\checkmark$} & \textcolor{green!50!black}{$\checkmark$} & \textcolor{green!50!black}{$\checkmark$} & \textcolor{yellow!50!black}{$\checkmark$} & & & \\
		\hline
	\end{tabular}
	\caption{Summary of the capabilities of some tools that are currently available. A green check mark indicates strong capabilities and full public availability. A yellow check mark indicates either missing features or inaccessibility. None of these tools can cover most dark sector models and current tools are generally focused on a handful of models. \texttt{Geant4} stands for a suite of tools developed by experimental collaborations, such as those within \texttt{sbncode} or \texttt{G4edep-sim}. The full labeling of the processes is dark bremsstrahlung, direct production (deep inelastic scattering), prompt meson decay, long-lived meson decay, flux, dark sector decay, electron scattering, nucleon elastic scattering, nucleon inelastic scattering (resonant and deep inelastic), detector simulation, reconstruction.}\label{tab:simulation}
\end{table}

Assessment of the reach of neutrino beam experiments to dark sector models requires detailed simulation of the signals. Several factors challenge efforts to develop such simulations. First, the target geometry can play a significant role in models where dark sector states arise from long-lived meson decays, which can be challenging to model accurately. The beam energies are high but not so high that nuclear effects in proton-target collisions are entirely negligible. Finally, the slowness of detector simulation for LArTPC experiments is a bottleneck on event generation for upcoming experiments. Given these challenges, several different tools have been developed that address the specific needs of given models. We describe several of these tools in \ref{sec:existing-tools}. 

Beyond tools for event generation, dark sector signals may introduce unique topologies that are not optimally detected by existing reconstruction tools, which focus primarily on neutrino-like signals. Ongoing work in the reconstruction of non-standard signals will require new tools, including potentially harnessing the power of machine learning. These issues are discussed more detail in \ref{sec:necessary-capabilities}.

\subsection{Event Generation Tools}\label{sec:existing-tools}

Event generation necessarily requires two components: generation of dark sector states as the protons of the beam collide on target and generation of a visible signal in the detector downstream. Each of these can pose unique challenges.

\textbf{Dark sector production}

The targets at neutrino experiments are typically thin, in the 1-2 radiation length range, leading to significant portions of the proton beam that go through to a downstream absorber. Furthermore, any hadronic particles produced in the collision may escape the target before decaying. Such decays can produce dark sector particles in some models, such as the Higgs Portal model (previously discussed in Section \ref{subsec:Benchmark:HiggsPortal}). Finally, magnetic horns are typically employed to enhance the neutrino or anti-neutrino content of the beam. For dark sector states that are produced promptly, either by short decays of, say, $\pi^0$ or direct production such as in dark bremsstrahlung, these challenges can be readily overcome. Tools like \texttt{BdNMC} \cite{deNiverville:2016rqh} and \texttt{MadDump} \cite{Buonocore:2018xjk} can be used to accurately simulate such prompt processes for a number of dark sector models.

For models such as the Higgs Portal model, where dark sector states are produced via decays of long-lived kaons, a more comprehensive, \texttt{Geant4}-based approach is typically required \cite{GEANT4:2002zbu}. This can be done by appropriately modifying the output of neutrino beam simulations. Unfortunately, this needs to be done on a beam-by-beam basis but has been successfully implemented for both the Booster Neutrino Beam (BNB) and Neutrinos at the Main Injector (NuMI) beam \cite{Batell:2019nwo}.

\textbf{Dark sector scattering and decay}

Subsequent to the production of dark sector states, these particles need to be propagated into the detector, which is a straightforward geometric exercise, provided the geometry is accurately known. They can leave a visible signal either by decay, which is generally straightforward to generate, or scattering, which may be more complicated. Event generation is typically straightforward if the dark sector state scatters electrons. Complications may arise if the dark sector topology is non-trivial, say if the scattering is inelastic and leads to a displaced decay. \texttt{BdNMC} and \texttt{MadDump} can handle straightforward electron scattering, and even inelastic scattering could and should be implemented without too much difficulty.

The more challenging situation is when the dark sector states scatter off of nucleons. The energy of particles produced in neutrino beams is typically around one to few GeV, leading to scattering with energy transfers around the QCD scale. This is a challenging energy scale to simulate, as there is no first-principles description of the physics. Instead, modeling and data-fitting are required. Furthermore, in this energy regime, any hadrons produced in the dark sector scattering can scatter as they exit the nuclear debris in a process known as Final State Interactions (FSI). These FSI can significantly distort both the particle IDs and spectra coming out of hadronic dark sector collisions. For a broad class of dark matter models, namely those with a spin-1 mediator, a tool has been developed to simulate hadronic scattering \cite{Berger:2018urf}. This tool is built within the \texttt{GENIE} software \cite{Andreopoulos:2009rq}, making it straightforward to test different nuclear models and to connect with detector simulation pipelines. Recently, a new tool (\texttt{ACHILLES}) was developed to increase the flexibility of including arbitrary models into neutrino event generators~\cite{Isaacson:2021xty}. \texttt{ACHILLES} separates the nuclear current from the leptonic current, providing an adaptable framework to quickly study new models. Furthermore, \texttt{ACHILLES} is the first code that correctly propagates spin correlations through the generation of $n$-body final state processes, instead of generating $2\rightarrow 2$ processes with subsequent on-shell decays.

\subsection{Necessary Experimental Capabilities}\label{sec:necessary-capabilities}

\textbf{Event timing}

Many models postulate BSM particles with longer time of flight than
neutrinos. In the usual case where the neutrino beam is pulsed,
the time window immediately after all the neutrinos have crossed the detector will offer
a sample without neutrino background.
Despite only containing a fraction of the signal, these kinds of samples exclude neutrinos which can mimic the signal topologies and cannot be easily eliminated.
In the search for heavy neutral leptons produced from kaon decays at rest in BNB, the MicroBooNE experiment adds a new configuration in the trigger system and thereby utilizes the data collected 0.6~\si{\micro\second} after the
arrival of the neutrinos produced in the same beam, which has an energy spectrum peaked at $\sim$800~MeV~\cite{MicroBooNE:2019izn}.
As a result, the remaining background of this analysis is the mis-reconstructed cosmic rays, which have different topologies from signal and are much more straightforward to get rid of.
In addition, it is possible to employ the time structure of the proton extraction in data collected in the arrival time window of neutrinos.
This typically requires \si{\nano\second} time resolution, and detectors designed for light detection usually meet the requirement.

Late signal events offer a particular challenge for the trigger system of LArTPC experiments. Trigger systems are extremely important to LArTPC-based experiments, whose millimeter spatial resolution results in a huge volume of data and requires real-time selection to make it manageable.
It is essential to have a flexible trigger system allowing different configurations optimized to different physics topics.
While the complete MicroBooNE experiment and the ongoing ICARUS experiment (the far detector of SBN) both have a trigger system up to a certain degree of flexibility, a more sophisticated trigger system is being developed.
For instance, based on the nanosecond scale of the beam structure, a trigger system with nanosecond time resolution will more efficiently select neutrinos or BSM particles with longer time-of-flight.
In a LArTPC, only the scintillation light of liquid argon have such a time resolution, and therefore light detectors required to have nanosecond time resolution, for both the real-time trigger logics and data analyses.

As LArTPC-based experiments utilize the scintillation light to obtain nanosecond-scale timing of interested activities and the ionization charge for the spatial and kinematic information,
matching the light and charge originating from the same activity is crucial. 
An accurate, efficient matching algorithm will allow us to determine the time-of-flight of a particle to the level of a few nanoseconds and therefore distinguish heavy BSM particles from SM neutrinos.

\textbf{Fast detector simulation and its challenges}

Unlike the particles produced in a collider experiment, the interactions or decays
of interests here can occur at every point inside (or outside) the active volume
of the neutrino detectors.
Given the finite volume,
it is important to have a correct detector geometry configuration to assess the
containment and acceptance of the particles produced in the signal events,
in particular in the cases with non-uniform beam intensities and with the beam
off the axis of the detector.
Further, a correct detector geometry will help to assess the impacts
or the systematic uncertainties from interactions with other materials in the
detectors, and from events occurring outside the active volume of the detectors,
while defining a fiducial volume can mitigate the impacts.

The official simulation toolkits used in an experiment,
such as \texttt{LArSoft} in SBN and DUNE,
typically have a driver to appropriately handle the detector geometry.
However, Monte Carlo generators for BSM signals are usually not fully integrated,
and the detector geometry hence needs to be carefully handled.
Instead of importing MC events into \texttt{LArSoft} and completing the full
detector simulation,
it would be helpful to design an interface for easier implementation of
event generator, taking into account the detector geometry in early stages.

Realistic sensitivity estimates are an important step in promoting and developing analyses with real data.
This typically requires kinematic distributions or simplified MC simulation, combined with parameters of detector efficiency and resolution.
Detector efficiency and resolution in neutrino experiments usually depend not only on the particle energy, but also on its direction and the relative location in the detector, and therefore it is complicated to obtain a parametrization matrix mapping all the phase space.
Nonetheless, it is desired to utilize realistic, approximate parametrization
matrices to quantify the sensitivity and compare with existing results. 
This parametrization needs to be consistent with an understanding of the detector geometry.

A fast detector response estimation will be an invaluable tool for both experimentalists and theorists, but in order to search for a model, a full modeling of particle trajectories through the detector volume and the resulting energy deposits is necessary. 
In particular, simulation of this response in LArTPCs with a multi-ten-ton scale is slow, limiting the number of signals that will be simulated.
Overcoming this bottleneck will be necessary to provide a broad program of dark sector searches at LArTPC experiments.

One intermediate possibility between a full detector simulation and a parametrized response is to use track and shower templates to model bypass the slow step of propagating the particles through the detector medium using \texttt{Geant4}. This could allow for study of more complicated in models in a realistic way and, perhaps, to determine experimental sensitivity to more models.

\textbf{MeV-scale signatures}

Many sub-GeV BSM particles can produce MeV-scale signatures.
The scope of BSM searches would be expanded if they were able to detect and reconstruct such MeV-scale deposits.
Furthermore, the search for exotic particles from their decay products, such as a muon-pion pair decaying from a hypothesized heavy neutral lepton, can also be separated from SM neutrino background by excluding MeV-scale particles around the decay vertex.
Lowering the detection threshold of a LArTPC can be achieved mainly by shortening the pitches of the charge detection system, and/or by enhancing the signal-to-noise ratio.
As a particle can be distinguished from noise by coincident energy depositions in two adjacent detection channels, smaller pitches (below the typical 3 or 4~\si{\milli\meter}) allow detection of particles with lower energy.
A large signal-to-noise ratio would further raise the likelihood to identify particles with smaller energy deposit than minimal ionization particles, which either leave a shorter track than the pitch width, or hypothetically carries fractional charge such as millicharged particles discussed in Sec.~\ref{ssec:mCP_Search}. 

A robust signal processing algorithm is required to disentangle the energy deposition and the effect of the induced charge from the adjacent channels, and provide more accurate information for hit finding and reconstruction algorithms.
It is also desired to increase the reconstruction efficiency of MeV-scale energy deposits.

\textbf{Reconstruction of complex topologies}

Several topologies are of particular interests in search for BSM signals.
They can also occur in a SM neutrino interaction, and reconstruction algorithms optimizing to such topologies will benefit the SM measurements.
For example, it would be very useful to characterize a electromagnetic shower containing an electron and a positron from the decay of a boosted, long-lived scalar particle, as illustrated in Sec.~\ref{subsec:Benchmark:HiggsPortal}. 
Such characterization will improve the reconstruction of $\pi^0$, the main background of $\nu_e$ appearance measurements.
In some cases, optimizing the reconstruction efficiency of particles coming from non-standard beam direction, such as MicroBooNE with respect to a particle produced at the hadron absorber of the NuMI beam (a nearly opposite direction of the BNB neutrino beam), would help improve the sensitivity.

For up-scattered BSM signals, an interaction and decay vertices occurring in coincidence is a distinctive feature, and, in LArTPCs, only the light signals have the required time resolution.
It is therefore crucial to have charge-light matching algorithms dealing with these cases. Reconstructing this kind of signature may be particularly important for surface LArTPCs.
For example, the interaction or decay point may only attach to a single track or an electromagnetic shower, and does not literally form a ``vertex,'' making the events similar to an interaction overlaid with a cosmic ray track.
Nonetheless, it is another example which can benefit a SM neutrino measurement.
A neutrino interaction producing an electromagnetic shower can have a radiated photon which travels a few tens of centimeters and then starts the pair production;
such a case would also have better resolution by correctly associate the pair
produced electrons and positrons with the original electromagnetic shower.

\subsection{Action Plan}\label{sec:action-plan}

\textbf{Simulation pipeline}

In order to streamline the process of simulating different dark sector models and increase coverage of model space, a seamless and modular simulation pipeline is necessary. As indicated in Table \ref{tab:simulation}, there are several stages of the pipeline and different tools for generation of the relevant physics at each. 

As an example, it would be highly useful to be able to interface the output of dark matter scattering in the detector with different models of detector simulation, from parametrized response to full detector simulation. Another use case is that nuclear final state interactions and modeling of low-energy inelastic scattering can have significant effects on sensitivity to hadronically interacting dark sector particles, so it is important to also have the flexibility to connect with different models of nuclear and hadronic physics. A framework to connect different models of dark sector physics with this challenging standard model physics is needed.  A prototype of how this framework could be expressed is outlined in Ref.~\cite{Isaacson:2021xty}, and implemented within the \texttt{ACHILLES} tool. This framework separates the nuclear current from the leptonic current, providing a realistic strategy for interfacing arbitrary dark matter models with arbitrary nuclear models.

At present, these tools do not easily interface with one another or with the pipeline used by neutrino experiment full simulation and reconstruction, including notably \texttt{LArSoft}. 

There are several ways in which this process could be streamlined.  
One option is a fully \texttt{Geant4}-compatible package for the dark matter simulation which is responsible for production of dark matter and its interaction inside (or outside) of the detector geometry. Collaborators of NA64 experiment at CERN developing such \texttt{Geant4} extension package called \texttt{DMG4} \cite{Bondi2021}, for 100 GeV electron beam and the newly added physics processes in this package are about production of dark matter via electron and muon bremsstrahlung off nuclei, resonant in-flight positron annihilation on atomic electrons and gamma to axion-like particles conversions on nuclei. Currently, the \texttt{DMG4} package support dark matter productions from only electron and muon beam but if we can extend this for proton beam, this will allow us to formulate an efficient simulation pipeline and also provide additional opportunity of cross check about the simulation results.

Similar full simulation from the production of new physics species to the experimental signature, albeit currently with less detector simulation details than \texttt{Geant4} could offer, can be performed for far-forward physics searches at the LHC with a dedicated \texttt{FORESEE} tool~\cite{Kling:2021fwx}. Currently, the package offers the possibility to compare expected signal rates in selected BSM models for different detector types, new physics models and experimental signatures that can be varied by the user. A modified version of \texttt{FORESEE} is also included in the FASER collaboration simulation pipeline, therefore making the tool an ideal interface between theory and experiment.

Perhaps more flexible would be to standardize the output of each step of the chain described by Table \ref{tab:simulation}. 
The standard format for collider high-energy experiments such as those at the Large Hadron Collider is the \texttt{HepMC3}~\cite{Buckley:2019xhk} format, which contains enough flexibility to describe the output at each step above given some common standards.

This does not necessarily solve all the problems, such as the problem of connecting to different nuclear models and fully tracking systematic and theoretical uncertainties through the process, but it would be an excellent start. The \texttt{HepMC3} format offers one way to handle the propagation of systematic and theoretical uncertainties through the simulation pipeline, though further work is needed to accurately quantify these uncertainties and build them into the pipeline.

A process should be started to assess how to best standardize the simulation pipeline and allow for easy interface between different models for the various stages of simulation. As a first step, a workshop should be held to bring together experts on the various stages of simulation, as well as experimentalists and theorists who use these tools, to recommend a set of standards and discuss their implementation.

It would also be helpful to ease connection of dark sector models into the pipeline. None of the tools currently available allow for simple implementation of new, fully general models. \texttt{MadDump} does allow for implementation of models using the standard \texttt{FeynRules} package and \texttt{UFO} model description format, but cannot cover models that have arbitrary kinds of interactions, such as those with multiple vertices in the detector. \texttt{ACHILLES} also allows for implementation of models using the standard \texttt{FeynRules} package and \texttt{UFO} model description format, but is currently unable to handle particles with spin $>$ 1.  Other tools generally focus on specific types of dark sector particles and interactions. In order to ensure that the dark sector models developed over the last several years are studied in upcoming searches, it is important that more models are implemented. While recent work has made large strides in this direction, there still exists a barrier to a robust dark matter search program with neutrino beams.

\textbf{Fast simulation and response}

Many of the tools that would be helpful for simulating more complicated production modes, such as via decay of long-lived mesons, are developed by experimental collaborations and have a high barrier to entry. Incorporating these processes in to the more theory-focused tools is challenging, as they are dependent on the specific beam and target geometries. Furthermore, complex analysis tools such as fine-grained timing of signals can be valuable and is also highly dependent on the beam geometry.  By standardizing the format, theorists could be free to implement their model using the tool of their choice of the underlying particle physics, with confidence that detailed predictions for experiments could be derived.

Detector simulation and reconstruction are particularly intensive for upcoming LArTPC experiments. Furthermore, the tools to perform these stages of the simulation chain are not accessible to theorists, leading to crude estimations of sensitivity. Development of fast simulation and reconstruction, either by a simplified \texttt{Geant4}-based approach or by parameterized efficiencies and resolutions, is needed by both the theory and experiment communities.

\textbf{New trigger and reconstruction algorithms}

On the experimental side, the wealth of complicated signals demands new reconstruction algorithms. The trigger system will need the flexibility to handle out-of-time signals. Timing can also help in distinguishing signals from backgrounds, but, in LArTPC-based experiments for example, this requires careful matching of light and charge signals. Low energy signals will require new reconstruction strategies and the full capabilities for going below ``standard'' reconstruction thresholds is unknown. Algorithms to pick out complex topologies with multiple, separated vertices, such as from dark sector up-scatter followed by decay, also need to be developed. Work on all of these is ongoing and should be a high priority as the scope of dark sector searches grows.



\section{Outlook
}
\label{sec:outlook}


During the next decade and beyond, a variety of neutrino beam experiments will address basic questions about the nature of neutrinos. As we have spotlighted in this whitepaper, these experiments also have the capability to pursue a broad program of BSM searches for new light, weakly coupled states that are part of a dark sector. 

Dark sector theories are motivated on several accounts, as they may address some of the outstanding mysteries in fundamental physics, including dark matter and neutrino masses, or explain certain experimental anomalies. A wide-ranging experimental program to probe the dark sector is emerging, and there has been a growing realization that neutrino beam experiments are a vital part of this enterprise. This is a consequence of their intense beams/sources, large neutrino and secondary fluxes, and cutting-edge sensitive detectors.

Substantial progress has been made in recent years in terms of the phenomenological and experimental developments needed to pursue a viable dark sector search program at neutrino beam facilities, but there is still work to be done. We have called attention to some of these needs, including accurate and fast simulation tools that can be easily integrated into experimental analyses, novel reconstruction and analyses methods, and improved modeling of neutrino-nucleus interactions. 

We anticipate this whitepaper will provide a broad snapshot of the state of this subfield along with guidance on the next steps needed to realize a robust dark sector search program at existing and future neutrino beam facilities. 



\section*{Acknowledgements}
\addcontentsline{toc}{section}{Acknowledgements}

B. Batell is supported by U.S. Department of Energy (DOE) Grant DE–SC0007914 and by PITT PACC.
F.~Kling acknowledges support by the Deutsche Forschungsgemeinschaft under Germany's Excellence Strategy - EXC 2121 Quantum Universe - 390833306.
WJ and JY are supported by the U.S. DOE Grant No. DE-SC0011686.
JCP is supported by the National Research Foundation of Korea (NRF) [NRF-2019R1C1C1005073 and NRF-2021R1A4A2001897].
S. Shin is supported by the National Research Foundation of Korea (NRF)
[NRF-2020R1I1A3072747 and NRF-2022R1A4A5030362].
TBS is supported by the National
Science Foundation Graduate Research Fellowship Program under Grant No. 1839285. 
VDR acknowledges financial support by the SEJI/2020/016 grant funded by
Generalitat Valenciana, by the Universitat de Val\`encia through the sub-programme “Attracci\'o de talent 2019” and by the Spanish grant PID2020-113775GB-I00 (AEI/10.13039/501100011033). S.~Trojanowski is supported by the grant ``AstroCeNT: Particle Astrophysics Science and Technology Centre'' carried out within the International Research Agendas programme of the Foundation for Polish Science financed by the European Union under the European Regional Development Fund, by the Polish Ministry of Science and Higher Education through its scholarship for young and outstanding scientists (decision no 1190/E-78/STYP/14/2019), and by the European Union's Horizon 2020 research and innovation programme under grant agreement No. 952480 (DarkWave 24 project).


\clearpage


\renewcommand{\refname}{References}

\printglossary

\bibliographystyle{utphys}

\bibliography{common/tdr-citedb}

\end{document}